\documentclass[12pt]{article}

\usepackage{graphicx}

\def\rf#1{(\ref{eq:#1})}
\def\lab#1{\label{eq:#1}}
\def\nonu{\nonumber}
\def\br{\begin{eqnarray}}
\def\er{\end{eqnarray}}
\def\be{\begin{equation}}
\def\ee{\end{equation}}
\def\({\left(}
\def\){\right)}

\def\u2{\mid u\mid^2}
\def\rlx{\relax\leavevmode}
\def\IR{\rlx\hbox{\rm I\kern-.18em R}}
\def\IZ{\rlx\hbox{\sf Z\kern-.4em Z}}

\def\vp{\varphi}
\def\ra{\rightarrow}

\newcommand{\sbr}[2]{\left\lbrack\,{#1}\, ,\,{#2}\,\right\rbrack}
\newcommand\bra[1]{\langle \, {#1}\, \mid}
\newcommand\ket[1]{\mid \, {#1} \, \rangle}

\def\FAaIA#1#2#3{{\sl Functional Analysis and Its Application} {\bf #1} (#2)
#3}

\def\PHSD#1#2#3{{\sl Physica} {\bf D#1} (#2) #3}
\def\PJA#1#2#3{{\sl Proc. Japan. Acad} {\bf #1A} (#2) #3}

\begin{document}

\begin{titlepage}
\vspace*{-1cm}

\vskip 3cm

\vspace{.2in}
\begin{center}
{\large\bf The Bullough-Dodd model coupled to matter fields}
\end{center}

\vspace{.5cm}

\begin{center}
P. E. G.  Assis~\footnote{Present address: Centre For Mathematical
  Science, City University, Northampton Square, London EC1V OHB, UK}
  and L. A. Ferreira

\vspace{.5 in}
\small

\par \vskip .2in \noindent
Instituto de F\'\i sica de S\~ao Carlos, IFSC/USP\\
Universidade de S\~ao Paulo  \\ 
Caixa Postal 369, CEP 13560-970, S\~ao Carlos-SP, Brazil\\

\normalsize
\end{center}

\vspace{.5in}

\begin{abstract}

The Bullough-Dodd model is an important two dimensional integrable
field theory which finds applications in physics and geometry. We
consider a conformally invariant extension of it, and 
study  its integrability properties using a zero curvature condition
based on the twisted Kac-Moody algebra $A_2^{(2)}$. The one and
two-soliton solutions as well as the breathers are 
constructed explicitly .   
We also consider integrable extensions of the Bullough-Dodd model by the
introduction of spinor (matter) fields. The resulting theories are
conformally invariant and present local internal symmetries. All the
one-soliton solutions, for two examples of 
those models, are constructed using an hybrid of the dressing and
Hirota methods. One 
model is of particular interest because it presents a confinement
mechanism for a given conserved charge inside the solitons.

\end{abstract} 
\end{titlepage}

\section{Introduction}
\setcounter{equation}{0}
 
Solitons play a role in many non-perturbative aspectos of field 
theories. They appear as classical solutions of the equations of
motion of some special non-linear theories. In general, their masses
and interaction strengths are inversely related to the coupling
constant for the original fields of the theory. Therefore, they are
weakly interacting and easy to excite at the strong coupling limit,
and so are natural candidates to describe the normal modes of the
theory in that regime. In the particular case of some supersymmetric
gauge theories the solitons (dyons) play a crucial role in
electromagnetic duality conjectures involving the weak and strong
coupling regimes \cite{duality}.

Two dimensional integrable field theories possessing soliton solutions 
constitute, besides their intrinsic beauty, a laboratory to test ideas
and develop exact methods on non-perturbative aspects of physical
theories. In addition, two dimensional models have direct applications in
many areas of physics and non-linear sciences, like condensed matter,
non-linear  optics, etc. Classical soliton theory in two dimensions is
 quite well developed \cite{babelonbook}. Practicaly all soliton
 solutions appear in 
 models that admit a representation of their equations of motion in
 terms of the so-called zero curvature condition or
 Lax-Zakharov-Shabat equation \cite{lax}, with the flat
 connections  living on Kac-Moody algebras
 \cite{kac}. That equation   leads to exact methods for 
 constructing  solutions and infinite number of conserved
 charges. An important point to emphasize is that the symmetries
 responsible for the appearance of solitons are symmetries of the
 zero curvature equation, and not necessarily of the equations of
 motions or of the Lagrangean. For that reason the conserved charges
 are not of the Noether type.

In this paper we consider special integrable extensions of the Bullough-Dodd
model \cite{bd}. That is a relativistic invariant
theory in two dimensions involving 
just one scalar field, which first appeared in the literature  in the context
of hyperbolic surfaces in $\IR^3$ \cite{firstbd}. It is an integrable
filed theory with solitons and some of their properties was considered
by Zhiber and Shabat \cite{zhiber} and  Mikhailov
\cite{mikhailov}. The integrabily properties follow from the fact that
the Bullough-Dodd equation admits a zero curvature representation with
flat connections taking values on the twisted affine Kac-Moody algebra
$A_2^{(2)}$, but with vanishing central term.  
The first extension we consider is on the lines of
\cite{bb,afgz,cfgz}, and which we call the  Conformal Bullough-Dodd
model. It involves the addition of two extra fields which makes the
theory conformally invariant, and imposes the zero curvature
condition to live on the algebra $A_2^{(2)}$, but
with non-trivial central term. That fact plays a crucial role in the
integrability properties of the model. The representation theory of
Kac-Moody algebras changes drastically when the central term is not zero. It
admits highest weight state representations which are an important
tool in the construction of exact solutions. We then construct the one and
two soliton solutions of the Bullough-Dodd using an hybrid of the
dressing method \cite{dressing} and the Hirota method \cite{hirota} as
proposed in \cite{bueno}. The physically interesting solitons exist
when the Bullough-Dood scalar field is taken to be complex. We also
discuss the construction of complex breather solutions. 

The second extension of the Bullough-Dood model that we consider
involves the addition of 
spinor (matter) fields following the ideas of \cite{matter}. That is
achieved by considering the principal (integer) gradation of the
algebra $A_2^{(2)}$ \cite{kac}. The flat connection corresponding to
the (Conformal) Bullough-Dodd model has components in the
eigensubspaces of grades $0$ and $\pm 1$. We extend that connection
by allowing it to have components with grades varying from $-l$ to $l$
with $l$ being a positive integer. The equations of motion of the
model are obtained by imposing the connections to satisfy the zero
curvature condition. Therefore, the models are integrable by construction. 
As shown in \cite{matter}, the
fields appearing in the non-zero grade components of the connections
transform as two 
dimensional Dirac spinors, and they couple to exponentials of the
Bullough-Dodd scalar field. The components of the connection with
grades $\pm l$ are taken to be constant, and they play a crucial role
in determining the physical properties of the resulting
model. We consider two models of that type corresponding to the
choices $l=3$ and $l=6$. The models with $l=2,4,5$ do not seem to
possess solitons. Again using an hybrid of the dressing and Hirota methods
we construct all the one-soliton solutions for those two models. 

The model corresponding to $l=6$ however is analysed in much more
detail because of its interesting physical properties. Besides, the
three scalar fields of the Conformal Bullough-Dood model, it possesses
seven Dirac spinor fields. The equations of motion for them are Dirac
equations with  interaction terms involving the product of
exponetials of the scalar fields with bilinear terms in the
spinors. For that reason the model does not possesses a 
Lagrangean which is local in those fields. The theory however, is
conformally 
invariant and possesses a local $U_L(1)\otimes U_R(1)$ chiral (gauge)
symmetry group. Although we can not use the Noether theorem, due to
the lack of Lagrangean, there exist two conserved currents
corresponding to that symmetry group. In addition, the conservation of
those currents is equivalent to the existence of two chiral currents,
depending on only one light cone variable each. A remarkable property
of that model is that all the one-soliton solutions belong to a
submodel where those two chiral currents vanish. On that submodel
there exists  a conserved current, depending only on the spinor
fields, 
which is exactly equal to the topological current involving the
Bullough-Dodd scalar field. Such an equivalence of matter and
topological currents leads to a confinement of  the corresponding
matter charge inside the solitons. As explained in \cite{matter} that
confinement mechanism is special to some Toda models coupled to matter
fields, and have been studied in some interesting cases
\cite{achic,blas,bueno}. In particular, in \cite{achic} it was shown
that it provides a very interesting way of understanding the quantum
equivalence between the sine-Gordon and Tirring models. From the
algebraic construction of \cite{matter}, the model for
$l=6$ that we discuss in this paper stands for the Bullough-Dodd model in
the same way as the model considered in \cite{achic} stands for the
sine-Gordon. We then hope that our work may help understanding if
there exist a theory which would be equivalent to the the
Bullough-Dodd model, in such way that the elementary excitations of its
fields would correspond to the Bullough-Dodd solitons. 

Our paper is organized as follows. In section \ref{sec:cbd} we
introduce the Conformal Bullough-Dodd model, discuss its integrability
properties and construct its one and two soliton solutions as well as
the breathers. The dressing tranformation methos is discussed in
section \ref{sec:dressing}. The  construction of the models involving
the matter (spinor) fields through the extension of the zero curvature
connections is discussed in section \ref{sec:coupling},
as well as the symmetries of the resulting theories.  The model
corresponding to $l=6$ is described in section \ref{sec:l=6}. The
symmetries and the equivalence of the matter and topological currents
is discussed in section \ref{subsec:symmetries}. The method for
constructing the soliton solutions for such model is described in
section  \ref{subsec:solutions}, and the particular one-soliton
solutions are given in sections \ref{sec:solvac1} and
\ref{sec:solvac2}. The model corresponding to $l=3$ is described in
section \ref{sec:l=3} as well as its soliton solutions.  Some basic
facts about the twisted Kac-Moody algebra $A_2^{(2)}$ is given in appendix
\ref{app:kacmoody}. The coefficients for the Hirota's tau functions
corresponding to the one-soliton of section \ref{subsec:mass1} is
given explicitly in appendix \ref{app:fullsol}, and some
simplifications of it in the appendix \ref{app:truncation}.

\section{The Conformal Bullough-Dodd model}
\label{sec:cbd}
\setcounter{equation}{0}

The Bullough-Dodd model is a $1+1$ dimensional relativistic 
field theory  described by the equation 
\be
\partial^2 \vp = - e^{ \vp } + e^{-2 \, \vp }
\lab{bdeqonly}
\ee
where $\partial^2 \equiv \partial_{\mu}\partial^{\mu}=\frac{1}{c^2}
\partial_t^2-\partial_x^2$. 
Together with the sine-Gordon model, it is one of the few relativistic 
invariant integrable
theories with just one scalar field. The integrability properties of
such theory were first discovered  by
Bullough and Dodd \cite{bd} who showed the existence of non-trivial
conservation laws. Zhiber and Shabat \cite{zhiber} found an infinite 
Lie-Backlund group of transformations for \rf{bdeqonly}, and Mikhailov
\cite{mikhailov} 
studied it in the context of the Toda models using inverse scattering
method. However, the first time eq. \rf{bdeqonly} appeared in the
literature was on a paper by Tzitzeica \cite{firstbd} in the context
of hyperbolic surfaces in $\IR^3$. See also \cite{sharipov,wang} for
more details.  

Here we consider an extension of the Bullough-Dodd model \rf{bdeqonly} on
the lines of \cite{bb,afgz,cfgz}, which we call the  Conformal
Bullough-Dodd model, and which  is defined by the equations of 
motion\footnote{A coupling constant $\beta$, mass parameter $m$, and
  relative coefficients for the exponential terms can be introduced
  into the equations by the redefinitions $\vp \ra \beta \vp - \ln
  \mu$, $x^{\mu} \ra m\, x^{\mu}$, in such way that the  eq. for $\vp$
  would read  
$\partial^2 \vp = - \frac{m^2}{\mu \beta}\; e^{\eta}\; \(  e^{\beta
    \vp } - \mu^{3} e^{-2 \, \beta\; \vp }\)$} 
\br
\partial^2 \vp &=& -e^{\eta}\, \( e^{ \vp } - e^{-2 \, \vp }\)  \nonu\\
\partial^2 \eta &=& 0 
\lab{cbdeqs}\\
\partial^2 \nu &=& - \frac{1}{2}\, e^{-2\, \vp +\eta}\nonu
\er
and corresponding Lagrangean 
\be
{\cal L} = \frac{1}{2}\, \partial_{\mu}\vp\, \partial^{\mu}\vp +
\partial_{\mu}\vp\, \partial^{\mu}\eta + 3\, \partial_{\mu}\eta\,
\partial^{\mu}\nu - \( e^{\vp+\eta}+\frac{1}{2}\,e^{-2\,\vp+\eta}\) 
\lab{cbdlag}
\ee 
The introduction of the field $\eta$ renders the theory conformally
invariant. Indeed, introducing the light cone coordinates 
\be
x_{\pm} \equiv \frac{c\, t\pm x}{2} \qquad\qquad \partial_{\pm} =
\frac{1}{c} \partial_t \pm \partial_x \qquad \qquad 
\partial_{+}\partial_{-} = \partial^2=\frac{1}{c^2} \partial_t^2 - \partial_x^2
\lab{lightcone}
\ee
one can check that \rf{cbdeqs} and \rf{cbdlag} are invariant under the
conformal transformations 
$x_{\pm}\rightarrow f_{\pm}\(x_{\pm}\)$ 
if the field $\vp$ is a scalar under the conformal group
and if $e^{-\eta}\rightarrow f^{\prime}_{+}
f^{\prime}_{-}e^{-\eta}$. The conformal weights of $\nu$ are
arbitrary \cite{bb,afgz,cfgz}. 

The field $\nu$ is just an expectant in the sense that it does not
really 
influence the dynamics of $\vp$ and $\eta$. However, it plays a
crucial role in the integrability properties of the model. As we
explain below, the construction of exact solutions under the dressing
method makes use of highest weight representations of the Kac-Moody algebra,
and that requires the existence of a non-trivial central
extension, and the consequent introduction the extra field $\nu$.  

The relevant algebra for the integrability properties of the
Bullough-Dodd model \rf{bdeqonly} and its conformal extension
\rf{cbdeqs} is the twisted affine Kac-Moody algebra $A_2^{(2)}$
  \cite{kac}. The commutation
  relations and a brief description of its properties are given in
  appendix \ref{app:kacmoody}. What makes the theories \rf{bdeqonly}
  and \rf{cbdeqs} integrable is the fact that they admit a
  representation of their eqs. of motion in terms of a zero curvature
  condition or the Lax-Zakharov-Shabat equation 
\be
\partial_{+}A_{-}- \partial_{-}A_{+} + \sbr{A_{+}}{A_{-}} = 0
\lab{zc}
\ee
where $\partial_{\pm}$ are derivatives w.r.t. the light cone
coordinates defined in \rf{lightcone}. 
For the Conformal Bullough-Dodd model \rf{cbdeqs} the potentials are given by 
\br
A_{+}&=& - B\, \Lambda_{+}\, B^{-1} = -\( \frac{\sqrt{2}}{2}\,
e^{\vp+\eta}\, T_3^0 + e^{-2\vp+\eta} \, L_{-2}^{1/2}\)\nonu\\ 
A_{-}&=& - \partial_{-} B\, B^{-1} +\Lambda_{-} = -\(
\partial_{-}\vp\, T_3^0+ \partial_{-}\eta\, Q+ \partial_{-}\nu\, C\) 
+\Lambda_{-}
\lab{zcpotbd}
\er
where 
\be
B= e^{\vp\, T_3^0+\eta\, Q+ \nu\, C}
\lab{bfield}
\ee
and
\be
\Lambda_{+}= \frac{\sqrt{2}}{2}\, T_+^0+ L_{-2}^{1/2} \qquad\qquad 
\Lambda_{-}= \frac{\sqrt{2}}{2}\, T_-^0+ L_{2}^{-1/2} 
\lab{lambdapm}
\ee
The operators $T_3^0$, $T_{\pm}^0$, $L_{\pm 2}^{\mp 1/2}$, $Q$, and
$C$ are generators of the  twisted affine Kac-Moody algebra
$A_2^{(2)}$ (see appendix
\ref{app:kacmoody}). $Q$ is the grading operator defined in
\rf{gradingop}, defining the so-called principal gradation of
$A_2^{(2)}$ (see \rf{gradation} and \rf{eigensub}). Notice that $B$ is
an element of the zero grade 
subgroup, i.e. the one obtained by exponentiating the zero grade
subalgebra ${\cal G}_0$ (see \rf{gradation} and \rf{eigensub}). The
elements $\Lambda_{+}$ and $\Lambda_{-}$ have grades $+1$ and $-1$
respectively, and satisfy
\be
\sbr{\Lambda_{+}}{\Lambda_{-}}= \frac{1}{2}\, C 
\ee
One can check that by replacing \rf{zcpotbd} into \rf{zc} all the
non-zero grade components vanish automatically, and the zero grade
component leads to the equation 
\be
\partial_{+}\( \partial_{-}B\, B^{-1}\) =
\sbr{\Lambda_{-}}{B\,\Lambda_{+}\,B^{-1}} 
\lab{cbdtodaeq}
\ee
The three components of \rf{cbdtodaeq}, in the direction of $T_3^0$, $Q$
and $C$, are the equations for the fields $\vp$, $\eta$, and $\nu$
respectively, given in \rf{cbdeqs}. 

The zero curvature representation for the usual Bullough-Dodd model
\rf{bdeqonly} is obtained from \rf{zcpotbd} by setting $\eta =0$ and
working with a representation of the algebra $A_2^{(2)}$ where
$C=0$. Then \rf{cbdtodaeq} will have just one component in the
direction of $T_3^0$ which corresponds to \rf{bdeqonly}. The algebra
$A_2^{(2)}$ with $C=0$, a so-called loop algebra, admits finite matrix
representations depending upon a complex  parameter (the so-called
spectral parameter). 

Notice that, if one allows the field $\vp$ to be complex, i.e. $\vp =
\vp_R+i\, \vp_I$, the
eqs. \rf{bdeqonly} and \rf{cbdeqs} are invariant  under the discrete
transformations 
\be
\vp_R \ra \vp_R \qquad\qquad \vp_I \ra \vp_I +  2 \, \pi\, n
\ee
with $n$ being an integer. Therefore, the theories have a degeneratre
vacua which allow the existence of non-trivial topological charges
defined by
\be
Q_{\rm top.} =  \int_{-\infty}^{\infty}dx\, j^0 =
\frac{1}{2\,\pi}\,\left[\vp_I\(x=\infty\)-\vp_I\(x=-\infty\)\right]
\lab{topolcharge}
\ee
where 
\be
j^{\mu}=\frac{1}{2\,\pi}\,\varepsilon^{\mu\nu}\,\frac{\partial\,
  \vp_I}{\partial x^{\nu}} 
\ee
with $\varepsilon^{\mu\nu}$ being antisymmetric, 
$\varepsilon^{01}=1$, $x^0=c\,t$, and $x^1=x$.

\subsection{The dressing method and the solitons solutions}
\label{sec:dressing}

The zero curvature \rf{zc} is invariant under gauge transformations of
the form: $A_{\mu} \ra g\, A_{\mu}\,g^{-1}-\partial_{\mu} g\,
g^{-1}$. The dressing transformations \cite{dressing} are special
types of gauge transformations that constitute maps among the
solutions of \rf{zc} or equivalently \rf{cbdeqs}. We shall use a
vacuum solution as the seed for the method, and construct the
corresponding orbit of
solutions under the dressing transformation group. The vacuum solution
of \rf{cbdeqs} we take is  
\be
\vp^{\rm vac} =0 \qquad \qquad \eta^{\rm vac} = 0 \qquad \qquad 
\nu^{\rm vac} = -\frac{1}{2}\, x_{+}\, x_{-} 
\lab{vacuumcbd}
\ee 
The zero curvature potentials evaluated on such solution can be written as
\be
A_{\mu}^{\rm vac} = - \partial_{\mu} \Psi_{\rm vac} \, \Psi_{\rm
  vac}^{-1}
\lab{vaccbdpot}
\ee
with 
\be
\Psi_{\rm vac} = e^{x_{+}\,\Lambda_{+}}\, e^{-x_{-}\,\Lambda_{-}}
\lab{psivaccbd}
\ee
with $\Lambda_{\pm}$ defined in \rf{lambdapm}. 
We now take a constant group element $h$ such that there exist the
Gauss type decomposition 
\be
\Psi_{\rm vac}\, h \, \Psi_{\rm vac}^{-1} = G_{-}\, G_{0}\, G_{+}
\lab{hdecomp}
\ee
where $G_{-}$, $ G_{0}$ and $G_{+}$ are groups elements obtained
by exponentiating the generators of $A_2^{(2)}$ with  negative, zero
and positive grades respectively, of the principal gradation
\rf{gradation} defined by $Q$ given in \rf{gradingop}. 
 We introduce the element
\be
\Psi_h \equiv G_{0}^{-1}\, G_{-}^{-1}\, \Psi_{\rm vac} \, h= G_{+}\,
\Psi_{\rm vac} 
\lab{psihdef}
\ee
and the transformed potentials
\be
A_{\mu}^{h} \equiv -\partial_{\mu}\Psi_h \, \Psi_h^{-1}
\lab{amuhdef}
\ee
The fact that $A_{\mu}^{h}$ is of the pure gauge form guarantees that
it is a solution of the zero curvature condition \rf{zc}. In addition,
the fact that $\Psi_h$ can be written in two different ways in terms
of $\Psi_{\rm vac}$, guarantees that $ A_{\mu}^{h}$ has the same
grading structure as the potentials \rf{zcpotbd}. Indeed, \rf{psihdef}
and \rf{amuhdef} implies that 
\br
A_{\mu}^{h} &=& G_{+}\, A_{\mu}^{\rm vac}\,G_{+}^{-1}-\partial_{\mu}
G_{+}\, G_{+}^{-1} 
\lab{postransf}\\
&=& \(G_{0}^{-1}\, G_{-}^{-1}\)\, A_{\mu}^{\rm vac}\,\(G_{0}^{-1}\,
G_{-}^{-1}\)^{-1}-\partial_{\mu} 
\(G_{0}^{-1}\, G_{-}^{-1}\)\, \(G_{0}^{-1}\, G_{-}^{-1}\)^{-1} 
\lab{negtransf}
\er
Eq. \rf{postransf} implies that $A_{\mu}^{h}$ has components of grades 
greater or equal than those of $ A_{\mu}^{\rm vac}$. On the other
hand, \rf{negtransf}  implies that $A_{\mu}^{h}$ has components of
grades 
smaller or equal than those of $ A_{\mu}^{\rm vac}$. Since
\rf{hdecomp} and \rf{psihdef} guarantee that both relations hold true
it follows 
that $A_{\mu}^{h}$ has the same grade components as $ A_{\mu}^{\rm
  vac}$, and so as the potentials \rf{zcpotbd}. By construction we
have $A_{\mu}^{h}$ given explicitly in terms of the space-time 
coordinates $x_{\pm}$. Therefore by equating it to \rf{zcpotbd}  we get the
solutions for the fields of the model, associated to the choice $h$ of
the constant group element, i.e. a point on the orbit of solutions of the
vacuum \rf{vacuumcbd}. 
In order to get the explicit solution we proceed as follows. From
\rf{vaccbdpot} and \rf{negtransf} we have that
\be
A_{-}^h = -\partial_{-} G_{0}^{-1}\, G_{0} -
G_{0}^{-1}\,\partial_{-}G_{-}^{-1}\, G_{-}\, G_{0} + \frac{1}{2}\,
x_{+}\, C +  
G_{0}^{-1}\,G_{-}^{-1}\, \Lambda_{-}\, G_{-}\, G_{0}
\lab{amhbd}
\ee
and so its zero grade part is
\be
\(A_{-}^h\)_0 = -\partial_{-} G_{0}^{-1}\, G_{0} + \frac{1}{2}\, x_{+}\, C = 
-\partial_{-} \( e^{-\frac{1}{2}x_{+}x_{-}C} \,G_{0}^{-1}\)
\;\(G_{0}\,e^{\frac{1}{2}x_{+}x_{-}C}\) 
\lab{zerogradeamh}
\ee
Comparing \rf{zerogradeamh} with the zero grade part of $A_{-}$ in
\rf{zcpotbd}, we get that 
\be
B= e^{-\frac{1}{2}x_{+}x_{-}C} \,G_{0}^{-1}
\ee
and so using \rf{bfield}, we have 
\be
G_{0} = e^{-\vp\, T_3^0- \(\nu+\frac{1}{2}x_{+}x_{-}\)\, C}
\lab{g0form}
\ee
Notice that, from the commutation relations of the appendix
\ref{app:kacmoody},  the operator $D$, and so $Q$, is never produced
by the commutator of any pair of generators of the algebra. Therefore,
if one starts with a vacuum solution with $\eta =0$, the dressing
method will never produce a solution with $\eta\neq 0$. Therefore,
$\eta$ does not appear in \rf{g0form}. 

In order to obtain the explicit expression for the solutions for the
fields we make use of highest weight representations of $A_2^{(2)}$
and introduce the Hirota's tau functions as the expectation value of
\rf{hdecomp}, i.e.  
\be
\tau_{\lambda} \equiv \bra{\lambda} \Psi_{\rm vac}\, h \, \Psi_{\rm
  vac}^{-1}\ket{\lambda} =  
\bra{\lambda} G_{0}\ket{\lambda}
\lab{hirtaudefcbd}
\ee
where in the second equality we have used the fact that the highest
weight state $\ket{\lambda}$ satisfies \rf{higheststate}, and so 
 $G_{+}\ket{\lambda}= \ket{\lambda}$, and $\bra{\lambda}\, G_{-}=
\bra{\lambda}$.  We shall use two of such representations and introduce
two tau functions. First $\tau_0$ associated to the choice
$\ket{\lambda}\equiv \ket{\lambda_0}$, and $\tau_1$ associated to the
choice $\ket{\lambda}\equiv \ket{\lambda_1}\otimes\ket{\lambda_1}$,
where $\ket{\lambda_0}$ and $\ket{\lambda_1}$ are defined in
\rf{fundrep}. Then,  from \rf{g0form}, \rf{hirtaudefcbd} and
\rf{fundrep}, we have that 
\be
\tau_0 = e^{- 2\(\nu+\frac{1}{2}x_{+}x_{-}\)} \qquad \quad 
\tau_1 = e^{-\vp - 2\(\nu+\frac{1}{2}x_{+}x_{-}\)}
\ee
or equivalently
\be
\vp =  \ln \frac{\tau_0}{\tau_1} \qquad \qquad \nu = -\frac{1}{2} \,
\ln \tau_0 - \frac{1}{2} \, x_{+}x_{-}
\lab{taudefbd}
\ee
Replacing into the eqs. of motion \rf{cbdeqs} (with $\eta=0$) we get
the Hirota's equations 
\br
\tau_0 \partial_{+}\partial_{-}\tau_0 - \partial_{+} \tau_0
\,\partial_{-}\tau_0 &=& \tau_1^2-\tau_0^2 \nonu\\ 
\tau_1 \partial_{+}\partial_{-}\tau_1 - \partial_{+} \tau_1
\,\partial_{-}\tau_1 &=& \tau_0 \,\tau_1-\tau_1^2 
\lab{hirotaeqcbd}
\er
The solutions on the orbit of the vacuum \rf{vacuumcbd} we are
interested in, are those obtained by the so-called solitonic
specialization procedure \cite{olivesolitonicspec}. We take the
constant group element $h$ to a product of exponentials of eigenvectors of
the operators $\Lambda_{\pm}$ defined in \rf{lambdapm}, i.e. 
\be
h = \prod_{j=1}^N e^{V_j}\qquad \qquad
\sbr{\Lambda_{\pm}}{V_j}=\beta_j^{(\pm)}\, V_j
\lab{hnsolitonbd}
\ee
Then, using \rf{psivaccbd}, one gets 
\be
\tau_{\lambda}=\bra{\lambda} \Psi_{\rm vac}\, h \, \Psi_{\rm
  vac}^{-1}\ket{\lambda} = \bra{\lambda}\prod_{j=1}^N
e^{e^{\Gamma_j}\,V_j} \ket{\lambda} 
\lab{tausolitonicspec}
\ee
with $\Gamma_j = \beta_j^{(+)}\, x_{+} - \beta_j^{(-)}\, x_{-}$. The
integer $N$ defines the $N$-soliton sector of the solution. For $N=1$
we get a solution traveling with constant velocity. If $\ket{\lambda}$
belongs to an integrable representation \cite{kac}, the operators
$V_j$ are nilpotent and the exponentials $e^{V_j}$ truncate at some
finite order. That explains the truncation of the Hirota's tau
functions in the Hirota's method \cite{hirota}. 

The eigenvectors of $\Lambda_{\pm}$ with non-zero eigenvalues are 
\be
V^{(\omega)}\(z\) = \sum_{n=-\infty}^{\infty} z^{-n}\, V^{(\omega)}_n 
\lab{eigenvectorcbd}
\ee
with
\br
V^{(\omega)}_{6n}&=& T_3^n -\frac{1}{3}\, \delta_{n,0}\,C\nonu\\
V^{(\omega)}_{6n+1}&=&-\,\omega^{-1}\,
\frac{\sqrt{3}}{3}\,\(\frac{\sqrt{2}}{2}\, T_{+}^n-2\, L_{-2}^{n+1/2}\)
\nonu\\
V^{(\omega)}_{6n+2}&=&\omega^{-2}\,\sqrt{2}\, L_{-1}^{n+1/2}
\nonu\\
V^{(\omega)}_{6n+3}&=&\,\omega^{-3}\,\sqrt{2}\,  L_{0}^{n+1/2}
\nonu\\
V^{(\omega)}_{6n+4}&=&\omega^{-4}\,\sqrt{2}\,  L_{1}^{n+1/2}
\nonu\\
V^{(\omega)}_{6n+5}&=&-\,\omega^{-5}\,
\frac{\sqrt{3}}{3}\,\(\frac{\sqrt{2}}{2}\, T_{-}^{n+1}-2\, L_{2}^{n+1/2}\)
\er
where $\omega^6=-1$. One can check that
\be
\sbr{\Lambda_{\pm}}{V^{(\omega)}_n} = \omega^{\pm 1}\,
\sqrt{3}\;V^{(\omega)}_{n\pm 1}
\lab{puregradeeigenvector}
\ee
and so  
\be
\sbr{\Lambda_{\pm}}{V^{(\omega)}\(z\)} =  \sqrt{3}\;
\(\omega\,z\)^{\pm 1}
\; V^{(\omega)}\(z\) 
\lab{eigenvectorvomegaz}
\ee
Therefore the eigenvalues depend upon the product of a complex
parameter $z$ and of a sixth root of $-1$. However, there is no
degeneracy because of the operator equality among degenerate
eigenvectors, i.e. $V^{(\omega)}\(z\)=V^{(\omega\, 
  \gamma)}\(z\,\gamma^{-1}\)$, with $\gamma^6=1$. 

Notice that the dependence of the eigenvalues on the free parameter
$z$ comes from a one parameter group acting on the
eigenvectors. Indeed, from \rf{puregradeeigenvector} one has the
eigenvalue equation
\be
\sbr{\Lambda_{\pm}}{\sum_{n=-\infty}^{\infty}V^{(\omega)}_n} = \omega^{\pm 1}\,
\sqrt{3}\;\sum_{n=-\infty}^{\infty} V^{(\omega)}_{n}
\lab{puregradeeigenvector1}
\ee 
But since $\Lambda_{\pm}$ and $V^{(\omega)}_n$ are eigenvectors of the
grading operator, i.e. $\sbr{Q}{\Lambda_{\pm}}=\pm \Lambda_{\pm}$, and
$\sbr{Q}{V^{(\omega)}_n}=n\,V^{(\omega)}_n$, we have that 
\be
V^{(\omega)}\(z\)=e^{-\ln z\,\,
  Q}\,\(\sum_{n=-\infty}^{\infty}V^{(\omega)}_n\,\) \, e^{\ln z\,\, Q}
\ee
and so, conjugating \rf{puregradeeigenvector1} with $e^{-\ln z\,\,
  Q}$, one gets \rf{eigenvectorvomegaz}. 

The final evaluation of the solutions requires the calculation of the
expectation values of products of the operators $V_j$ in the heighest
weight states $\ket{\Lambda}$ (see \rf{tausolitonicspec}). In many
cases that may prove to be a laborious task. One can then use an
hybrid method (see \cite{bueno} for details) in which one uses the
algebraic dressing method to obtain the 
relation among the tau functions and the fields (where the Hirota's
method is helpless), and then apply the Hirota's method on the
equations for the tau functions, with the ansatz provided by the
relation \rf{tausolitonicspec}, i.e.
\be
\tau_{\lambda}= 1+ \sum_{j=1}^N \delta_j^{\lambda}\, e^{\Gamma_j} + 
\sum_{j\geq i=1}^N \delta_{i,j}^{\lambda}\, e^{\Gamma_i+\Gamma_j} +
\ldots 
\lab{hiransatzcbd}
\ee
with $\delta_j^{\lambda}=\bra{\lambda}V_j\ket{\lambda}$,
$\delta_{i,j}^{\lambda}\sim\bra{\lambda}V_i\, V_j\ket{\lambda}$,  
etc.  We shall use such hybrid procedure in
this paper, to determine the coefficients  $\delta_j^{\lambda}$,
$\delta_{i,j}^{\lambda}$, by the Hirota's method through a computer
algorithm. Since the normalization of the eingenvectors $V_j$ are
not fixed by the above procedure, the coefficients
$\delta^{\lambda}$'s will be determined up to some rescaling
constants.

\subsection{One soliton solution}

The solutions in the one-soliton sector is obtained by taking the
constant group element, introduced in \rf{hdecomp}, as
$h=e^{V^{(\omega)}\(z^{\prime}\)}$, with $V^{(\omega)}\(z^{\prime}\)$
given in  \rf{eigenvectorcbd}. Then solving \rf{hirotaeqcbd} with the
ansatz \rf{hiransatzcbd} for $N=1$, one gets the solution
\br
\tau_0&=& 1 -4\, a\, e^{\Gamma} + a^2\,e^{2\,\Gamma}\nonu\\
\tau_1&=& \(1 + a\, e^{\Gamma}\)^2 
\lab{onesolbd}
\er
with ($z\equiv z^{\prime} \omega$) 
\be
\Gamma = \sqrt{3}\,\( z\, x_{+}-\frac{x_{-}}{z}\)=
\frac{\sqrt{3}}{\sqrt{1-\frac{v^2}{c^2}}}\,
\left[\cos\theta\,\(x-v\,t\) + i\, \sin\theta\(c\,t-\frac{v}{c}\,x\)\right]
\ee
where we have parametrized $z$ as $z=e^{-\alpha+i\,\theta}$, and defined 
\be
v=c\, \tanh\alpha
\ee
We now write $a=e^{\beta+i\, \xi}$ and define ${\widetilde \Gamma}\equiv
\Gamma +\beta+i\, \xi\equiv \Gamma_R+i\, \Gamma_I$ with
\br
\Gamma_R = \frac{\sqrt{3}\, \cos\theta}{\sqrt{1-\frac{v^2}{c^2}}}\,
\(x-v\,t\)+ \beta \qquad\qquad \quad 
\Gamma_I =\frac{\sqrt{3}\, \sin\theta}{\sqrt{1-\frac{v^2}{c^2}}}\,
\(c\,t-\frac{v}{c}\,x\)+ \xi
\er
Then, using \rf{taudefbd} and \rf{onesolbd} one gets
$\vp=\vp_R+i\,\vp_I$ with 
\br
\vp_R &=& \frac{1}{2}\, \ln \left[ 1+3\,\frac{\(1-2\,\cosh\Gamma_R\,
    \cos\Gamma_I\)}{\(\cosh\Gamma_R+\cos\Gamma_I\)^2}\right] 
\nonu\\ 
 \vp_I &=& {\rm ArcTan}\left[\frac{3\sinh\Gamma_R\, \sin\Gamma_I}{
\(\cosh\Gamma_R+\cos\Gamma_I\)^2-3\(1+\cosh\Gamma_R\,\cos\Gamma_I\)}\right]
\er

We have the following particular types of solutions (we shall set
$v=0$ since it can be recovered by a Lorentz boost): 
\begin{enumerate}
\item There are two types of real solutions ($\vp_I=0$), but both singular:
\begin{enumerate}
\item Take $\cos\theta=\pm 1$, $\xi=0$, $\beta=0$ and then
\be
\vp_R = \frac{1}{2}\, \ln \left[
  1+3\,\frac{\(1-2\,\cosh\(\sqrt{3}\, x\)  
    \)}{\(\cosh\(\sqrt{3}\, x\)+1\)^2}\right] 
\ee
and so $\vp_R\ra 0$ for $x\ra \pm\infty$, and $\vp_R\ra-\infty$ for
$x= \pm {\rm ArcCosh} (2)/\sqrt{3}$. 
\item Take $\sin\theta=\pm 1$, $\xi=0$, $\beta =0$ and so
\be
\vp_R = \frac{1}{2}\, \ln \left[
  1+3\,\frac{\(1-2\,\cos\(\sqrt{3}\, c\,t\)  
    \)}{\(1+\cos\(\sqrt{3}\, c\,t\)\)^2}\right]
\ee
which is singular whenever $t=(2n+1)\pi/(c\,\sqrt{3})$. 

\end{enumerate}

\item The regular static one-soliton solution is obtained when
  $\cos\theta=\varepsilon =\pm 1$, $\beta=0$, $\xi\neq 0, \pi$, and then
\br
\vp_R &=& \frac{1}{2}\, \ln \left[ 1+3\,\frac{\(1-2\,\cosh\(\sqrt{3}\,
  x\) \,
    \cos\xi\)}{\(\cosh\(\sqrt{3}\, x\)+\cos\xi\)^2}\right] 
\lab{trueonesolbd}\\ 
 \vp_I &=& {\rm
  ArcTan}\left[\frac{3\sinh\(\sqrt{3}\,x\)\, \varepsilon\,\sin\xi}{ 
\(\cosh\(\sqrt{3}\,x\)+\cos\xi\)^2-3\(1+\cosh\(\sqrt{3}\,x\)\,\cos\xi\)}\right]
\nonu
\er
One can check that, as $x$ varies from $-\infty$ to $\infty$, $\vp_I$
  varies continuously from $-\pi$ to $\pi$ for $\varepsilon\sin\xi<0$, and from
  $\pi$ to $-\pi$ for $\varepsilon\sin\xi>0$. Therefore, the
  topological charge \rf{topolcharge} is given by 
\be
 Q_{\rm top.} = -{\rm sign}\(\varepsilon\,\sin\xi\)
\ee
We give in figure \ref{fig:onesolcbd} the plots of the one-soliton
solution \rf{trueonesolbd} for two values of the parameters
$\varepsilon$ and $\xi$. 

\begin{figure}
\scalebox{1.5}{
\includegraphics{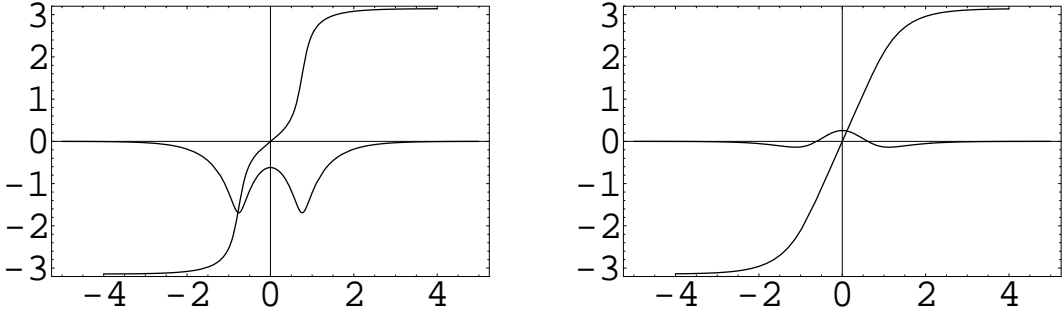}}
\caption{Plots of the real and imaginary parts of $\vp$ against $x$,
  for the one-soliton solution \rf{trueonesolbd}, with 
   $\(\varepsilon,\xi\)=\(-1,\pi/10\)$,  (on the left), and 
  $\(\varepsilon,\xi\)=\(-1,2\pi/5\)$ (on the right). The real part of
  $\vp$ corresponds to the curve 
  that goes to zero as $x\ra \pm \infty$.
\label{fig:onesolcbd} }   
\end{figure}

\item Denote $\omega\equiv \sin\theta$, and set $\beta=\xi=0$ to get
  $\Gamma_R= \sqrt{3\(1-\omega^2\)}\;x$ and
  $\Gamma_I=\sqrt{3}\,\omega\,c\,t$. That is a solution which
  oscillates among the possible forms of the one-soliton given in
  \rf{trueonesolbd}, and develops a singularity whenever
  $t=n\,\pi/\(\sqrt{3}\,\omega\, c\)$. At those values of time the topological
  charge \rf{topolcharge} flips sign. 

\end{enumerate}

\subsection{Two soliton solution}

The solutions on the two-soliton sector are obtained by taking 
$h=e^{V^{(\omega_1)}\(z^{\prime}_1\)}\,e^{V^{(\omega_2)}\(z^{\prime}_2\)}$,
with $V^{(\omega_i)}\(z^{\prime}_i\)$ 
given in  \rf{eigenvectorcbd}. Then solving \rf{hirotaeqcbd} with the
ansatz \rf{hiransatzcbd} for $N=2$, one gets the solution
($z_i=\omega_i\,z^{\prime}_i$) 
\br
\tau_0&=& 1 -4\, a_1\, e^{\Gamma_1}-4\, a_2\, e^{\Gamma_2}  
+a_1^2\,e^{2\,\Gamma_1}
+a_2^2\,e^{2\,\Gamma_2}\nonu\\
&+&8\,a_1\,a_2\,\frac{2\,z_1^4-z_1^2\,z_2^2+2\,z_2^4}{
\(z_1+z_2\)^2\(z_1^2+z_1\,z_2+z_2^2\)} \, e^{\Gamma_1+\Gamma_2} 
\nonu\\
&-&4\,a_1^2\,a_2\,
\frac{\(z_1-z_2\)^2\(z_1^2-z_1\,z_2+z_2^2\)}{
\(z_1+z_2\)^2\(z_1^2+z_1\,z_2+z_2^2\)} \, e^{2\,\Gamma_1+\Gamma_2}
\nonu\\
&-&4\,a_1\,a_2^2\,
\frac{\(z_1-z_2\)^2\(z_1^2-z_1\,z_2+z_2^2\)}{
\(z_1+z_2\)^2\(z_1^2+z_1\,z_2+z_2^2\)} \, e^{\Gamma_1+2\,\Gamma_2}
\nonu\\
\nonu\\
&+&\,a_1^2\,a_2^2\,
\frac{\(z_1-z_2\)^4\(z_1^2-z_1\,z_2+z_2^2\)^2}{
\(z_1+z_2\)^4\(z_1^2+z_1\,z_2+z_2^2\)^2} \, e^{2\,\Gamma_1+2\Gamma_2}
\nonu\\
\tau_1&=& 1 + 2\, a_1\, e^{\Gamma_1} + 2\, a_2\, e^{\Gamma_2}
+a_1^2\,e^{2\,\Gamma_1}
+a_2^2\,e^{2\,\Gamma_2}\nonu\\
&+&4\, a_1\,a_2\,\frac{z_1^4+4\,z_1^2\,z_2^2+z_2^4}{
\(z_1+z_2\)^2\(z_1^2+z_1\,z_2+z_2^2\)}  \, e^{\Gamma_1+\Gamma_2}
\nonu\\
&+&2\,a_1^2\,a_2\,
\frac{\(z_1-z_2\)^2\(z_1^2-z_1\,z_2+z_2^2\)}{
\(z_1+z_2\)^2\(z_1^2+z_1\,z_2+z_2^2\)} \, e^{2\,\Gamma_1+\Gamma_2}
\nonu\\
&+&2\,a_1\,a_2^2\,
\frac{\(z_1-z_2\)^2\(z_1^2-z_1\,z_2+z_2^2\)}{
\(z_1+z_2\)^2\(z_1^2+z_1\,z_2+z_2^2\)} \, e^{\Gamma_1+2\,\Gamma_2}
\nonu\\
&+&\,a_1^2\,a_2^2\,
\frac{\(z_1-z_2\)^4\(z_1^2-z_1\,z_2+z_2^2\)^2}{
\(z_1+z_2\)^4\(z_1^2+z_1\,z_2+z_2^2\)^2} \, e^{2\,\Gamma_1+2\Gamma_2}
\lab{twosolcbd}
\er
where, as before, 
\be
\Gamma_i = \sqrt{3}\,\( z_i\, x_{+}-\frac{x_{-}}{z_i}\)=
\frac{\sqrt{3}}{\sqrt{1-\frac{v_i^2}{c^2}}}\,
\left[\cos\theta_i\,\(x-v_i\,t\) 
+ i\, \sin\theta_i\(c\,t-\frac{v_i}{c}\,x\)\right]
\ee
and we have parametrized $z_i=e^{-\alpha_i+i\,\theta_i}$, and defined 
$v_i=c\, \tanh\alpha_i$.

Solutions describing  the scattering of two one-solitons 
can be obtained from \rf{twosolcbd} by setting the  parameters $a_i$
and $z_i$, $i=1,2$, to values corresponding to the ones chosen for the
one-soliton solutions. For instance, to get the scattering of two
regular one-solitons given in  \rf{trueonesolbd} one has to take
$z_i=e^{-\alpha_i+i\theta_i}$ and $a_i=e^{i\xi_i}$, with $\cos
\theta_i =\pm 1$, $\xi\neq 0,\pi$, and $\alpha_i$ setting their
velocities by $v_i=c\, \tanh\alpha_i$. However, we do not explore in
this paper such solutions further.

\subsection{Breathers}

Let us now consider the two-soliton solution \rf{twosolcbd} for the
case where $\alpha_1=\alpha_2=0$,  
$\theta_1=-\theta_2\equiv \theta$, and $a_2=-a_1\equiv -a$, with 
\be
a=\sqrt{-\frac{
\(z_1+z_2\)^2\(z_1^2+z_1\,z_2+z_2^2\)}{\(z_1-z_2\)^2\(z_1^2-z_1\,z_2+z_2^2\)}}
= \mid {\rm cotan} \theta\mid \sqrt{\frac{2\cos 2\theta +1}{2\cos 2\theta -1}}
\ee
Using
\rf{taudefbd} on then gets that
\be
\vp=\ln\frac
{\cosh 2\,\Gamma_R-i\, 8\,a\, \cosh\Gamma_R\,  \sin\Gamma_I+
a^2\cos 2\,\Gamma_I -  b}{\cosh 2\,\Gamma_R+i\, 4\,a\,
  \cosh\Gamma_R\,  \sin\Gamma_I+ a^2\cos 2\,\Gamma_I -  c}
\lab{breathersolphi}
\ee
with
\be
\Gamma_R=\sqrt{3}\,\cos\(\theta\)\,\, x \qquad \qquad 
\Gamma_I=\sqrt{3}\,\sin\(\theta\)\,\, c\,t
\ee
and
\be
b=\frac{4\,\cos 4\theta -1}{\sin^2\theta\(2\cos 2\theta -1\)} \qquad\qquad 
c=\frac{\cos 4\theta +2}{\sin^2\theta\(2\cos 2\theta -1\)}
\ee
Therefore, if $a$ is pure imaginary we have
that the argument of the logarithm in \rf{breathersolphi} is real,
but it is either positive or negative. Therefore, the solution for
$\vp$ is not regular since it has an imaginary part that jumps from
$0$ to $\pi$.    

On the other hand if $a$ is real
we have that the solution is complex, and writing
$\vp=\vp_R+i\,\vp_I$, we have  
\br
\vp_R&=& \frac{1}{2}\ln 
\frac{\left[\cosh 2\,\Gamma_R+a^2\cos 2\,\Gamma_I - b\right]^2+
\left[8\,a\, \cosh\Gamma_R\,  \sin\Gamma_I\right]^2}
{\left[\cosh 2\,\Gamma_R+a^2\cos 2\,\Gamma_I -c\right]^2+
\left[4\,a\, \cosh\Gamma_R\,  \sin\Gamma_I\right]^2}\nonu\\
\vp_I&=&{\rm ArcTan}\,\frac{{\cal X}}{{\cal Y}}\nonu
\er
with
\br
{\cal X}&=&4\,a\,\cosh\Gamma_R\,\sin\Gamma_I\left[2\,c+b-3\(\cosh
  2\,\Gamma_R+a^2\cos 2\,\Gamma_I\)\right]  
\nonu\\
{\cal Y}&=&-32\,a^2\,\cosh^2\Gamma_R\,\sin^2\Gamma_I\nonu\\
&+&\left[\cosh 2\,\Gamma_R+a^2\cos 2\,\Gamma_I -  b\right]
\left[\cosh 2\,\Gamma_R+a^2\cos 2\,\Gamma_I -  c\right]
\er
Notice that $\vp_R\ra 0$ and $\tan \vp_I\ra 0$, as $x\ra \pm
\infty$. It is a localized solution that oscillates in time, and so it
is like a breather, however with a structure much more complex than that
of the sine-Gordon breather.  

\section{The coupling to matter fields}
\label{sec:coupling}
\setcounter{equation}{0}

We now consider a generalization of the Conformal Bullough-Dood model
\rf{cbdeqs}, on the lines of \cite{matter,achic,bueno}, by the
introduction of extra fields. That is done by enlarging the zero
curvature potentials \rf{zcpotbd} with components having grades greater
than $1$ and smaller than $-1$. We then introduce 
\be
A_{+}= - B\, \( E_{+l}+F_{+}\)\, B^{-1} \qquad\qquad 
A_{-}= - \partial_{-} B\, B^{-1} +E_{-l}+F_{-} 
\lab{zcpotbdgen}
\ee 
where $B$ is the same group element as in \rf{bfield}, i.e. the
subgroup obtained by exponentiating the subalgebra of grade zero
w.r.t. to the grading operator $Q$ defined in \rf{gradingop}. $E_{\pm
  l}$ are constant elements of the algebra $A_2^{(2)}$ with grades
$\pm l$, and $F_{\pm}$ have components with grades varying from $\pm
1$ to $\pm\(l-1\)$, i.e.
\be
F_{\pm} = \sum_{m=1}^{l-1} F_{\pm}^{m} \qquad \qquad\qquad
\sbr{Q}{F_{\pm}^{m}}=\pm m \, F_{\pm}^{m}
\lab{fpmdef}
\ee
The extra fields are contained in the $F_{\pm}^{m}$, and their
explicit expression is given in the examples we discuss below. 

The equations of motion of the theory is obtained by imposing that the
potentials $A_{\pm}$ in \rf{zcpotbdgen}, satisfy the zero curvature condition
\rf{zc}. Replacing \rf{zcpotbdgen} into \rf{zc} and splitting it into
their grading components one gets the equations of motion
\br
\partial_{+}\(\partial_{-}B\, B^{-1}\) &=& \sbr{E_{-l}}{B\,
  E_{l}\,B^{-1}} + \sum_{n=1}^{l-1}\sbr{F_{-}^n}{B\,F_{+}^n\,B^{-1}}
\lab{eqmotgen1}\\
\partial_{-}F_{+}^m&=& \sbr{E_{l}}{B^{-1}\,F_{-}^{l-m}\, B} +
\sum_{n=1}^{l-m-1}\sbr{F_{+}^{n+m}}{B^{-1}\,F_{-}^n\,B}
\lab{eqmotgen2}\\
\partial_{+}F_{-}^m&=&- \sbr{E_{-l}}{B\,F_{+}^{l-m}\, B^{-1}} -
\sum_{n=1}^{l-m-1}\sbr{F_{-}^{n+m}}{B\,F_{+}^n\,B^{-1}}
\lab{eqmotgen3}
\er

The equations \rf{eqmotgen1}-\rf{eqmotgen3} have a large symmetry
group. Indeed,  
one can check they are invariant under the conformal transformations
\be
x_{+} \ra f_{+}\(x_{+}\) \qquad \qquad x_{-}\ra f_{-}\(x_{-}\)
\ee
with $f_{+}$ and $f_{-}$ being analytic functions, and with the fields transforming as
\br
\vp\(x_{+},x_{-}\) &\ra & {\tilde \vp}\({\tilde x}_{+},{\tilde x}_{-}\) =
\vp\(x_{+},x_{-}\)\nonu\\ 
e^{-\nu\(x_{+},x_{-}\)} &\ra &e^{-{\tilde \nu}\({\tilde x}_{+},{\tilde
    x}_{-}\)}= \(f_{+}^{\prime}\)^{\delta}\,\(f_{-}^{\prime}\)^{{\bar \delta}}  
 e^{-\nu\(x_{+},x_{-}\)}
\lab{conformaltransf}\\
e^{-\eta\(x_{+},x_{-}\)} &\ra &e^{-{\tilde \eta}\({\tilde x}_{+},{\tilde
    x}_{-}\)}= \(f_{+}^{\prime}\)^{1/l}\,\(f_{-}^{\prime}\)^{1/l} \,  
 e^{-\eta\(x_{+},x_{-}\)}\nonu\\
F_{+}^m\(x_{+},x_{-}\)&\ra & {\tilde F}_{+}^m\({\tilde x}_{+},{\tilde
  x}_{-}\) = \(f_{+}^{\prime}\)^{-1+m/l}F_{+}^m\(x_{+},x_{-}\)\nonu\\
F_{-}^m\(x_{+},x_{-}\)&\ra & {\tilde F}_{-}^m\({\tilde x}_{+},{\tilde
  x}_{-}\) = \(f_{-}^{\prime}\)^{-1+m/l}F_{-}^m\(x_{+},x_{-}\)\nonu
\er
where the conformal weights of $e^{-\nu}$, namely $\delta$ and ${\bar
  \delta}$,  are arbitrary. 

In addition we have local symmetries associated to the generators of
the zero grade subalgebra ${\cal G}_0$ (see \rf{gradation}) which
commute with $E_{\pm l}$. Indeed, suppose we have group elements
satisfying
\be
h_R\(x_{+}\) \, E_l\, h_R^{-1}\(x_{+}\) = E_l \qquad \qquad \qquad
h_L\(x_{-}\) \, E_{-l}\, h_L^{-1}\(x_{-}\) = E_{-l}
\ee
with $h_R$ and $h_L$ being exponentiations of generatros of ${\cal
  G}_0$. Then the transformations
\br
B\(x_{+},x_{-}\)&\ra& h_L\(x_{-}\) \,B\(x_{+},x_{-}\)\,h_R\(x_{+}\) \nonu\\
F_{+}^m\(x_{+},x_{-}\)&\ra&h_R^{-1}\(x_{+}\)\, F_{+}^m\(x_{+},x_{-}\)\, h_R\(x_{+}\)
\lab{localtransf}\\
F_{-}^m\(x_{+},x_{-}\)&\ra&h_L\(x_{-}\)\, F_{-}^m\(x_{+},x_{-}\)\, h_L^{-1}\(x_{-}\)\nonu
\er
leave the equations \rf{eqmotgen1}-\rf{eqmotgen3} invariant.

We are interested in models that possess soliton solutions. One of the
key ingredients for the appearance of solitons, as explained in
\cite{fmg}, is the existence of vacuum solutions such that the zero
curvature potentials $A_{\pm}$ evaluated on them, lie on an oscillator
(Heisenberg) subalgebra the Kac-Moody algebra. That is an abelian
subalgebra up to central terms, i.e. it has generators $b_n$ such that
their commutation relations is of the form
\be
\sbr{b_m}{b_n} = \beta  m \, \delta_{m+n,0}\, C
\ee
Notice that the constant element $E_{-l}$ will always be present in
$A_{-}$, when evaluated on any solution, since its coefficient is
unity. In addition, the term  $B\, E_{+l}\, B^{-1}$ can never
vanish. Therefore, when evaluated on the vacuum solution its
commutator with $E_{-l}$  has to produce at most a central term. Since
the vacuum value of $B$ can always be absorbed in to the definition of
$E_{+l}$, i.e. $B_{\rm vac.}\, E_{+l}\, B_{\rm vac.}^{-1}\ra E_{+l}$,
we conclude that we need 
\be
\sbr{E_l}{E_{-l}} \sim C
\ee
If one looks at the eigensubspaces \rf{eigensub} and the commutation
relations \rf{a22comrel} for the $A_2^{(2)}$ algebra, one notices that
the only possibilities for $E_{\pm l}$ are 
\br
E_{6n} &\sim& T_3^n \qquad\qquad \qquad\qquad\qquad E_{-6n}\sim T_3^{-n}\nonu\\
E_{6n+1} &\sim& \frac{\sqrt{2}}{2}\, T_+^n+ L_{-2}^{n+1/2} \qquad\qquad 
E_{-6n-1}\sim \frac{\sqrt{2}}{2}\, T_-^{-n}+ L_{2}^{-n-1/2} \nonu\\
E_{6n+3} &\sim& L_0^{n+1/2}\qquad\qquad \qquad\qquad E_{-6n-3} \sim
L_0^{-n-1/2}
\nonu
\er
The gradation \rf{gradation} has a period six, and the simplest models
occur on the first period. The elements $E_{\pm\(6n+1\)}$, for $n=0$
correspond to $\Lambda_{\pm}$, given in \rf{lambdapm}, leading to the
usual Bullough-Dodd model. We will then consider in this paper the
models corresponding to $E_{\pm 6}$,  and $E_{\pm 3}$. The first one
is more interesting from the physical point of view and we discuss it
in detail in section \ref{sec:l=6}. The second model is discussed in
section \ref{sec:l=3}.

\section{The model for  $l=6$}
\label{sec:l=6}
\setcounter{equation}{0}

We now take the potentials \rf{zcpotbdgen} with $l=6$, and choose 
\be
E_{\pm 6} \equiv m\; T_3^{\pm 1}
\lab{epm6def}
\ee
where $m$ is a parameter which will set the mass scale for the
particles and solitons of the theory. We then have 
\be
A_{+}= - B\, \( E_{+6}+\sum_{m=1}^{5} F_{+}^{m}\)\, B^{-1} \qquad\qquad 
A_{-}= - \partial_{-} B\, B^{-1} +E_{-6}+ \sum_{m=1}^{5} F_{-}^{m}
\lab{zcpotbdl=6}
\ee 
The matter fields are contained in $F_{\pm}^{m}$, and are defined as 
\br
F_{+}^5 &=& \frac{1}{2}\, \sqrt{m} \, {\tilde \psi}_R^1 T_{-}^1 -
\sqrt{2 m} \, {\tilde \psi}_R^2 L_2^{1/2} \qquad \quad 
F_{-}^{5} = \frac{1}{2}\, \sqrt{m} \, \psi_L^1 T_{+}^{-1} -
\sqrt{2 m} \, \psi_L^2 L_{-2}^{-1/2}\nonu\\ 
F_{+}^4 &=& \sqrt{m} \, {\tilde \psi}_R^3  L_1^{1/2} \qquad \qquad
\qquad \qquad \quad \quad 
F_{-}^{4} = - \sqrt{m} \, \psi_L^3 L_{-1}^{-1/2} \nonu\\
F_{+}^3 &=& \psi_R^0 L_0^{1/2} \qquad \qquad \qquad \qquad \qquad
\quad \; \quad 
F_{-}^{3} = \psi_L^0 L_0^{-1/2}
\lab{fpm6def}\\
F_{+}^2 &=& \sqrt{m} \,\psi_R^3  L_{-1}^{1/2} \qquad \qquad \qquad
\qquad \quad \quad 
F_{-}^{2} = \sqrt{m} \, {\tilde \psi}_L^3 L_{1}^{-1/2}\nonu\\
F_{+}^1 &=&  \frac{1}{2}\, \sqrt{m} \, \psi_R^1  T_{+}^0 +
\sqrt{2 m} \, \psi_R^2 L_{-2}^{1/2} \qquad \quad 
F_{-}^{1} = -\frac{1}{2}\, \sqrt{m} \, {\tilde \psi}_L^1 T_{-}^0 -
\sqrt{2 m} \, {\tilde \psi}_L^2  L_{2}^{-1/2}\nonu 
\er
Replacing into the equations \rf{eqmotgen1}-\rf{eqmotgen3} we get the
equations of motion of the theory
\br
\partial^2 \varphi &=& 
\frac{i}{2}\, m\, {\bar \psi}^1\, W\(\eta\)\, \gamma_5\, V^1\, \psi^1
- i\, 2 m \, {\bar \psi}^2\, W\(\eta\)\, \gamma_5\, V^2\, \psi^2 \nonu\\
&-&\frac{i}{2}\, m\, {\bar \psi}^3\, W\(\eta\)\, \gamma_5\, V^3\,
\psi^3
\lab{eqmotl=61}\\
\partial^2 \nu &=& 
-2 \, m^2 \, e^{6\, \eta}  
-\frac{1}{2}\, e^{3\, \eta} \, \psi_L^0\, \psi_R^0 
+  i\, m\, e^{6\, \eta}\, {\bar \psi}^1 \frac{\(1-\gamma_5\)}{2}\,
V^1\, \psi^1 \nonu\\ 
&+& i\, m\, {\bar \psi}^2\, W\(\eta\)\, V^2\, \psi^2 
+ \frac{i}{2}\, m\, {\bar \psi}^3\, W\(\eta\)\, V^3\, \psi^3
\lab{eqmotl=62}\\
\partial^2 \eta &=& 0
\lab{eqmotl=63}\\
i\, \gamma^{\mu}\, \partial_{\mu} \, \psi^i &=& m_i\, W\(\eta\)\,
V^i\, \psi^i + U^i
\lab{eqmotl=64}\\ 
i\, \gamma^{\mu}\, \partial_{\mu} \, {\tilde \psi}^i &=& m_i\, {\tilde
  W}\(\eta\)\, {\tilde V}^i\, {\tilde \psi}^i  
+ {\tilde U}^i  
\lab{eqmotl=65}\\
i\, \gamma^{\mu}\, \partial_{\mu} \, \psi^0 &=& U^0
\lab{eqmotl=66}
\er
where in \rf{eqmotl=64} and \rf{eqmotl=65} we have $i=1,2,3$, with 
\be 
m_1=m_3=m \qquad  {\rm and}\qquad  m_2=2\, m
\lab{midef}
\ee  
We have introduced a spinor notation for the matter fields as 
\br
\psi \equiv \(
\begin{array}{c}
\psi_R\\
\psi_L
\end{array}\)\qquad \qquad \qquad 
U^i \equiv \(
\begin{array}{c}
U^i_R\\
U^i_L
\end{array}\)
\er
with similar notation for ${\tilde \psi}$ and ${\tilde U}^i$. We have 
defined ${\bar \psi}$ and ${\bar U}$ as 
\be
{\bar \psi}^i \equiv \({\tilde \psi}^i\)^T \, \gamma_0 \qquad\qquad
{\bar U}^i \equiv \({\tilde U}^i\)^T \, \gamma_0 \qquad\qquad {\rm for}\;\;\; i=1,2,3
\ee
and
\be 
{\bar \psi}^0 \equiv \({\psi}^0\)^T \, \gamma_0 \qquad\qquad
{\bar U}^0 \equiv \({U}^0\)^T \, \gamma_0
\ee  
Notice that in general ${\tilde \psi}$ is not the complex conjugate of
$\psi$. The representation for the Dirac's $\gamma$-matrices is 
\br
\gamma_0=-i\, \(
\begin{array}{cc}
0&-1\\
1&0
\end{array}\)\qquad \quad 
\gamma_1=-i\, \(
\begin{array}{cc}
0&1\\
1&0
\end{array}\)\qquad \quad 
\gamma_5\equiv \gamma_0 \, \gamma_1 = 
\(
\begin{array}{cc}
1&0\\
0&-1
\end{array}\)
\er
In addition, we have introduced the quantities 
\be
W\(\eta\)\equiv \frac{\(1+\gamma_5\)}{2} + e^{6\, \eta}\,
\frac{\(1-\gamma_5\)}{2} \qquad \quad  
{\tilde W}\(\eta\)\equiv e^{6\, \eta}\, \frac{\(1+\gamma_5\)}{2} +
\frac{\(1-\gamma_5\)}{2} 
\ee
and the potentials involving the scalar fields\footnote{
Notice that, $W\(0\)={\tilde W}\(0\) = 1$, 
${\tilde W}\(\eta\)\, \gamma_0= \gamma_0\, W\(\eta\)$, and 
${\tilde V}^i\, \gamma_0= \gamma_0\, V^i$}
\br
V^1&=& e^{\(\eta + \varphi\)\,\gamma_5} \qquad \qquad{\tilde V}^1=
e^{-\(\eta + \varphi\)\,\gamma_5}\nonu\\ 
V^2&=& e^{\(\eta -2 \varphi\)\,\gamma_5} \qquad \qquad{\tilde V}^2=
e^{-\(\eta -2 \varphi\)\,\gamma_5}
\lab{scalarpot}\\ 
V^3&=& e^{\(2\eta - \varphi\)\,\gamma_5} \qquad \qquad{\tilde V}^3=
e^{-\(2\eta - \varphi\)\,\gamma_5} 
\er
We have also introduced  potentials quadractic in the matter fields 
\br
U_R^0&=&  
\sqrt{\frac{3}{2}}\, \, m\,\(  e^{\eta +\varphi} \psi_L^3 \psi_R^1 +
 e^{2 \eta  -\varphi  } \psi_L^1  \psi_R^3 \)
\nonu\\
U_L^0&=& 
\sqrt{\frac{3}{2}}\, \, m\,\(  e^{2 \eta -\varphi} {\tilde \psi}_L^3  
{\tilde \psi}_R^1 + e^{\eta  +\varphi} 
{\tilde \psi}_L^1  {\tilde \psi}_R^3 \)
\nonu\\
U^1_L&=&- \sqrt{2 \,m}\,  
\( e^{\eta - 2\varphi}\psi_R^3 \, {\tilde \psi}_L^2 + e^{4\eta + \varphi}
    \psi_L^3 {\tilde \psi}_R^2\)
-\sqrt{\frac{3}{2}} \,\( e^{2\eta - \varphi} 
\psi_R^0\, {\tilde \psi}_L^3
   - e^{3\eta} \psi_L^0 {\tilde \psi}_R^3 \)
\nonu\\
U^2_L&=& \sqrt{\frac{m}{2}}\,
\( e^{\eta  +\varphi  }\psi_R^3 {\tilde \psi}_L^1 
 - e^{4 \eta +\varphi  } \psi_L^3 {\tilde \psi}_R^1 \)
\nonu\\
U^3_R&=& -\sqrt{2\, m}\,
\( e^{\eta  +\varphi  } \psi_L^2 \psi_R^1
+ e^{\eta -2 \varphi} \psi_L^1  \psi_R^2 \)
\lab{matterpot}\\
U^3_L&=&
\sqrt{\frac{3}{2}}\; \( e^{\eta  +\varphi} \psi_R^0  {\tilde \psi}_L^1
+ e^{3 \eta} \psi_L^0 {\tilde \psi}_R^1\)
\nonu\\
{\tilde U}^3_R&=&
\sqrt{\frac{3}{2}}\;\(e^{3 \eta  }\psi_R^0 \psi_L^1
- e^{\eta +\varphi} \psi_L^0 \psi_R^1 \)
\nonu\\
{\tilde U}^3_L&=& -\sqrt{2 \, m}\,
\(e^{\eta  -2 \varphi} {\tilde \psi}_L^2  {\tilde \psi}_R^1
+ e^{\eta +\varphi  } {\tilde \psi}_L^1  {\tilde \psi}_R^2 \)
\nonu\\
{\tilde U}^2_R&=& \sqrt{\frac{m}{2}}\,
\( e^{\eta  +\varphi}\psi_R^1 {\tilde \psi}_L^3
- e^{4 \eta +\varphi} \psi_L^1 {\tilde \psi}_R^3\)
\nonu\\
{\tilde U}^1_R&=& -\sqrt{2 \, m}\,
\(e^{\eta-2 \varphi  }\psi_R^2  {\tilde \psi}_L^3
+e^{4 \eta  +\varphi  } \psi_L^2 {\tilde \psi}_R^3 \)
+\sqrt{\frac{3}{2}}\; \(e^{3 \eta}\psi_R^0 \psi_L^3 
+ e^{2 \eta  -\varphi} \psi_L^0 \psi_R^3 \)
\nonu
\er
and
$$
U_R^1=U_R^2={\tilde U}_L^1={\tilde U}_L^2=0
$$

\subsection{Symmetries} 
\label{subsec:symmetries}

As we have discussed in \rf{conformaltransf} the model
\rf{eqmotl=61}-\rf{eqmotl=66} is invarinat under conformal
transformations on the two dimensional space-time. In addition, it is
invariant under local transfomrations of the type \rf{localtransf},
since $T_3^0\in {\cal G}_0$, commutes with $E_{\pm 6}$ given in
\rf{epm6def}. Denoting $h_R\(x_{+}\)=\exp\(-\xi_R\(x_{+}\)\, T_3^0\)$,
and $h_L\(x_{-}\)=\exp\(\xi_L\(x_{-}\)\, T_3^0\)$, we have that the
fields transform, under such $U_R\(1\)\otimes U_L\(1\)$ group, as  
\br  
\vp &\ra& \vp - \xi_R\(x_{+}\)+\xi_L\(x_{-}\) \qquad \qquad\;\;
\quad\eta \ra \eta  
\qquad \qquad \nu \ra \nu \nonu\\
\psi^0 &\ra& \psi^0
\lab{localtransfl=6}\\  
\psi^i &\ra&
e^{q_i\left[\xi_R\,\frac{1}{2}\(1+\gamma_5\)+\xi_L\,\frac{1}{2}\(1-\gamma_5\)\right]}\,\psi^i 
\qquad \qquad\;\;  
{\tilde \psi}^i \ra
e^{-q_i\left[\xi_R\,\frac{1}{2}\(1+\gamma_5\)+\xi_L\,\frac{1}{2}\(1-\gamma_5\)\right]}\,{\tilde 
  \psi}^i  
\nonu
\er
with the charges being $q_1=1$, $q_2=-2$ and $q_3=-1$. 
Notice that we can take two special global subgroups of the local $U_R\(1\)\otimes U_L\(1\)$ symmetry group. Taking $\xi_R=\xi_L\equiv \theta$ with $\theta$ constant, we have
\br 
\vp&\ra&\vp \qquad\qquad \qquad\quad\eta \ra \eta \qquad\qquad \qquad\quad \nu \ra \nu
\nonu\\ 
\psi^i &\ra& e^{q^i\, \theta}\, \psi^i 
\qquad \qquad{\quad\tilde \psi}^i \ra e^{-q^i\, \theta}\, {\tilde \psi}^i
\qquad\qquad\quad\psi^0 \ra \psi^0 
\er 
Taking now  $\xi_R=-\xi_L\equiv {\bar \theta}$ with ${\bar \theta}$ constant, we have
\br
\vp&\ra&\vp -2\, {\bar \theta}\qquad \qquad\qquad\eta \ra \eta \qquad \qquad \qquad\nu \ra \nu 
\nonu\\
 \psi^i &\ra& e^{q^i\, {\bar \theta}\,\gamma_5}\, \psi^i 
\qquad \qquad{\tilde \psi}^i \ra e^{-q^i\, {\bar \theta}\,\gamma_5}\, {\tilde \psi}^i\qquad\qquad\psi^0 \ra \psi^0 
\er 
Therefore, if the theory had a Lagrangean we would expect, due to the
Noether theorem, one vector and one axial conserved currents. Those
two currents do exist but not quite in the form one would expect. The
field $\psi^0$, that has charge zero, do contribute to the current as
we explain below.  
 
Using the eqs. of motion \rf{eqmotl=61}-\rf{eqmotl=66} one gets that
\br 
\partial_{\mu}\({\bar \psi}^a\,\gamma^{\mu}\,\psi^a\)&=& 
i\,\({\bar U}^a\,\psi^a-{\bar \psi}^a\, U^a\) \qquad \qquad \qquad
\qquad \qquad \qquad \quad a=0,1,2,3\nonu\\
\partial_{\mu}\({\bar \psi}^i\,\gamma^{\mu}\,\gamma_5\,\psi^i\)&=& 
2\,i\,m_i\,{\bar \psi}^i\,W\(\eta\)\,V^i\,\gamma_5\, \psi^i+ i\,\({\bar U}^i\,\gamma_5\,\psi^i+{\bar \psi}^i\, \gamma_5\,U^i\) \qquad 
i=1,2,3\nonu\\
\partial_{\mu}\({\bar \psi}^0\,\gamma^{\mu}\,\gamma_5\,\psi^0\)&=& 
i\,\({\bar U}^0\,\gamma_5\,\psi^0+{\bar \psi}^0\, \gamma_5\,U^0\)
\er 
with $m_i$ as in \rf{midef}. 
One can check that the following identities hold true without the use of the equations of motion
\br
\({\bar U}^1\, \psi^1 - {\bar \psi}^1\, U^1\) 
&-&
2\,\({\bar U}^2\, \psi^2 - {\bar \psi}^2\, U^2\)-
\({\bar U}^3\, \psi^3 - {\bar \psi}^3\, U^3\)\nonu\\
&+&
\frac{1}{2\,m}\,\({\bar U}^0\,\gamma_5\, \psi^0 + 
{\bar \psi}^0\, \gamma_5\,U^0\)=0
\er 
\br 
\({\bar U}^1\,\gamma_5\, \psi^1 + {\bar \psi}^1\,\gamma_5\, U^1\) 
&-&
2\,\({\bar U}^2\,\gamma_5\, \psi^2 + {\bar \psi}^2\, \gamma_5\,U^2\)-
\({\bar U}^3\,\gamma_5\, \psi^3 + {\bar \psi}^3\,\gamma_5\, U^3\)\nonu\\
&+&
\frac{1}{2\,m}\,\({\bar U}^0\, \psi^0 - 
{\bar \psi}^0\, U^0\)=0
\er 
Therefore, we have that the currents
\be 
J_{\mu}= {\bar \psi}^1\,\gamma^{\mu}\,\psi^1
-2\,  {\bar\psi}^2\,\gamma^{\mu}\,\psi^2 
-{\bar \psi}^3\,\gamma^{\mu}\,\psi^3 
+ \frac{1}{2\,m}\, {\bar\psi}^0\,\gamma^{\mu}\,\gamma_5\,\psi^0
\ee 
and
\be 
J_{\mu}^5= -4\,\partial_{\mu}\vp
+{\bar \psi}^1\,\gamma^{\mu}\,\gamma_5\,\psi^1
-2\, {\bar \psi}^2\,\gamma^{\mu}\,\gamma_5\,\psi^2
-{\bar \psi}^3\,\gamma^{\mu}\,\gamma_5\,\psi^3
+ \frac{1}{2\,m}\, {\bar\psi}^0\,\gamma^{\mu}\,\psi^0 
\ee
are conserved
\be 
\partial^{\mu} J_{\mu}=0\qquad \qquad \qquad 
\partial^{\mu} J_{\mu}^5=0
\lab{conservcurrnice}
\ee 
as a consequence of the eqs. of motion \rf{eqmotl=61}-\rf{eqmotl=66}. 

The conservation laws \rf{conservcurrnice} imply that the currents 
\be 
{\cal J}= {\tilde \psi}_R^1\, \psi_R^1 
- 2\, {\tilde \psi}_R^2\, \psi_R^2 
- {\tilde \psi}_R^3\, \psi_R^3 
+\frac{1}{2\,m}\, \(\psi_R^0\)^2 - 2\,\partial_{+}\vp 
\lab{chiralcurrj}
\ee
and 
\be 
{\bar {\cal J}}= {\tilde \psi}_L^1\, \psi_L^1 
- 2\, {\tilde \psi}_L^2\, \psi_L^2 
- {\tilde \psi}_L^3\, \psi_L^3 
-\frac{1}{2\,m}\, \(\psi_L^0\)^2 + 2\,\partial_{-}\vp 
\lab{chiralcurrjb}
\ee
are chiral
\be 
\partial_{-}{\cal J}=0 \qquad\qquad\qquad 
\partial_{+}{\bar {\cal J}} =0
\lab{chiralconserv}
\ee
Indeed, if one introduces the currents $J^{(\pm)}_{\mu}=\(J_{\mu}\pm J_{\mu}^5\)/4$ one gets that their light cone components (see \rf{lightcone}) are 
\br 
J^{(+)}_{+}&=& {\tilde \psi}_R^1\, \psi_R^1 
- 2\, {\tilde \psi}_R^2\, \psi_R^2 
- {\tilde \psi}_R^3\, \psi_R^3 
+\frac{1}{2\,m}\, \(\psi_R^0\)^2 - \partial_{+}\vp 
\nonu\\
J^{(+)}_{-}&=& - \partial_{-}\vp
\\
J^{(-)}_{+}&=& \partial_{+}\vp
\nonu\\
J^{(-)}_{-}&=&{\tilde \psi}_L^1\, \psi_L^1 
- 2\, {\tilde \psi}_L^2\, \psi_L^2 
- {\tilde \psi}_L^3\, \psi_L^3 
-\frac{1}{2\,m}\, \(\psi_L^0\)^2 + \partial_{-}\vp
\nonumber
\er
The conservation of $J^{(\pm)}_{\mu}$, namely $\partial^{\mu}J^{(\pm)}_{\mu}=\partial_{+}J^{(\pm)}_{-}+\partial_{-}J^{(\pm)}_{+}=0$, implies \rf{chiralconserv}. 

The soliton solutions we construct in section \ref{subsec:solutions} belong to a submodel of the theory \rf{eqmotl=61}-\rf{eqmotl=66} where the chiral currents \rf{chiralcurrj} and \rf{chiralcurrjb} vanish, i.e.
\be 
{\cal J} = 0 \qquad\qquad\qquad  {\bar {\cal J}} =0
\ee
Such conditions can be written in the form of an equivalence between currents 
\be 
\frac{1}{8\,\pi}\,j^{{\rm matter}}_{\mu}=j^{{\rm top.}}_{\mu}
\lab{niceequiv}
\ee
with
\br 
j^{{\rm matter}}_{\mu}&=& 
-{\bar \psi}^1\,\gamma_{\mu}\, \psi^1 
+ 2\, {\bar \psi}^2\,\gamma_{\mu}\, \psi^2
+ {\bar \psi}^3\,\gamma_{\mu}\, \psi^3
+\frac{1}{2\,m}\,\varepsilon_{\mu\nu}\, 
{\bar \psi}^0\,\gamma^{\nu}\, \psi^0
\nonu\\
j^{{\rm top.}}_{\mu}&=& \frac{1}{2\,\pi}\, 
\varepsilon_{\mu\nu}\,\partial^{\nu}\vp
\lab{topmattercurr}
\er 
with $\varepsilon_{\mu\nu}$ being an antisymmetric symbol such that
$\varepsilon_{01}=1$.  The fact that $j^{{\rm top.}}_{\mu}$ is
trivially conserved, i.e. it is a topological current, then the
equivalence \rf{niceequiv} implies that $j^{{\rm matter}}_{\mu}$ is
also conserved. So, we have  
\be 
\partial^{\mu} j^{{\rm matter}}_{\mu}=0 \qquad\qquad\qquad 
\partial^{\mu}j^{{\rm top.}}_{\mu} =0
\ee
The equivalence \rf{niceequiv} between matter and topological currents
has interesting physical consequence, as already pointed out in
similar models in \cite{matter,achic,bueno}. From \rf{niceequiv} it
follows that the charge density $j^{{\rm matter}}_{0}$ is proportional
to the space derivative of the scalar field $\vp$. Therefore, for
those solutions where $\vp$ is constant everywhere except for a small
region in space, like in a kink type solution,  the charge density
will also have to be localized. That means that the charge gets
confined inside the soliton, and outside it we can have only zero
charge  states. The equivalence \rf{niceequiv} thus provides a
confinement mechanism for the charge associated to the current
$j^{{\rm matter}}_{\mu}$.  

\subsection{Soliton solutions}
\label{subsec:solutions}

We construct the solutions for the theory
\rf{eqmotl=61}-\rf{eqmotl=66} using the dressing method
\cite{dressing}, in a manner similar to that used in section
\ref{sec:dressing} for the Bullough-Dood model. We shall consider two
vacuum solutions as the seeds for the dressing method. The first and
simplest vacuum is a solution of \rf{eqmotl=61}-\rf{eqmotl=66} where 
\be 
\vp_{{\rm vac}_1} = 0 \qquad \eta_{{\rm vac}_1} = 0 \qquad 
\nu_{{\rm vac}_1} = -2\,m^2\, x_{+}\, x_{-} \qquad 
\psi^a_{{\rm vac}_1} = 0 \qquad {\tilde \psi}^i_{{\rm vac}_1} =0
\lab{vacuum1}
\ee
with $a=0,1,2,3$ and $i=1,2,3$. The second vacuum solution we consider is
\br  
\vp_{{\rm vac}_2} &=& 0 \qquad \quad\eta_{{\rm vac}_2} = 0 
\lab{vacuum2}\\
\nu_{{\rm vac}_2} &=& -2\,m^2\, x_{+}\, x_{-} -\frac{1}{2}\,
\rho_{+}\(x_{+}\)\,\rho_{-}\(x_{-}\) + \sigma_{+}\(x_{+}\) +
\sigma_{-}\(x_{-}\)\nonu\\ 
\psi^i_{{\rm vac}_2} &=& 0 \qquad \quad{\tilde \psi}^i_{{\rm vac}_2} =0
\qquad \quad \(\psi^0_R\)_{{\rm vac}_2}=\rho_{+}^{\prime}\(x_{+}\)\qquad\quad 
\(\psi^0_L\)_{{\rm vac}_2}=\rho_{-}^{\prime}\(x_{-}\)\nonu
\er 
The vacuum \rf{vacuum1} can be considered as a particular case of
vacuum \rf{vacuum2}, where one takes $\rho_{\pm}=\sigma_{\pm}=0$. The
spectrum of soliton solutions  obtained, through the dressing method,
from the vacuum \rf{vacuum1}, using the solitonic specialization
\cite{olivesolitonicspec},  is however much richer than that obtained
from vacuum \rf{vacuum2}. But  the general analysis of the dressing
method can be done for vacuum \rf{vacuum2}, and only when the explicit
evaluation of the solution is needed, one specializes to one of the
vacua.  

The zero curvature potentials \rf{zcpotbdl=6} evaluated on the
va\-cuum \rf{vacuum2} takes the form 
\be 
A_{+}^{{\rm vac}} = -\Omega_{+} \qquad \qquad 
A_{-}^{{\rm vac}} = \Omega_{-} -\partial_{-}\nu_{{\rm vac}_2}\, C
\lab{vacl=6pot}
\ee
where we have denoted
\be 
\Omega_{+} = E_{6}+\rho_{+}^{\prime}\(x_{+}\)\,L_0^{1/2}\qquad \qquad 
\Omega_{-} = E_{-6}+\rho_{-}^{\prime}\(x_{-}\)\,L_0^{-1/2}
\lab{omegapmdef}
\ee
The potentials \rf{vacl=6pot} can be written in the pure gauge form as 
\be 
A_{\mu}^{{\rm vac}} = - \partial_{\mu}\Psi_{{\rm vac}}\, \Psi_{{\rm vac}}^{-1}
\ee
with
\be 
\Psi_{{\rm vac}} = e^{x_{+}\,E_{6}+\rho_{+}\(x_{+}\)\,L_0^{1/2}}\, 
e^{-x_{-}\,E_{-6}-\rho_{-}\(x_{-}\)\,L_0^{-1/2}}\, e^{\sigma_{-}\,C}
\lab{psivacl=6def}
\ee
The dressing method works for the vacuum solutions \rf{vacuum1} and
\rf{vacuum2}, and potentials \rf{zcpotbdl=6}, in the very same way as
in section \ref{sec:dressing}. Indeed, we consider a constant group
element $h$ such that $\Psi_{\rm vac}$, given in \rf{psivacl=6def},
admitis the Gauss type decomposition  
\be
\Psi_{\rm vac}\, h \, \Psi_{\rm vac}^{-1} = G_{-}\, G_{0}\, G_{+}
\lab{hdecompl=6}
\ee
such that $G_{-}$, $ G_{0}$ and $G_{+}$ are group elements obtained
by exponentiating the generators of $A_2^{(2)}$ with  negative, zero
and positive grades respectively, of the principal gradation
\rf{gradation} defined by $Q$ given in \rf{gradingop}. Then, we define
the group element  
\be
\Psi_h \equiv G_{0}^{-1}\, G_{-}^{-1}\, \Psi_{\rm vac} \, h= G_{+}\,
\Psi_{\rm vac} 
\lab{psihdefl=6}
\ee
and the transformed potentials
\be
A_{\mu}^{h} \equiv -\partial_{\mu}\Psi_h \, \Psi_h^{-1}
\lab{amuhdefl=6}
\ee
Since $A_{\mu}^{h}$ is of the pure gauge it is a solution of the zero
curvature condition \rf{zc} for the potentials
\rf{zcpotbdl=6}. Replacing \rf{psihdefl=6} into \rf{amuhdefl=6} one
gets  
\br
A_{\mu}^{h} &=& G_{+}\, A_{\mu}^{\rm vac}\,G_{+}^{-1}-\partial_{\mu}
G_{+}\, G_{+}^{-1} 
\lab{postransfl=6}\\
&=& \(G_{0}^{-1}\, G_{-}^{-1}\)\, A_{\mu}^{\rm vac}\,\(G_{0}^{-1}\,
G_{-}^{-1}\)^{-1}-\partial_{\mu} 
\(G_{0}^{-1}\, G_{-}^{-1}\)\, \(G_{0}^{-1}\, G_{-}^{-1}\)^{-1} 
\lab{negtransfl=6}
\er
Notice that. \rf{postransfl=6} implies that $A_{\mu}^{h}$ has
components of grades  
greater or equal than those of $ A_{\mu}^{\rm vac}$. In addition,
\rf{negtransfl=6}  implies that $A_{\mu}^{h}$ has components of 
grades 
smaller or equal than those of $ A_{\mu}^{\rm vac}$. Since
\rf{hdecompl=6} and \rf{psihdefl=6} guarantee that both relations hold
true, it follows  
that $A_{\mu}^{h}$ has the same grade components as $ A_{\mu}^{\rm
  vac}$, and so as the potentials \rf{zcpotbdl=6}. By construction we
have $A_{\mu}^{h}$ given explicitly in terms of the space-time 
coordinates $x_{\pm}$. Therefore by equating it to \rf{zcpotbdl=6}  we get the
solutions for the fields of the model, associated to the choice $h$ of
the constant group element, i.e. a point on the orbit of solutions of the
vacuum \rf{vacuum1} or \rf{vacuum2}. However, the procedure of
equating \rf{zcpotbdl=6} to $A_{\mu}^{h}$ can reveal complex and
requires some care, as we now explain.  

Taking the zero grade part of the $x_{-}$ component of
\rf{negtransfl=6}, and using \rf{vacl=6pot}, we have  
\be 
\(A_{-}^h\)_0 = -\partial_{-}G_0^{-1}\,G_0-\partial_{-}\nu_{{\rm vac}_2}\, C
\ee
Comparing with the zero part of $A_{-}$ given in \rf{zcpotbdl=6}, 
\be 
\(A_{-}\)_0 =- \partial_{-} B\, B^{-1}
\ee
we get
\be
G_0^{-1}= B\,e^{-\nu_{{\rm vac}_2}\, C}=e^{\vp\, T_3^0+
  \(\nu-\nu_{{\rm vac}_2}\)\, C} 
\lab{bg0rell=6}
\ee
where we used \rf{bfield}, and have set $\eta$ to zero. The reason is
that we have started with vacua \rf{vacuum1}-\rf{vacuum2} where the
$\eta$ field is zero. Since the grading operator $Q$, defined in
\rf{gradingop}, can not be the result of any commutator (since it
contais $D$), it follows that the dressing transformation does not
excite $\eta$.  

Comparing the positive grade part of  $A_{+}^h$ in  \rf{negtransfl=6},
with $A_{+}$ in \rf{zcpotbdl=6}, which has only positive parts, and
using  \rf{vacl=6pot} and \rf{bg0rell=6}, we get,  
\be 
\(G_{-}^{-1}\, \Omega_{+}\,G_{-}\)_{>0} = E_{+6}+\sum_{m=-1}^5 F_{+}^m
\lab{gminustofields1}
\ee
Comparing the negative part of $A_{-}^{h}$ in  \rf{postransfl=6},  with the negative part of $A_{-}$ in \rf{zcpotbdl=6} we get 
\be 
\(G_{+}\, \Omega_{-}\,G_{+}^{-1}\)_{<0}=E_{-6}+ \sum_{m=1}^{5} F_{-}^{m}
\lab{gplustofields1}
\ee
Now, comparing the components of grades   $1$ and $2$ of $A_{+}^h$ in \rf{postransfl=6}, with those same grade components of $A_{+}$ in \rf{zcpotbdl=6} we obtain
\be 
\(\partial_{+}G_{+}\,G_{+}^{-1}\)_m = B\,  F_{+}^{m}\, B^{-1} \qquad\qquad\qquad m=1,2
\lab{gplustofields2}
\ee
We do not include the grades $3$ and higher, because $A_{+}^{{\rm vac}}$ (see \rf{vacl=6pot}) contains $L_0^{1/2}$ which has grade $3$  and so the term $G_{+}\, A_{+}^{\rm vac}\,G_{+}^{-1}$ in \rf{postransfl=6} would have to be considered. 

Similarly, comparing the components of grades   $-1$ and $-2$ of $A_{-}^h$ in \rf{negtransfl=6}, with those same grade components of $A_{-}$ in \rf{zcpotbdl=6}, and using \rf{bg0rell=6}, we obtain
\be 
\(\partial_{-}G_{-}^{-1}\,G_{-}\)_{-m} = -B^{-1}\,  F_{-}^{m}\, B \qquad\qquad\qquad m=1,2
\lab{gminustofields2}
\ee
From \rf{psivacl=6def} and \rf{hdecompl=6} we have that, given the choice of $h$,   $G_0$, $G_{+}$ and $G_{-}$ are given explicitly as functions of the the space-time coordinates. Therefore, the relations \rf{bg0rell=6}-\rf{gminustofields2},  provide  ways to get the explicit solutions for the fields contained in $B$, $F_{+}^m$ and $F_{-}^m$. Notice that if we write 
\be 
G_{+}=\exp\(\sum_{n=1}^{\infty} t^{(n)}\) \qquad \qquad 
G_{-}^{-1}=\exp\(\sum_{n=1}^{\infty} t^{(-n)}\)
\ee
with $t^{(\pm n)}$ being gnenerators of grades $\pm n$, we have from \rf{gminustofields1} and \rf{gplustofields1} that 
\br
F_{+}^{6-n} &=& \sbr{t^{(-n)}}{E_{+6}} + \mbox{\rm terms involving
  $t^{(-l)}$ with $l<n$}\nonu\\ 
F_{-}^{6-n} &=& \sbr{t^{(n)}}{E_{-6}} + \mbox{\rm terms involving
  $t^{(l)}$ with $l<n$}\nonu 
\er 
Therefore starting with $n=1$, we can recursively  relate
$F_{+}^{6-n}$ to $t^{(-l)}$,  and $F_{-}^{6-n}$ to $t^{(l)}$, with
$l\leq n$. However, at grades $\pm 3$ we have $t^{(-3)} \sim
L_0^{-1/2}$ and $t^{(3)} \sim L_0^{1/2}$, which commute with $E_{\pm
  6}\sim T_3^{\pm 1}$, and so  the recursion relations break
down. Therefore, our strategy is to use \rf{gminustofields1} and
\rf{gplustofields1} to relate $F_{\pm}^5$, $F_{\pm}^4$ and $F_{\pm}^3$
to the $t^{(\pm l)}$'s, and them to use \rf{gplustofields2} and
\rf{gminustofields2}  to relate $F_{\pm}^1$ and $F_{\pm}^2$ to the
same $t^{(\pm l)}$'s, but now involving derivatives of them.  

In order to extract the explicit space-time dependence of the $t^{(\pm
  l)}$'s we consider matrix elements of both sides of relation
\rf{hdecompl=6} in  highest weight state representations of the
$A_2^{(2)}$ algebra. In fact we use the two fundamental
representations of $A_2^{(2)}$ (see appendix \ref{app:kacmoody}). The
relevant matrix elements which we need to express the  $t^{(\pm l)}$'s
are  
\br
\tau_0 &=& \bra{\lambda_0} \Psi_{\rm vac}\, h\, \Psi_{\rm
  vac}^{-1}\ket{\lambda_0} 
\qquad\qquad\quad
\tau_1 = \bra{2\lambda_1} \Psi_{\rm vac}\, h\, \Psi_{\rm
  vac}^{-1}\ket{2\lambda_1} 
\lab{taudefl=6}\\
\tau_{R,1}&=& \bra{2\lambda_1} T_{+}^0\, \Psi_{\rm vac}\, h\,
\Psi_{\rm vac}^{-1}\ket{2\lambda_1} 
\qquad \quad\quad
{\tilde \tau}_{L,1}= \bra{2\lambda_1}  \Psi_{\rm vac}\, h\, \Psi_{\rm
  vac}^{-1}\, T_{-}^0\,\ket{2\lambda_1} 
\nonu\\
\tau_{R,2}&=& \bra{\lambda_0} L_{-2}^{1/2}\, \Psi_{\rm vac}\, h\,
\Psi_{\rm vac}^{-1}\ket{\lambda_0} 
\qquad\quad\quad
{\tilde \tau}_{L,2}=\bra{\lambda_0}  \Psi_{\rm vac}\, h\, \Psi_{\rm
  vac}^{-1}\,L_{2}^{-1/2}\,\ket{\lambda_0} 
\nonu\\
\tau_{R,3(0)}&=& \bra{\lambda_0} L_{-1}^{1/2}\, \Psi_{\rm vac}\, h\,
\Psi_{\rm vac}^{-1}\ket{\lambda_0} 
\qquad\quad\quad
{\tilde \tau}_{L,3(0)}=\bra{\lambda_0}  \Psi_{\rm vac}\, h\, \Psi_{\rm
  vac}^{-1}\,L_{1}^{-1/2}\,\ket{\lambda_0} 
\nonu\\
\tau_{R,3(1)}&=& \bra{2\lambda_1} L_{-1}^{1/2}\, \Psi_{\rm vac}\, h\,
\Psi_{\rm vac}^{-1}\ket{2\lambda_1} 
\quad\quad
{\tilde \tau}_{L,3(1)}=\bra{2\lambda_1}  \Psi_{\rm vac}\, h\,
\Psi_{\rm vac}^{-1}\,L_{1}^{-1/2}\,\ket{2\lambda_1} 
\nonu
\er
Then, using relations \rf{bg0rell=6}-\rf{gminustofields2} in the way
explained above, we obtain the fileds in terms of the tau-functions
(or equivalently of the $t^{(\pm l)}$'s) 
\br
\vp&=& 
\log \frac{\tau_0}{\tau_1}
\nonu\\
\nu&=&-\frac{1}{2} \log \tau_0 -2 m^2 x_{+}x_{-} -\frac{1}{2} \, 
\rho_{+}\(x_{+}\)\rho_{-}\(x_{-}\)+\sigma_{+}\(x_{+}\) +\sigma_{-}\(x_{-}\)
\nonu\\
\psi_R^1&=& \frac{1}{\sqrt{m}}\, 
\frac{\tau_1}{\tau_0}\; \partial_{+}\(\frac{{\tilde \tau}_{L,1}}{\tau_1}\)
\nonu\\
\psi_L^1&=& 
-\sqrt{m}\; \frac{ {\tilde \tau}_{L,1}}{\tau_1 }
\nonu\\
\psi_R^2&=& \frac{1}{\sqrt{2 m}}\;
\frac{\tau_0^2}{\tau_1^2}\; \partial_{+}\(\frac{{\tilde \tau}_{L,2}}{\tau_0}\)
\nonu\\
\psi_L^2&=& -\sqrt{2 m}  \;\frac{ {\tilde \tau}_{L,2}}{\tau_0}
\nonu\\
\psi_R^3&=& \frac{1}{\sqrt{m}}\;\frac{\tau_0}{\tau_1}\; \left[
\frac{{\tilde \tau}_{L,1}}{\tau_1}\; \partial_{+}\(\frac{{\tilde
    \tau}_{L,2}}{\tau_0} \) 
-\partial_{+}\(\frac{{\tilde \tau}_{L,3(0)}}{\tau_0}\)\right]
\nonu\\
\psi_L^3&=&
-\sqrt{m}\; 
\left(\frac{{\tilde \tau}_{L,3(0)}}{\tau_0}-
\frac{4 \, {\tilde \tau}_{L,3(1)}}{\tau_1}\right)
\nonu\\
{\tilde \psi}_R^1 &=&-\sqrt{m}\; \frac{\tau_{R,1}}{ \tau_1}
\lab{taufieldsrell=6}\\
{\tilde \psi}_L^1&=& -\frac{1}{\sqrt{m}}\; 
\frac{\tau_1}{\tau_0}\; \partial_{-}\(\frac{\tau_{R,1}}{\tau_1}\)
\nonu\\
{\tilde \psi}_R^2&=& -\sqrt{2 m}\; \frac{ \tau_{R,2}}{ \tau_0}
\nonu\\
{\tilde \psi}_L^2&=& -\frac{1}{\sqrt{2 m}}\; 
\frac{\tau_0^2}{\tau_1^2}\;\partial_{-}\(\frac{\tau_{R,2}}{\tau_0}\)
\nonu\\
{\tilde \psi}_R^3&=& \sqrt{m}\; 
\left(\frac{\tau_{R,3(0)}}{\tau_0}-\frac{4 \,\tau_{R,3(1)}}{\tau_1}\right)
\nonu\\
{\tilde \psi}_L^3&=& -\frac{1}{\sqrt{m}}\; \frac{\tau_0}{\tau_1}\;\left[ 
\frac{\tau_{R,1}}{\tau_1}\;\partial_{-}\(\frac{\tau_{R,2}}{\tau_0} \) 
+ \partial_{-}\(\frac{\tau_{R,3(0)}}{\tau_0}\)\right]
\nonu\\
\psi_R^0&=& \rho_{+}^{\prime}\(x_{+}\)+\sqrt{6}\, m\; 
\frac{ \tau_{R,1} \tau_{R,3(1)}}{\tau_1^2}
\nonu\\
\psi_L^0&=& \rho_{-}^{\prime}\(x_{-}\)-
\sqrt{6}\, m\; \frac{ {\tilde \tau}_{L,1} {\tilde \tau}_{L,3(1)}}{\tau_1^2}
\nonu
\er

Now comes the point to make a distinction between the two vacua
\rf{vacuum1} and \rf{vacuum2}. As explained in \rf{hnsolitonbd} we
construct the soliton solutions using the so-called solitonic
specialization procedure \cite{olivesolitonicspec}. According to that,
we take $h$ to be a product of exponentials of eigenvectors of the
operators $\Omega_{\pm}$ appearing in $A_{\mu}^{{\rm vac}}$, given in
\rf{vacl=6pot}. Now, for the vacuum \rf{vacuum1}, $\Omega_{\pm}$
reduces to $E_{\pm 6}$. Therefore,  the set of  of eigenvectors
associated to the vacua \rf{vacuum1} and \rf{vacuum2} are different,
and so the spectrum of soliton solutions one obtains through the
dressing is also different.  

The eigenvectors of $E_{\pm 6}$, with non-vanishing eigenvalues, are
given by\footnote{The free parameter $\zeta$ has the same origin as the
  one explained in \rf{puregradeeigenvector1}.}
\br 
V_{\pm 1}^{(T)}\(\zeta\) &=& \sum_{n=-\infty}^{\infty} \zeta^{-(6\, n\pm 1)}\,
T_{\pm}^n \ 
\nonu\\
V_{\pm 1}^{(L)}\(\zeta\) &=& \sum_{n=-\infty}^{\infty} \zeta^{-(6\, n+3\pm 1)
}\, L_{\pm 1}^{n+1/2}  
\lab{eigenvectorepm6}\\
V_{\pm 2}^{(L)} \(\zeta\)&=& \sum_{n=-\infty}^{\infty} \zeta^{-(6\, n+3\pm 2)
}\, L_{\pm 2}^{n+1/2} \nonu 
\er 
and satisfying 
\br 
\sbr{E_{+6}}{V_{\pm 1}^{(T)}\(\zeta\)} &=&\pm m\, \zeta^6\, V_{\pm
  1}^{(T)}\(\zeta\) 
\qquad \quad 
\sbr{E_{-6}}{V_{\pm 1}^{(T)}\(\zeta\)} =\pm m\, \zeta^{-6}\, V_{\pm
  1}^{(T)}\(\zeta\) 
\nonu\\
\sbr{E_{+6}}{V_{\pm 1}^{(L)}\(\zeta\)} &=&\pm m\, \zeta^6\, V_{\pm
  1}^{(L)}\(\zeta\) 
\qquad \quad 
\sbr{E_{-6}}{V_{\pm 1}^{(L)}\(\zeta\)} =\pm m\, \zeta^{-6}\, V_{\pm
  1}^{(L)}\(\zeta\) 
\nonu\\
\sbr{E_{+6}}{V_{\pm 2}^{(L)}\(\zeta\)} &=&\pm 2\,m\, \zeta^6\, V_{\pm
  2}^{(L)}\(\zeta\) 
\qquad \quad 
\sbr{E_{-6}}{V_{\pm 2}^{(L)}\(\zeta\)} =\pm 2\,m\, \zeta^{-6}\, V_{\pm
  2}^{(L)}\(\zeta\) 
\nonu
\er 
Notice that the eigenvalues $m\, \zeta^{\pm 6}$ of $E_{\pm 6}$, are four
fold degenerate, with the corresponding degenerate eigenvectors being  
$V_{+ 1}^{(T)}\(\zeta\)$, $V_{- 1}^{(T)}\(\omega\, \zeta\)$,
$V_{+1}^{(L)}\(\zeta\)$, and $V_{- 1}^{(L)}\(\omega\, \zeta\)$, with $\omega$
a sixth root of $-1$, i.e. $\omega^6=-1$. Similarly, the eigenvalues
$2\, m\, \zeta^{\pm 6}$ of $E_{\pm 6}$, are two fold degenerate, and the
corresponding degenerate eigenvectors are $V_{+2}^{(L)}\(\zeta\)$, and
$V_{- 2}^{(L)}\(\omega\, \zeta\)$. Notice that the six possible 
choices for $\omega$ do increase the degeneracy. Indeed, if $\omega$
and $\gamma\,\omega$, with $\gamma^6=1$, are two sixth roots of $-1$,
then $V_{- 1}^{(T)}\(\gamma\, \omega\, \zeta\)= \gamma^{\mp 1}\, V_{-
  1}^{(T)}\( \omega\, \zeta\)$, and so one gets the same eigenvector. The
same holds true for the other type of eigenvectors. Clearly, the
eigenvalues $-m\, \zeta^{\pm 6}$ and $-2\,m\, \zeta^{\pm 6}$ are also four
fold and two fold degenerate respectively, with the set of degenerate
eigenvectors being $\(V_{+ 1}^{(T)}\(\omega\, \zeta\), V_{- 1}^{(T)}\( \zeta\),
V_{+1}^{(L)}\(\omega\, \zeta\), V_{- 1}^{(L)}\( \zeta\)\)$, and
$\(V_{+2}^{(L)}\(\omega\, \zeta\),V_{- 2}^{(L)}\( \zeta\)\)$, respectively. 

Notice that $L_0^{\pm 1/2}$ commutes with $L_{\pm
  2}^{n+1/2}$. Therefore, $V_{\pm 2}^{(L)} \(\zeta\)$ given in
\rf{eigenvectorepm6}, has the same eigenvalue of $\Omega_{\pm}$, given
in \rf{omegapmdef}, as that of $E_{\pm 6}$, i.e. 
\be
\sbr{\Omega_{+}}{V_{\pm 2}^{(L)}\(\zeta\)} =\pm 2\,m\, \zeta^6\, V_{\pm 2}^{(L)}\(\zeta\)
\qquad \quad 
\sbr{\Omega_{-}}{V_{\pm 2}^{(L)}\(\zeta\)} =\pm 2\,m\, \zeta^{-6}\, V_{\pm
  2}^{(L)}\(\zeta\) 
\ee
As for the subspace spanned by  $V_{\pm 1}^{(T)}\(\zeta\)$ and $V_{\pm
  1}^{(L)}\(\zeta\)$, we get four non-degenerate eigenvectors of
$\Omega_{\pm}$. Introducing 
\be
V_{\varepsilon_1}^{(\varepsilon_2)}\(\zeta\)= V_{\varepsilon_1}^{(T)}\(\zeta\)+
\varepsilon_2 V_{\varepsilon_1}^{(L)}\(\zeta\)
\lab{eigenvectore1e2}
\ee
with $ \varepsilon_1,\varepsilon_2=\pm 1$, one gets
\br
\sbr{\Omega_{+}}{V_{\varepsilon_1}^{(\varepsilon_2)}\(\zeta\)} &=& 
\left[\varepsilon_1\, m\, \zeta^6-\varepsilon_2\, \sqrt{\frac{3}{2}}\;
  \zeta^3\, \rho_{+}^{\prime}\(x_{+}\)\right]\;
V_{\varepsilon_1}^{(\varepsilon_2)}\(\zeta\) \\
\sbr{\Omega_{-}}{V_{\varepsilon_1}^{(\varepsilon_2)}\(\zeta\)} &=& 
\left[\varepsilon_1\, m\, \zeta^{-6}-\varepsilon_2\, \sqrt{\frac{3}{2}}\;
  \zeta^{-3}\, \rho_{-}^{\prime}\(x_{-}\)\right]\;
V_{\varepsilon_1}^{(\varepsilon_2)}\(\zeta\)\nonu
\er

The soliton solutions are constructed taking the constant group
element of the dressing method, introduced in \rf{hdecompl=6}, as the
exponential, or product of exponentials, of the eigenvectors of
$E_{\pm 6}$, in the case of vacuum \rf{vacuum1}, or of $\Omega_{\pm}$, in
the case of vacuum \rf{vacuum2}. That is the same solitonic
specialization procedure \cite{olivesolitonicspec}  we used for the
Bullough-Dodd solitons in \rf{hnsolitonbd}. The explicit soliton solutions are
then obtained by evaluating the matrix elements, given in
\rf{taudefl=6}, in the two fundamental representations of the
Kac-Moody algebra $A_2^{(2)}$. 

The evaluation of those matrix elements are in general very laborious,
and we use here an hybrid of the dressing and Hirota methods as
explained in \cite{bueno}. The dressing method is very powerful and
reveals the  mathematical structure behind soliton solutions, namely
integrable highest weight state representations of Kac-Moody
algebras \cite{kac}. The Hirota method is a very direct one, and the
calculations 
in it are easy to implement in a computer algorithm using an algebraic
manipulation program like Mathematica or Maple. However, it has a serious
drawback, since it does not give a concrete way of finding the
transformation among the tau functions and the fields. The main
property of the Hirota's tau functions is that they truncate at some
finite order when the Hirota's ansatz is used. The dressing method
provides the algebraic explanation of why that truncation occurs. The
matrix elements  \rf{taudefl=6} correspond to the Hirota's tau
functions and their truncation comes from the nilpotency of the
operators, in the integrable representations of the Kac-Moody algebra,
corresponding to the eigenvectors of $E_{\pm 6}$ or of
$\Omega_{\pm}$. In addition, the dressing method does give the
concrete relations among the tau functions and the fields. For the
model under consideration, those are the relations
\rf{taufieldsrell=6}. Therefore our hybrid method consists of using the
dressing method up to the point of obtaining the relations
\rf{taufieldsrell=6}, then we replace those into the equations of
motion \rf{eqmotl=61}-\rf{eqmotl=66} to obtain the Hirota's equations
for the tau functions.  We then solve those equations using Hirota's 
recurrence  method implemented in a computer algorithm  using the Mathematica
program. In addition, we use the Hirota's ansatz for the tau
functions, that follows from the choice of $h$ as exponentials of the
eigenvectors of $E_{\pm 6}$ or $\Omega_{\pm}$, in a way similar to
that done in \rf{hnsolitonbd}, \rf{tausolitonicspec} and \rf{hiransatzcbd}

We point out however that the spectrum of soliton solutions we obtain
through such hybrid method is richer than that obtained by the
solitonic specialization procedure. The Hirota's recurrence method
presents some degeneracies at higher order that makes the solutions to
depend upon a number of parameters higher than that expect from the
analysis of the degenracy of the eigenvectors of $E_{\pm 6}$ and
$\Omega_{\pm}$, done above.  That happens for instance in the
soliton solution discussed in section \ref{subsec:mass1}. The one
soliton solutions for the abelian affine Toda models also present an
enlarged spectrum as reported in \cite{hirotanpb}. 

We give in the
following sections the various types of soliton solutions. It is worth
mentioning that all soliton solutions have the property that the
chiral currents  \rf{chiralcurrj} and \rf{chiralcurrjb} vanish when
evaluated on them. Therefore, in all those solutions it is valid the
equivalence \rf{niceequiv} between  the
matter and topological currents. Consequently, it holds true the confinement
mechanism of the charge associated to the current $j^{{\rm
    matter}}_{\mu}$ given in \rf{topmattercurr}, and 
discussed at the end of section \ref{subsec:symmetries}.

\subsection{One-soliton solutions from vacuum \rf{vacuum1}}
\label{sec:solvac1}

\subsubsection{One-soliton associated to  $V_{\pm 2}^{(L)}$}
\label{subsec:mass2}

We construct here the solution associated to the choice of the
constant group element in \rf{psihdefl=6} as $h=e^{V}$ with $V$ being
a linear combination of the degenerate eigenvectors $V_{+2}^{(L)}\(\zeta\)$ and
$V_{-2}^{(L)}\(\omega\,\zeta\)$, defined in \rf{eigenvectorepm6},
  associated to the eigenvalues 
  $2\,m\,\zeta^{\pm 6}$ of $E_{\pm 6}$. Therefore, the
  solution depends upon two parameters. By replacing that $h$ into
  \rf{taudefl=6} with $\Psi_{\rm vac}$ given by \rf{psivacl=6def} with
  $\rho_{\pm}=\sigma_{\pm}=0$,  we get the following
   ansatz for the tau functions 
\be
\tau_{*} = \sum_{l=0}^N \delta_{*}^{(l)}\, e^{l\, \Gamma_2} 
\ee
with 
\be
\Gamma_2= 2\; m \( z\, x_{+}- \frac{x_{-}}{z}\)= \varepsilon\,
\gamma_2\, \( x - v\, t\)  
\ee
where we have denoted $\zeta^6\equiv z=\varepsilon\, e^{-\theta}$, with
$\varepsilon=\pm 1$, and $\theta$ real, and so 
\be
\gamma_2= 2\, m\, \cosh \theta \qquad \qquad v = c\, \tanh \theta
\ee
As explained at the end of section \ref{subsec:solutions}, we solve
the equations for the tau functions, using the Hirota's method
implemented in a computer
algorithm, and find that the expansion
truncates at order $N=4$, with the solution given by
\br
\tau_0&=&
\(1+\frac{a_2 \; {\bar a}_2}{4} \; e^{2\; \Gamma_2 } \)
\(1-\frac{a_2 \; {\bar a}_2}{4} \; e^{2\; \Gamma_2 } \)
\nonu\\
\tau_1&=& 
\(1-\frac{a_2 \; {\bar a}_2}{4} \; e^{2\; \Gamma_2 } \)^2
\nonu\\
{\tilde \tau}_{L,2}&=& 
a_2\; e^{\Gamma_2 }\(1-\frac{a_2 \; {\bar a}_2}{4} \; e^{2\; \Gamma_2 } \)
\nonu\\
\tau_{R,2}&=& 
{\bar a}_2\; e^{\Gamma_2 }\(1-\frac{a_2 \; {\bar a}_2}{4} \; e^{2\;
  \Gamma_2 } \) 
\er
where $a_2$ and ${\bar a}_2$ are the two parameters of the solution mentioned
above and associated to the degeneracy of the eigenvalues
$2\,m\,\zeta^{\pm 6}$, of $E_{\pm 6}$.  

Using relations \rf{taufieldsrell=6}, one can write the solution in
terms of the fileds as 
\br
\vp&=& \log \(\frac{1+\frac{a_2 \; {\bar a}_2}{4} \; e^{2\; \Gamma_2
}}{1-\frac{a_2 \; {\bar a}_2}{4} \; e^{2\; \Gamma_2 }}\) \nonu\\ 
\nu&=& -\frac{1}{2}\, \log \( 1- \frac{a_2^2 \; {\bar a}_2^2}{16} \;
e^{4\; \Gamma_2 }\) -2 m^2 x_{+}x_{-}  \nonu
\er

\br
\psi^2= \sqrt{2 m}\, a_2\, e^{\Gamma_2} \; \( 
\begin{array}{c}
\frac{z}{1-\frac{a_2 \; {\bar a}_2}{4} \; e^{2\; \Gamma_2 }}\\
\frac{-1}{1+\frac{a_2 \; {\bar a}_2}{4} \; e^{2\; \Gamma_2 }}
\end{array}\)
\qquad 
{\tilde \psi}^2= \sqrt{2 m}\, {\bar a}_2\, e^{\Gamma_2} \; \( 
\begin{array}{c}
\frac{-1}{1+\frac{a_2 \; {\bar a}_2}{4} \; e^{2\; \Gamma_2 }}\\
\frac{1/z}{1-\frac{a_2 \; {\bar a}_2}{4} \; e^{2\; \Gamma_2 }}
\end{array}\)
\er

$$
\eta=\psi^1={\tilde \psi}^1=\psi^3={\tilde \psi}^3=\psi^0 =0
$$

From the fact that the solutions excites only the fields $\vp$, $\nu$,
$\psi^2$, and ${\tilde \psi}^2$, it is associated to the $sl(2)$
Kac-Moody subalgebra generated by $T_3^n$, $L_{\pm 2}^{n+1/2}$ and
$C$ (see apendix \ref{app:kacmoody}). In fact that solution is the
same as the one constructed in 
section 10.1 of ref. \cite{matter}, and also studied in
\cite{achic,blas}. The solution for $\vp$ is the same as that for the
sine-Gordon model. 

The chiral currents \rf{chiralcurrj} and \rf{chiralcurrjb} vanish when
evaluated on such solution, and so one has the equivalence
\rf{niceequiv} of the
matter and topological currents, i.e.  
\be
\varepsilon_{\mu\nu}\,\partial^{\nu}\vp = \frac{1}{2}\, 
 {\bar  \psi}^2\,\gamma_{\mu}\, \psi^2 
\ee
Therefore, we have here a confinement mechanism as discussed at the end of
section \ref{subsec:symmetries}, and proposed in \cite{matter}.

\subsubsection{One-soliton associated to $V_{\pm 1}^{(T)}$ and 
$V_{\pm 1}^{(L)}$} 
\label{subsec:mass1}

We now consider the solution obtained by taking the element $h$,
defined in \rf{psihdefl=6}, to be an exponentiation of a linear
combination of the four degenerated eigenvectors \rf{eigenvectorepm6}
of $E_{\pm 6}$, namely $V_{+ 1}^{(T)}\(\zeta\)$, 
$V_{- 1}^{(T)}\(\omega\, \zeta\)$, 
$V_{+1}^{(L)}\(\zeta\)$, and $V_{- 1}^{(L)}\(\omega\, \zeta\)$, with
$\omega^6=-1$.  By replacing that $h$ into
  \rf{taudefl=6} with $\Psi_{\rm vac}$ given by \rf{psivacl=6def} with
  $\rho_{\pm}=\sigma_{\pm}=0$,  we get the following
   ansatz for the tau functions 
\be
\tau_{*} = \sum_{l=0}^N \delta_{*}^{(l)}\, e^{l\, \Gamma_1} 
\ee
with 
\be
\Gamma_1=  m \( z\, x_{+}- \frac{x_{-}}{z}\)= \varepsilon\, \gamma_1\,
\( x - v\, t\)  
\lab{gamma1def}
\ee
where we have denoted $\zeta^6\equiv z=\varepsilon\, e^{-\theta}$, with
$\varepsilon=\pm 1$, and $\theta$ real, and so 
\be
\gamma_1=  m\, \cosh \theta \qquad \qquad v = c\, \tanh \theta
\ee
Since the solution  is associated to four degenerated eigenvectors, it
 should then depend upon four parameters. However, it depends in fact
 upon six parameters. The reason is that 
the Hirota's recursion method works in a such a way in this case that at
second order not all coefficients $\delta_{*}^{(2)}$ are determined
uniquely in terms of the  coefficients $\delta_{*}^{(1)}$. There remain two of
them free. Such solution with six free parameters truncate at order
$N=16$, and the coefficients $\delta_{*}^{(l)}$ of the tau functions
are given explicitly in appendix \ref{app:fullsol}. 

Despite the complexity of such solution, the chiral currents
\rf{chiralcurrj} and \rf{chiralcurrjb} vanish when 
evaluated on it. Therefore,  the equivalence
\rf{niceequiv} between the
matter and topological currents holds true, and so does the confinement
mechanism discussed at the end of section \ref{subsec:symmetries}.

Such solution takes some simpler forms for some special values of the
parameters. For instance, the tau functions, with coefficients given
in appendix \ref{app:fullsol}, truncate at order $8$ if one takes 
\be
b= \varepsilon_1 \;\sqrt{z} \; a \qquad \qquad 
{\bar b} = i\; \varepsilon_2 \;\frac{{\bar a}}{\sqrt{z}} 
\ee
with $\varepsilon_a=\pm 1$, $a=1,2$. 

However, the solution takes an even simpler form if one takes the
parameter to satisfy
\be
b= \varepsilon_1 \;\sqrt{z} \; a \; \qquad  \quad
{\bar b} = i\; \varepsilon_2 \;\frac{{\bar a}}{\sqrt{z}} \; \qquad \quad
c=-2\, \varepsilon_1\, a^2\, \sqrt{z}\; \qquad \quad 
{\bar c}= i\,2 \, \varepsilon_2 \, \frac{{\bar a}^2}{\sqrt{z}}
\ee
with $\varepsilon_a=\pm 1$, $a=1,2$. In this case the tau functions
truncate at order $4$ and are given by 
\br
\tau_0&=&
1 -8\, a \,{\bar a}\, e^{2\,\Gamma_1}+8 \,a^2 \,{\bar a}^2\, e^{4\,\Gamma_1}
\nonumber\\
\tau_1&=&
\left(1+2 \,\sqrt{2}\; e^{i\, \varepsilon_1
  \,\varepsilon_2\,\frac{3\,\pi}{4}}\,a \,{\bar a}
\,e^{2\,\Gamma_1}\right)^2 
\nonumber\\
{\tilde \tau}_{L,1}&=&
4 \,a \,e^{\Gamma_1}\, \left(1+2 \,\sqrt{2}\; e^{i\, \varepsilon_1
  \,\varepsilon_2\,\frac{3\,\pi}{4}} 
 \, a \,{\bar a}\, e^{2\,\Gamma_1}\right)
\nonumber\\
{\tilde \tau}_{L,2}&=&
4 \,i \,\varepsilon_2 \,\frac{{\bar a}^2}{\sqrt{z}}\; e^{2\,\Gamma_1}
\\
{\tilde \tau}_{L,3(0)}&=&
4 \,i \,\varepsilon_2 \,\frac{{\bar a}}{\sqrt{z}}\,  e^{\Gamma_1} \left(1+2
\,\sqrt{2}\; e^{-i\, \varepsilon_1
  \,\varepsilon_2\,\frac{3\,\pi}{4}}\, a \,{\bar a}\,  \,
e^{2\,\Gamma_1}\right) 
\nonumber\\
{\tilde \tau}_{L,3(1)}&=&
2 \,i\,\varepsilon_2\, \frac{{\bar a}}{\sqrt{z}}\,  e^{\Gamma_1}\, \left(1+2\, 
\,\sqrt{2}\; e^{i\, \varepsilon_1 \,\varepsilon_2\,\frac{3\,\pi}{4}}\,
a\, {\bar a} \, 
 \,e^{2\,\Gamma_1}\right)
\nonumber\\
\tau_{R,1}&=& 
4\, {\bar a}\, e^{\Gamma_1} \,\left(1+ 2 \,\sqrt{2}\; e^{i\,
  \varepsilon_1 \,\varepsilon_2\,\frac{3\,\pi}{4}}\, 
a \,{\bar a}\,  \
e^{2\,\Gamma_1}\right)
\nonumber\\
\tau_{R,2}&=&
-4\, \varepsilon_1 \,\sqrt{z}\; a^2 \, e^{2\,\Gamma_1} 
\nonumber\\
\tau_{R,3(0)}&=&
4\, \varepsilon_1\, \sqrt{z}\; a \, e^{\Gamma_1} \left(1+2\, 
\,\sqrt{2}\; e^{-i\, \varepsilon_1
  \,\varepsilon_2\,\frac{3\,\pi}{4}}\, a\, {\bar a}\,
e^{2\,\Gamma_1}\right)  
\nonumber\\
\tau_{R,3(1)}&=&
2\, \varepsilon_1\, \sqrt{z}\; a \, e^{\Gamma_1}\, \left(1+2\, 
\,\sqrt{2}\; e^{i\, \varepsilon_1 \,\varepsilon_2\,\frac{3\,\pi}{4}}\,
a\, {\bar a}\, e^{2\,\Gamma_1}\right)  
\nonumber
\er

\begin{figure}
\scalebox{1}{
\includegraphics{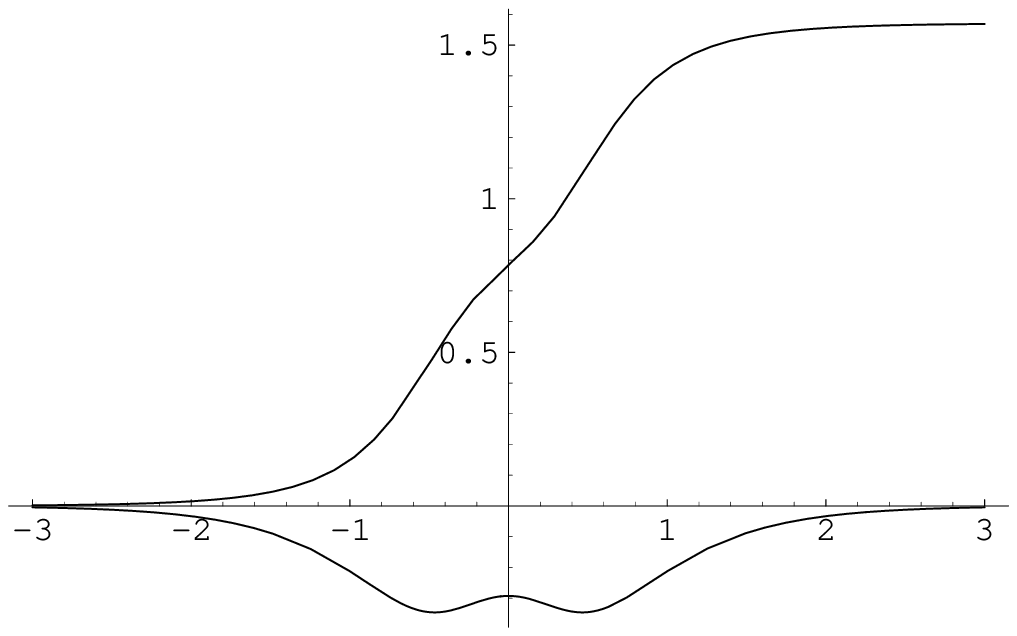}}
\caption{Plots of the real and imaginary parts of $\vp$ against $x$,
  for the one-soliton solution \rf{onesolvpmass1}, with 
   $\varepsilon=\varepsilon_1\,\varepsilon_2=1$,
  $\xi=-\frac{3\pi}{8}$. The real part of $\vp$ corresponds to the curve
  that goes to zero as $x\ra \pm \infty$. 
\label{fig:onesolmass1a} }   
\end{figure}

\begin{figure}
\scalebox{1}{
\includegraphics{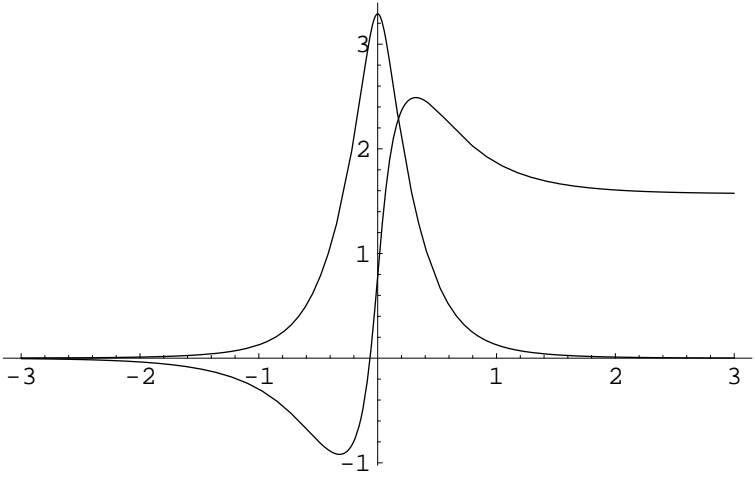}}
\caption{Plots of the real and imaginary parts of $\vp$ against $x$,
  for the one-soliton solution \rf{onesolvpmass1}, with
  $\varepsilon=\varepsilon_1\,\varepsilon_2=1$,
  $\xi=\frac{\pi}{3}$. The real part of $\vp$ corresponds to the curve
  that goes to zero as $x\ra \pm \infty$. 
\label{fig:onesolmass1b} }   
\end{figure}

The solution for $\vp$ can be obtained using the relation between
$\vp$ and the tau functions given in \rf{taufieldsrell=6}. Writing
$\vp=\vp_R+i\,\vp_I$, and choosing $a\, {\bar
  a}=e^{i\xi}/(2\sqrt{2})$,  we have
\br
\vp_R&=& \frac{1}{2}\,
\ln\frac{\cosh\(4\Gamma_1\)+4+\cos\(2\xi\)-4\sqrt{2}\,
  \cos\(\xi\)\cosh\(2\Gamma_1\)}
{\cosh\(4\Gamma_1\)+2+\cos\(2\(\xi+\varepsilon_1\,\varepsilon_2\,
\frac{3\pi}{4}\)\)+4\, \cos\(\xi+\varepsilon_1\,\varepsilon_2\,
\frac{3\pi}{4}\)\cosh\(2\Gamma_1\)} \nonu\\
\vp_I&=&{\rm ArcTan}\,\frac{{\cal A}}{{\cal B}}
\lab{onesolvpmass1}
\er
with
\br
{\cal A}&=&\varepsilon_1\,\varepsilon_2\,e^{4\Gamma_1}+
4\,\varepsilon_1\,\varepsilon_2\,+\sin\(2\xi\)+
\varepsilon_1\,\varepsilon_2\,\cos\(2\xi\)\nonu\\
&-&
\sqrt{2}\left[\(\sin\(\xi\)+
3\,\varepsilon_1\,\varepsilon_2\,\cos\(\xi\)\)\,e^{2\,\Gamma_1}+
\(\sin\(\xi\)+
\varepsilon_1\,\varepsilon_2\,\cos\(\xi\)\)\,e^{-2\,\Gamma_1}\right] \nonu\\
{\cal B}&=&e^{-4\Gamma_1}+4+\,\varepsilon_1\,\varepsilon_2\,\sin\(2\xi\)+
\cos\(2\xi\)\nonu\\
&-&
\sqrt{2}\left[\(\varepsilon_1\,\varepsilon_2\,\sin\(\xi\)+
\cos\(\xi\)\)\,e^{2\,\Gamma_1}+
\(\varepsilon_1\,\varepsilon_2\,\sin\(\xi\)+
3\,\cos\(\xi\)\)\,e^{-2\,\Gamma_1}\right]  \nonu
\er
The real part $\vp_R$ goes to zero as $x\ra \pm \infty$. The imaginary
part $\vp_I$  is very sensitive to variations in $\xi$. We give in
  figures   \ref{fig:onesolmass1a} and \ref{fig:onesolmass1b} the
  plots of the real and imaginary parts for $\xi=-\frac{3\pi}{8}$ and
  $\xi=\frac{\pi}{3}$, respectively, and in both we have
  $\varepsilon=\varepsilon_1\,\varepsilon_2=1$.

Some other simpler forms of the solution \ref{app:fullsol} are given
in the appendix \ref{app:truncation}.

\subsection{One-soliton solutions from  vacuum \rf{vacuum2}}
\label{sec:solvac2}

In this case we take the element $h$ introduced in \rf{psihdefl=6} to
be an exponential of one of the four (non-degenerated) eigenvectors
$V_{\varepsilon_1}^{(\varepsilon_2)}\(\zeta\)$, given in
\rf{eigenvectore1e2}.   
 By replacing that $h$ into
  \rf{taudefl=6} with $\Psi_{\rm vac}$ given by \rf{psivacl=6def},  we
  get the following 
   ansatz for the tau functions 
\be
\tau_{*} = \sum_{l=0}^N \delta_{*}^{(l)}\, e^{l\,
  \Gamma_{\varepsilon_1,\varepsilon_2} } 
\ee
with 
\be
\Gamma_{\varepsilon_1,\varepsilon_2}= \varepsilon_1\; m\; \(z \; x_{+}
-\frac{x_{-}}{z}\) +  
\varepsilon_2 \;\sqrt{\frac{3}{2}}\;\(\rho_{+}^{\prime}\(x_{+}\)
\;\sqrt{z}-\frac{\rho_{-}^{\prime}\(x_{-}\)}{\sqrt{z}}\) 
\ee
where we have denoted $\zeta^6\equiv z$. 

We solve the equations for the tau functions by the Hirota's method
implement on a computer (see end of section \ref{subsec:solutions}), and
find that the expansion 
truncates at order $N=2$. However, the dressing method in this case
does not excite the fields $\vp$, $\nu$, and $\psi^0$ (and also $\eta$
for the reasons explained below \rf{bg0rell=6}). Therefore, in
all four solutions these fields have their vacuum values, i.e.  
\br
\vp&=&0\nonu\\ 
\eta &=&0\nonu\\
\nu&=&-2\,m^2\, x_{+}\, x_{-} - \frac{1}{2} \rho_{+}\(x_{+}\)\,
\rho_{-}\(x_{-}\)  + \sigma_{-}\(x_{-}\) +  
\sigma_{+}\(x_{+}\)\nonu\\
\psi^0&=&\( \begin{array}{c}
\rho_{+}^{\prime}\(x_{+}\)\\
\rho_{-}^{\prime}\(x_{-}\)
\end{array}\)
\er
The non-vanishing tau functions in the four solutions are given by
\begin{enumerate}
\item For $\varepsilon_1=-1$ , $\varepsilon_2=-1$
\br
\tau_0&=& 1 \qquad\qquad \qquad \qquad\qquad \qquad \qquad\;\;
\tau_1= 1
\nonu\\
{\tilde \tau}_{L,2} &=& 
-\frac{1}{4} \, \sqrt{z}\, a_{--}^2\; e^{2 \;\Gamma_{--} }
\qquad\qquad \qquad
{\tilde \tau}_{L,3(0)} = 
 a_{--}\; e^{\Gamma_{--} }
\nonu\\
{\tilde \tau}_{L,3(1)} &=& 
\frac{1}{2} \, a_{--}\; e^{\Gamma_{--} }
\qquad\qquad \qquad\qquad\quad\;\;
\tau_{R,1} =
- \sqrt{z}\, a_{--}\; e^{\Gamma_{--} }
\nonu
\er

\item For $\varepsilon_1=-1$ , $\varepsilon_2=1$
\br
\tau_0&=& 1
\qquad\qquad \qquad \qquad\qquad \qquad \qquad\;\;
\tau_1= 1
\nonu\\
{\tilde \tau}_{L,2} &=& 
\frac{1}{4} \, \sqrt{z}\, a_{-+}^2 \; e^{2 \Gamma_{-+} }
\qquad\qquad \qquad \qquad\qquad 
{\tilde \tau}_{L,3(0)} = 
 a_{-+}\; e^{\Gamma_{-+} }
\nonu\\
{\tilde \tau}_{L,3(1)} &=& 
\frac{1}{2}\,  a_{-+}\; e^{\Gamma_{-+} }
\qquad\qquad \qquad \qquad\qquad \qquad 
\tau_{R,1} = 
 \sqrt{z}\, a_{-+}\; e^{\Gamma_{-+} }
\nonu
\er

\item For $\varepsilon_1=1$ , $\varepsilon_2=-1$
\br
\tau_0&=& 1
\qquad\qquad \qquad \qquad\qquad \qquad \qquad\qquad \;\;
\tau_1= 1
\nonu\\
{\tilde \tau}_{L,1} &=& 
-\frac{2}{\sqrt{z}}\, a_{+-}\; e^{\Gamma_{+-} }
\qquad\qquad \qquad \qquad\qquad 
\tau_{R,2} =
\frac{1}{\sqrt{z}}\, a_{+-}^2\; e^{2 \Gamma_{+-} }
\nonu\\
\tau_{R,3(0)} &=& 
2 \, a_{+-}\; e^{\Gamma_{+-} }
\qquad\qquad \qquad \qquad\qquad \qquad
\tau_{R,3(1)} = 
 a_{+-}\; e^{\Gamma_{+-} }
\nonu
\er

\item For $\varepsilon_1=1$ , $\varepsilon_2=1$
\br
\tau_0 &=& 1
\qquad\qquad \qquad \qquad\qquad \qquad \qquad\qquad \qquad\;\;
\tau_1 = 1
\nonu\\
{\tilde \tau}_{L,1} &=& 
\frac{2}{\sqrt{z}}\, a_{++}\; e^{\Gamma_{++} }
\qquad\qquad \qquad \qquad\qquad \qquad \quad
\tau_{R,2} = 
-\frac{1}{\sqrt{z}}\, a_{++}^2 \; e^{2 \Gamma_{++} }
\nonu\\
\tau_{R,3(0)} &=& 
2 \, a_{++}\; e^{\Gamma_{++} }
\qquad\qquad \qquad \qquad\qquad \qquad \quad
\tau_{R,3(1)} = 
 a_{++} \; e^{\Gamma_{++} }
\nonu
\er
\end{enumerate}

\section{The model for $l=3$}
\label{sec:l=3}

As discussed at the end of section \ref{sec:coupling} we have two
types of models defined by the potentials \rf{zcpotbdgen} which are of
more interest. In section \ref{sec:l=6} we have discussed the first
case, and here we present the second one. In this case we take the
constant elements $E_{\pm l}$ in \rf{zcpotbdgen}, with $l=3$ and given by 
\begin{equation}
E_{\pm 3} \equiv m \, L_0^{\pm 1/2}
\end{equation}
where again $m$ is a parameter setting the mass scale of the particles
and solitons of the theory. The potentials \rf{zcpotbdgen} are 
\begin{equation}
A_+ = -B\left[E_{+3} + \sum_{n=1}^{2} F_+^{n}\right]B^{-1}
\;\;\;\;\;\;\;\;\;\;\;\;\;\; 
A_-=-\partial_{-}B\, B{-1} + E_{-3} + \sum_{n=1}^{2} F_-^{n}
\lab{potl=3}
\end{equation}
with $B$ being the same group element as in \rf{bfield}, and 
\begin{center}
\begin{tabular}{ l l l l l l l l l}
$F_+^{1} = \chi_R L_{-2}^{1/2} + \(\sqrt{\beta_1}/2\)\psi_R T_+^0$ &
& & & & & & $F_-^{1} = \chi_L
L_{2}^{-1/2} + \(\sqrt{\beta_1}/2\)\tilde{\psi}_L T_-^0$ \\
$F_+^{2} = \sqrt{\beta_1}\tilde{\psi}_R L_{-1}^{1/2}$ & & & &
& & & $F_-^{2} = \sqrt{\beta_1}\psi_L L_{1}^{-1/2}$
\end{tabular}
\end{center}
where $\beta_1$ is a free parameters, rescaling the spinors. 
By imposing the zero curvature condition \rf{zc} on the potentials
\rf{potl=3} we get the equations of motion 
\br
&& \partial^2 \varphi - \frac{i}{2}\,e^{\eta-2\varphi}\,{\bar
  \chi}\,\gamma_5\,\chi   +
\frac{i}{2}\, \beta_1 {\bar \psi}\, W\(\eta\)\, \gamma_5\, V\,\psi =0
\nonu\\
&& \partial^2 \nu + \frac{1}{2} m^2 e^{3 \eta} +
\frac{i}{4}\,e^{\eta-2\varphi} \,{\bar \chi}\,\gamma_5\,\chi  + 
\frac{i}{2}\beta_1 e^{2\eta-\varphi} {\bar \psi}\frac{(1-\gamma_5)}{2}\psi =0
\nonu\\
&& 
\partial^2 \eta =0
\\
&& 
\imath \gamma^\mu \partial_\mu \psi + \sqrt{6}\, m \, W(\eta)\,
V \psi + U =0
\nonu\\
&& 
\imath \gamma^\mu \partial_\mu \tilde{\psi} + \sqrt{6}\, m\, 
\tilde{W}(\eta)\, \tilde{V} \tilde{\psi} + \tilde{U} =0
\nonu\\
&& 
\imath \gamma^\mu \partial_\mu \chi + \tilde{\mathcal{U}} =0
\nonu
\er
where
\begin{equation}
W(\eta) \equiv \frac{(1+\gamma_5)}{2} + e^{3\eta}\frac{(1-\gamma_5)}{2}
\;\;\;\;\;\;\;\;\;\;\;\;\;\;\; \tilde{W}(\eta) \equiv
e^{3\eta}\frac{(1+\gamma_5)}{2} + \frac{(1-\gamma_5)}{2}
\end{equation}
\begin{equation}
V = e^{(\eta + \varphi)\gamma_5} \;\;\;\;\;\;\;\;\;\;\;\;\;\;\;
\tilde{V} = e^{-(\eta + \varphi)\gamma_5}
\end{equation}
and
\begin{equation}
U = \left( \begin{array}{c}
  0 \\
  e^{\eta-2\varphi}\tilde{\psi}_R \chi_L \\
\end{array} \right)
\qquad
\tilde{U} = \left( \begin{array}{c}
  e^{\eta-2\varphi}\psi_L \chi_R \\
  0 \\
\end{array} \right)
\qquad 
\mathcal{U} = \left( \begin{array}{c}
  \beta_1e^{\eta+\varphi}\psi_R \psi_L \\
  \beta_1e^{\eta+\varphi}\tilde{\psi}_R \tilde{\psi}_L  \\
\end{array} \right)
\end{equation}
The spinors have the form 
\begin{equation}
\psi = \left( \begin{array}{c}
  \psi_R \\
  \psi_L  \\
\end{array} \right)
\;\;\;\;\;\;\;\;\;\;\;\quad
{\tilde \psi} = \left( \begin{array}{c}
  {\tilde \psi}_R \\
  {\tilde \psi}_L  \\
\end{array} \right)
\;\;\;\;\;\;\;\;\;\;\;\quad
\chi = \left( \begin{array}{c}
  \chi_R \\
  \chi_L  \\
\end{array} \right)
\end{equation}
and we have defined 
\begin{equation}
\bar{\psi} = \tilde{\psi}^T \gamma_0 \;\;\;\;\;\;\;\;\;\;\;\;\;
\bar{\chi} = \chi^T \gamma_0
\;\;\;\;\;\;\;\;\;\;\;\;\;
\bar{U} = \tilde{U}^T \gamma_0 \;\;\;\;\;\;\;\;\;\;\;\;\;
\bar{\mathcal{U}} = \mathcal{U}^T \gamma_0
\end{equation}

\subsection{$\tau$-functions and solutions}

We construct the soliton solutions using the dressing method, as
described in sections \ref{sec:dressing} and \ref{subsec:solutions},
starting from the vacuum solutions
\be
\vp=\eta=\psi={\tilde \psi}=\chi=0 \qquad \qquad
\nu=-\frac{1}{2}\,m^2\,x_{+}\,x_{-}
\ee
Then we select a set of $\tau$-functions appropriate to express the
solutions for the fields. The and the field
The $\tau$-functions are chosen as 
\begin{center}
\begin{tabular}{ l l l l l l l l l}
$\tau_0 = \langle \lambda_0 | \Psi_{\rm vac} h \Psi_{\rm vac}^{-1} |
\lambda_0 \rangle$ & & & & & & & $\tau_1 = \langle 2 \lambda_1 |
\Psi_{\rm vac} h \Psi_{\rm vac}^{-1} |
2 \lambda_1 \rangle$ \\
$\tau_1^R =  \langle 2 \lambda_1 | T_+^0 \Psi_{\rm vac} h
\Psi_{\rm vac}^{-1} | 2 \lambda_1 \rangle$ & & & & & & & $\tau_1^L =
\langle 2 \lambda_1 | \Psi_{\rm vac} h \Psi_{\rm vac}^{-1} T_-^0 | 2
\lambda_1 \rangle$ \\
$\tau_0^R =  \langle \lambda_0 | L_{-2}^{\frac{1}{2}} \Psi_{\rm vac} h
\Psi_{\rm vac}^{-1} | \lambda_0 \rangle$ & & & & & & & $\tau_0^L =
\langle \lambda_0 | \Psi_{\rm vac} h \Psi_{\rm vac}^{-1}
L_{2}^{-\frac{1}{2}} | \lambda_0 \rangle$
\end{tabular}
\end{center}
The relations among the $\tau$-functions and the fileds are 
\br
\varphi &=& \log\frac{\tau_0}{\tau_1}
\qquad\qquad \qquad\qquad\qquad \qquad\quad
\nu = -\frac{1}{2}\log \tau_0-\frac{1}{2}\,m^2\,x_{+}\,x_{-}
\nonu\\
\psi_R &=& \frac{1}{ \sqrt{\beta_1}}\frac{\tau_1}{\tau_0} \partial_+
\left( \frac{\tau_1^L}{\tau_1} \right)
\qquad\qquad\qquad\qquad
\psi_L = \sqrt{\frac{3}{2}}\frac{m}{\sqrt{\beta_1}}\frac{\tau_1^L}{\tau_1}
\nonu\\
\tilde{\psi}_R &=& -
\sqrt{\frac{3}{2}}\frac{m}{\sqrt{\beta_1}}\frac{\tau_1^R}{\tau_1} 
\qquad \qquad\qquad\qquad\qquad
\tilde{\psi}_L = \frac{1}{ \sqrt{\beta_1}}\frac{\tau_1}{\tau_0}
\partial_+ \left( \frac{\tau_1^R}{\tau_1} \right)
\nonu\\
\chi_R &=& \frac{\tau_0^2}{\tau_1^2} \partial_+ \left(
\frac{\tau_0^L}{\tau_0} \right)
\qquad \qquad\qquad\qquad\qquad
\chi_L = \frac{\tau_0^2}{\tau_1^2} \partial_- \left(
\frac{\tau_0^R}{\tau_0} \right)
\er
Replacing those relations in to the equations of motion we get the
Hirota's equations for the $\tau$-functions. We then solve those using
the Hirota's recurrence method to get the one-soliton solution  
\begin{eqnarray}
\nonumber \tau_0 &=& 1 - \frac{1}{4} a b e^{2\,\Gamma} 
\qquad \qquad\qquad \qquad\;\;
\nonumber \tau_1 = 1\nonu \\
 \tau_0^R &=& -\frac{1}{4}z b^2 e^{2\,\Gamma} 
\qquad \qquad\qquad \qquad\quad
 \tau_0^L = \frac{1}{4z} a^2 e^{2\,\Gamma} \\
\nonumber \tau_1^R &=& a e^{\Gamma} 
\qquad \qquad\qquad \qquad\qquad \qquad
\nonumber \tau_1^L = b e^{\Gamma} \nonu
\end{eqnarray}
where
\be
\Gamma = \sqrt{\frac{3}{2}}m(x_+ z - x_- z^{-1})
\ee
The solution in terms of the fields read
\begin{eqnarray}
\nonumber \varphi &=& \log{\left(1 - \frac{1}{4} a b 
e^{2\,\Gamma}\right)} \\ 
\nonumber \nu &=& -\frac{1}{2}\log{\left(1 - \frac{1}{4} a b
  e^{2\,\Gamma}\right)}-\frac{1}{2}m^2 x_+ x_- \\ 
\nonumber \chi_R &=& \frac{1}{2}\sqrt{\frac{3}{2}}m a^2\,
e^{2\,\Gamma} \\ 
 \chi_L &=&  \frac{1}{2}\sqrt{\frac{3}{2}}m b^2\,
e^{2\,\Gamma} \\ 
\nonumber \psi_R &=& - 2\sqrt{6} z \frac{m}{\sqrt{\beta_1}} \frac{b
  e^{\Gamma} }{ (-4+ a b
  e^{2\,\Gamma})} \\ 
\nonumber \psi_L &=& \sqrt{\frac{3}{2}}\frac{m }{\sqrt{\beta_1}} b
e^{\Gamma} \\ 
\nonumber \tilde{\psi}_R &=& -\sqrt{\frac{3}{2}}\frac{m}{\sqrt{\beta_1}} a
e^{\Gamma} \\ 
\tilde{\psi}_L &=& 2\,\sqrt{6}\, \frac{1}{z} \frac{m}{\sqrt{\beta_1}} \frac{a
e^{\Gamma} }{ (-4+ a b
e^{2\,\Gamma})}\nonu
\end{eqnarray}

\newpage 

\appendix

\section{The twisted affine Kac-Moody algebra $A_2^{(2)}$}
\label{app:kacmoody}
\setcounter{equation}{0}

The generators of the algebra $A_2^{(2)}$ are given by $T_3^m$,
$T_{\pm}^m$, and $L_j^{r}$, 
with $m,n \in \IZ$, $r,s\in \IZ+\frac{1}{2}$ and $j,k=0,\pm 1,\pm 2$,
and the commutation relations are 
\br
\sbr{T_3^m}{T_3^n}&=& 2\, m\, \delta_{m+n,0}\, C\; ; \qquad \qquad
\sbr{T_{+}^m}{T_{-}^n}= 2\, T_3^{m+n} + 4\, m \, \delta_{m+n,0}\, C \nonu\\
\sbr{T_3^m}{T_{\pm}^n}&=& \pm\, T_{\pm}^{m+n}   \nonu\\
\sbr{T_3^m}{L_k^r}&=& k\, L_k^{m+r}\; ; \qquad \qquad 
\sbr{T_{\pm}^m}{L_k^r} = \sqrt{6-k\(k\pm 1\)}\; \; L_{k\pm 1}^{m+r} \nonu\\
\sbr{L_k^r}{L_{-k}^s} &=& \(-1\)^k\(\frac{k}{2}\, T_3^{r+s} + r\,
\delta_{r+s,0}\, C\)\nonu\\ 
\sbr{L_0^r}{L_{\pm 1}^s}&=& -\frac{\sqrt{6}}{4}\, T_{\pm}^{r+s}\; ;
\qquad \qquad\sbr{L_0^r}{L_{\pm 2}^s} = 0 \nonu\\ 
\sbr{L_{1}^r}{L_{-2}^s}&=& \frac{1}{2}\, T_{-}^{r+s}\; ; \qquad \qquad 
\sbr{L_{-1}^r}{L_{2}^s}= \frac{1}{2}\, T_{+}^{r+s}\nonu\\ 
\sbr{L_{1}^r}{L_{2}^s}&=& 0 \; ; \qquad \qquad\qquad
\sbr{L_{-1}^r}{L_{-2}^s}= 0  
\lab{a22comrel}
\er
Notice that those commutation relations are compatible with the
hermiticity conditions ${T_3^m}^{\dagger}=T_3^{-m}$,  
${T_{+}^m}^{\dagger}=T_{-}^{-m}$, ${L_0^r}^{\dagger}=L_0^{-r}$,
${L_1^r}^{\dagger}=-L_{-1}^{-r}$, and ${L_2^r}^{\dagger}=L_{-2}^{-r}$.  

In the text we have made use of the so-called principal gradation of
the algebra $A_2^{(2)}$, i.e. 
\be
{\cal G} = \oplus_{n=-\infty}^{\infty} \, {\cal G}_n \qquad \qquad
\sbr{Q}{{\cal G}_n} = n \, {\cal G}_n \qquad \qquad \sbr{{\cal
    G}_m}{{\cal G}_n}\subset {\cal G}_{m+n}
\lab{gradation}
\ee
and where the grading operator $Q$ is 
\be
Q\equiv T_3^0 + 6 D 
\lab{gradingop}
\ee
The operator $D$ measures the upper index of the generators, i.e.
\be
\sbr{D}{T_j^m}=m\,T_j^m \qquad \qquad \sbr{D}{L_p^{m+1/2}}=
  \(m+\frac{1}{2}\)\, L_p^{m+1/2} 
\ee 
with $m\in \IZ$, $j=3,\pm$, and $p=0,\pm 1,\pm 2$. Therefore the
eigensubspaces of $Q$ are
\br
{\cal G}_0 &=& \{ T_3^0, Q, C\}\nonu\\
{\cal G}_{6n} &=& \{ T_3^n\} \qquad n\neq 0\nonu\\
{\cal G}_{6n+1} &=& \{ T_+^n, L_{-2}^{n+1/2}\}\nonu\\
{\cal G}_{6n+2} &=& \{ L_{-1}^{n+1/2}\}
\lab{eigensub}\\
{\cal G}_{6n+3} &=& \{ L_{0}^{n+1/2}\}\nonu\\
{\cal G}_{6n+4} &=& \{ L_{1}^{n+1/2}\}\nonu\\
{\cal G}_{6n+5} &=& \{ T_-^{n+1}, L_{2}^{n+1/2}\}\nonu
\er

A highest weight representation of ${\cal G}$ is one possessing a
highest weight state $\ket{\lambda}$ satisfying
\be
{\cal G}_n\, \ket{\lambda} = 0 \qquad \qquad\bra{\lambda}\, {\cal G}_{-n}=0
\qquad \qquad\quad {\rm for} \;\;\; n>0 
\lab{higheststate}
\ee
The highest weight states of the two fundamental representations of
$A_2^{(2)}$ satisfy 
\br
T_3^0 \,\ket{\lambda_0} &=& 0 \qquad \qquad \qquad \; T_3^0
\,\ket{\lambda_1} = \frac{1}{2} \, \ket{\lambda_1}\nonu\\ 
C\,\ket{\lambda_0} &=& 2\, \ket{\lambda_0} \qquad \qquad
C\,\ket{\lambda_1} = \ket{\lambda_1} 
\lab{fundrep}
\er

\section{Coefficients of $\tau$-functions of soliton of sec. 
  \ref{subsec:mass1} }
\label{app:fullsol}
\setcounter{equation}{0}


The $\tau$-functions for the solution described in section
\ref{subsec:mass1} truncate at  at order $16$. We express the
$\tau$-functions as 
\be
\tau_{*} = \sum_{l=0}^{16} \delta_{*}^{(l)} \; e^{l\, \Gamma_1}
\ee
where $*$ stands for any of the labels of the $\tau$-functions defined
in \rf{taudefl=6}. The
quantity $\Gamma_1$ is defined in \rf{gamma1def}. The following 
coefficients $ \delta_{*}^{(l)}$ vanish
\br
\delta_{0}^{(2n+1)}&=&\delta_{1}^{(2n+1)}=
{\tilde  \delta}_{L,2}^{(2n+1)}=\delta_{R,2}^{(2n+1)}=0
\qquad\qquad \qquad \quad n=1,2,\ldots 7
\\
{\tilde \delta}_{L,1}^{(2n)}&=&{\tilde \delta}_{L,3(0)}^{(2n)}=
{\tilde \delta}_{L,3(1)}^{(2n)}=\delta_{R,1}^{(2n)}= 
\delta_{R,3(0)}^{(2n)}= \delta_{R,3(1)}^{(2n)}=0
\qquad\quad n=0,1,2,\ldots 8
\nonu
\er
The remaining coefficients $ \delta_{*}^{(l)}$ are given by 
\br
\delta_{0}^{(0)}
&=&
1
\nonumber\\
\delta_{0}^{(
2
)}&=&
-8 a {\bar a}
\nonumber\\
\delta_{0}^{(
4
)}&=&
\frac{4}{z} \left(\left(7 a^2 z-3 b^2\right) {\bar a}^2+{\bar b} (4 a b+c) z \
{\bar a}+z \left(-b^2 {\bar b}^2+3 a^2 z {\bar b}^2-a b \
{\bar c}\right)\right)
\nonumber\\
\delta_{0}^{(
6
)}&=&
-\frac{8}{z} \left(7 {\bar a} z \left({\bar a}^2+{\bar b}^2 z\right) a^3+2 b \
z \left(2 z {\bar b}^3+4 {\bar a}^2 {\bar b}-{\bar a} {\bar c}\right) \
a^2
\right. \nonumber\\
&+&\left.\left(-7 b^2 {\bar a}^3+2 {\bar b} c z {\bar a}^2-3 b^2 {\bar b}^2 z \
{\bar a}+2 {\bar b} z \left({\bar b}^2 c z-b^2 {\bar c}\right)\right)
a \right.
\nonumber\\
&-& \left. b (2 {\bar a} {\bar b}-{\bar c}) \left(2 {\bar a} b^2-{\bar b} c \
z\right)\right)
\nonumber\\
\delta_{0}^{(
8
)}&=&
\frac{1}{z^2}\(2 \left(35 z^2 a^4-54 b^2 z a^2+19 b^4\right) {\bar
  a}^4+24 {\bar b} \ 
(4 a b+c) z \left(a^2 z-b^2\right) {\bar a}^3
\right.\nonumber\\
&+& \left. 4 z \left(b^2-a^2 z\right) \
\left(7 b^2 {\bar b}^2-27 a^2 z {\bar b}^2+6 a b {\bar c}\right) \
{\bar a}^2
\right.\nonumber\\
&+& \left. 4 {\bar b} z \left(4 {\bar c} b^4-24 a {\bar b}^2 z b^3-6 \
{\bar b}^2 c z b^2
+4 a z \left(6 a^2 z {\bar b}^2+c {\bar c}\right) b
\right. \right. \nonumber\\
&+& \left. \left. c z \
\left(6 a^2 z {\bar b}^2+c {\bar c}\right)\right) {\bar a}
\right.\nonumber\\
&+& \left. z 
\left(\left(6 {\bar b}^4 z-4 {\bar c}^2\right) b^4+24 a {\bar b}^2 \
{\bar c} z b^3-28 a^2 {\bar b}^4 z^2 b^2
\right.\right.\nonumber\\
&-& \left. \left.
4 a z \left(6 a^2 {\bar c} z \
{\bar b}^2+c \left({\bar c}^2-4 {\bar b}^4 z\right)\right) b
+  z \left(38 \
a^4 {\bar b}^4 z^2-c^2 \left({\bar c}^2-4 {\bar b}^4 \
z\right)\right)\right)\)
\nonumber\\
\delta_{0}^{(
10
)}&=&
-\frac{8}{z^2} \left(b^2-a^2 z\right) \left({\bar a}^2+{\bar b}^2 z\right) \
\left(-7 {\bar a} z \left({\bar a}^2+{\bar b}^2 z\right) a^3
\right. \nonumber\\
&+& \left. 2 b z \
\left(-2 z {\bar b}^3-4 {\bar a}^2 {\bar b}+{\bar a} {\bar c}\right) \
a^2
\right.\nonumber\\
&+&\left. \left(7 b^2 {\bar a}^3-2 {\bar b} c z {\bar a}^2+3 b^2 {\bar b}^2 z \
{\bar a}+2 {\bar b} z \left(b^2 {\bar c}-{\bar b}^2 c z\right)\right) \
a \right. \nonumber\\
&+& \left. b (2 {\bar a} {\bar b}-{\bar c}) \left(2 {\bar a} b^2-{\bar b} c \
z\right)\right)
\nonumber\\
\delta_{0}^{(
12
)}&=&
-\frac{4}{z^3} \left(b^2-a^2 z\right)^2 \left({\bar a}^2+{\bar b}^2
z\right)^2 \ 
\left(\left(3 b^2-7 a^2 z\right) {\bar a}^2-{\bar b} (4 a b+c) z \
{\bar a}
\right. \nonumber\\
&+& \left. z \left(b^2 {\bar b}^2-3 a^2 z {\bar b}^2+a b \
{\bar c}\right)\right)
\nonumber\\
\delta_{0}^{(
14
)}&=&
-\frac{8 a {\bar a} \left(a^2 z-b^2\right)^3 \left({\bar a}^2+{\bar b}^2 \
z\right)^3}{z^3}
\nonumber\\
\delta_{0}^{(
16
)}&=&
\frac{\left(b^2-a^2 z\right)^4 \left({\bar a}^2+{\bar b}^2 \
z\right)^4}{z^4}
\nonumber\\
\delta_{1}^{(0)}
&=&
1
\nonumber\\
\delta_{1}^{(
2
)}&=&
4 b {\bar b}-4 a {\bar a}
\nonumber\\
\delta_{1}^{(
4
)}&=&
4 a^2 {\bar a}^2+4 {\bar b} c {\bar a}+4 b^2 {\bar b}^2-4 a b \
{\bar c}-2 c {\bar c}
\nonumber\\
\delta_{1}^{(
6
)}&=&
\frac{4}{z} \left({\bar a} z \left({\bar a}^2+{\bar b}^2 z\right) a^3+b z \
\left(z {\bar b}^3-3 {\bar a}^2 {\bar b}+2 {\bar a} {\bar c}\right) \
a^2
\right.\nonumber\\
&+&\left. \left(-b^2 {\bar a}^3-2 {\bar b} c z {\bar a}^2+\left(3 b^2 \
{\bar b}^2+c {\bar c}\right) z {\bar a}-2 b^2 {\bar b} {\bar c} \
z\right) a
\right. \nonumber\\
&-& \left. b {\bar b} \left({\bar a}^2 b^2+{\bar b}^2 z b^2-2 {\bar a} \
{\bar b} c z+c {\bar c} z\right)\right)
\nonumber\\
\delta_{1}^{(
8
)}&=&
\frac{1}{z^2}\(-2 \left(5 z^2 a^4-6 b^2 z a^2+b^4\right) {\bar a}^4
\right. \nonumber\\
&-& \left. 4     {\bar b}^2 z \ 
\left(3 b^4-10 a^2 z b^2-4 a c z b+z \left(3 a^4 z-c^2\right)\right) \
{\bar a}^2
\right.\nonumber\\
&-&\left. 4 {\bar b} (2 a b+c)^2 {\bar c} z^2 {\bar a}
\right. \nonumber\\
&+& \left. z^2 \left(-10 \
b^4 {\bar b}^4-2 a^4 z^2 {\bar b}^4+c^2 {\bar c}^2+4 a b c {\bar c}^2+4 \
a^2 b^2 \left(3 z {\bar b}^4+{\bar c}^2\right)\right)\)
\nonumber\\
\delta_{1}^{(
10
)}&=&
-\frac{4}{z^2} \left(b^2-a^2 z\right) \left({\bar a}^2+{\bar b}^2 z\right) \
\left({\bar a} z \left({\bar a}^2+{\bar b}^2 z\right) a^3-b z \left(z \
{\bar b}^3-3 {\bar a}^2 {\bar b}+2 {\bar a} {\bar c}\right) \
a^2
\right.\nonumber\\
&-&\left. \left(b^2 {\bar a}^3-2 {\bar b} c z {\bar a}^2-3 b^2 {\bar b}^2 z \
{\bar a}+c {\bar c} z {\bar a}+2 b^2 {\bar b} {\bar c} z\right) a
\right. \nonumber\\
&+& \left. b \
{\bar b} \left({\bar a}^2 b^2+{\bar b}^2 z b^2+2 {\bar a} {\bar b} c \
z-c {\bar c} z\right)\right)
\nonumber\\
\delta_{1}^{(
12
)}&=&
\frac{2 \left(2 a^2 {\bar a}^2+2 b^2 {\bar b}^2+2 a b {\bar c}+c \
({\bar c}-2 {\bar a} {\bar b})\right) \left(b^2-a^2 z\right)^2 \
\left({\bar a}^2+{\bar b}^2 z\right)^2}{z^2}
\nonumber\\
\delta_{1}^{(
14
)}&=&
-\frac{4 (a {\bar a}+b {\bar b}) \left(a^2 z-b^2\right)^3 \
\left({\bar a}^2+{\bar b}^2 z\right)^3}{z^3}
\nonumber\\
\delta_{1}^{(
16
)}&=&
\frac{\left(b^2-a^2 z\right)^4 \left({\bar a}^2+{\bar b}^2 \
z\right)^4}{z^4}
\nonumber\\
{\tilde \delta}_{L,1}^{(0)}
&=&
0
\nonumber\\
{\tilde \delta}_{L,1}^{(
1
)}&=&
4 a
\nonumber\\
{\tilde \delta}_{L,1}^{(
3
)}&=&
-20 {\bar a} a^2-4 {\bar b} c+\frac{12 {\bar a} b^2}{z}
\nonumber\\
{\tilde \delta}_{L,1}^{(
5
)}&=&
\frac{4}{z} \left(3 z \left(3 {\bar a}^2+{\bar b}^2 z\right) a^3+b (4 \
{\bar a} {\bar b}-3 {\bar c}) z a^2
\right.\nonumber\\
&-&\left. \left(9 {\bar a}^2 b^2+7 {\bar b}^2 \
z b^2-4 {\bar a} {\bar b} c z+c {\bar c} z\right) a
\right. \nonumber\\
&+& \left. 4 {\bar a} b^3 \
{\bar b}+b^3 {\bar c}-2 b {\bar b}^2 c z\right)
\nonumber\\
{\tilde \delta}_{L,1}^{(
7
)}&=&
-\frac{1}{z^2}\(4 \left(\left(5 z^2 a^4-6 b^2 z a^2+b^4\right) {\bar a}^3+2 \
{\bar b} (4 a b+3 c) z \left(a^2 z-b^2\right) {\bar a}^2
\right.\right.\nonumber\\
&+&\left.\left. z \left(5 \
{\bar b}^2 b^4+8 a {\bar c} b^3+\left(3 c {\bar c}-2 a^2 {\bar b}^2 \
z\right) b^2+\left(8 a {\bar b}^2 c z-8 a^3 {\bar c} z\right) b
\right.\right.\right.\nonumber\\
&+&\left.\left. \left. z \
\left({\bar b}^2 \left(5 z a^4+2 c^2\right)-3 a^2 c {\bar c}\right)\right) \
{\bar a}-{\bar b} z \left(2 {\bar c} b^4-8 a {\bar b}^2 z b^3+2 a^2 \
{\bar c} z b^2
\right.\right.\right.\nonumber\\
&+&\left.\left.\left. 4 a z \left(2 a^2 z {\bar b}^2+c {\bar c}\right) b+c^2 \
{\bar c} z\right)\right)\)
\nonumber\\
{\tilde \delta}_{L,1}^{(
9
)}&=&
-\frac{1}{z^2}\(4 \left(a^2 z-b^2\right) \left(z \left(5 {\bar a}^4+6
{\bar b}^2 z \ 
{\bar a}^2+{\bar b}^4 z^2\right) a^3+6 b {\bar c} z \
\left({\bar a}^2+{\bar b}^2 z\right) a^2
\right.\right.\nonumber\\
&+&\left.\left. \left(-5 b^2 {\bar a}^4-4 \
{\bar b} c z {\bar a}^3+\left(3 c {\bar c} z-10 b^2 {\bar b}^2 z\right) \
{\bar a}^2-4 {\bar b} z \left({\bar b}^2 c z-2 b^2 {\bar c}\right) \
{\bar a}
\right.\right.\right.\nonumber\\
&+&\left.\left.\left. z \left(\left(3 {\bar b}^4 z-2 {\bar c}^2\right) b^2+3 \
{\bar b}^2 c {\bar c} z\right)\right) a
\right.\right.\nonumber\\
&-&\left.\left. b \left(8 b^2 {\bar b} \
{\bar a}^3+2 {\bar b}^2 c z {\bar a}^2+\left(8 b^2 {\bar b}^3 z-4 \
{\bar b} c {\bar c} z\right) {\bar a}+c z \left({\bar c}^2-2 {\bar b}^4 \
z\right)\right)\right)\)
\nonumber\\
{\tilde \delta}_{L,1}^{(
11
)}&=&
\frac{4}{z^3} \left(b^2-a^2 z\right)^2 \left({\bar a}^2+{\bar b}^2 z\right) \
\left(-3 \left(b^2-3 a^2 z\right) {\bar a}^3+{\bar b} (8 a b-c) z \
{\bar a}^2
\right.\nonumber\\
&+&\left. z \left(b^2 {\bar b}^2+9 a^2 z {\bar b}^2+c {\bar c}\right) \
{\bar a}+{\bar b} z \left(-2 {\bar c} b^2+8 a {\bar b}^2 z b+{\bar b}^2 \
c z\right)\right)
\nonumber\\
{\tilde \delta}_{L,1}^{(
13
)}&=&
\frac{4 \left(b^2-a^2 z\right)^3 \left({\bar a}^2+{\bar b}^2 z\right)^2 \
\left(5 a {\bar a}^2+4 b {\bar b} {\bar a}-b {\bar c}+3 a {\bar b}^2 \
z\right)}{z^3}
\nonumber\\
{\tilde \delta}_{L,1}^{(
15
)}&=&
\frac{4 {\bar a} \left(b^2-a^2 z\right)^4 \left({\bar a}^2+{\bar b}^2 \
z\right)^3}{z^4}
\nonumber\\
{\tilde \delta}_{L,2}^{(0)}
&=&
0
\nonumber\\
{\tilde \delta}_{L,2}^{(
2
)}&=&
2 {\bar c}
\nonumber\\
{\tilde \delta}_{L,2}^{(
4
)}&=&
8 \left(b \left(\frac{{\bar a}^3}{z}-{\bar b}^2 {\bar a}+{\bar b} \
{\bar c}\right)+a \left(-z {\bar b}^3+{\bar a}^2 {\bar b}-{\bar a} \
{\bar c}\right)\right)
\nonumber\\
{\tilde \delta}_{L,2}^{(
6
)}&=&
-\frac{2}{z} \left(-c {\bar a}^4+16 b^2 {\bar b} {\bar a}^3+\left(2 \
{\bar b}^2 c z-6 b^2 {\bar c}\right) {\bar a}^2+4 {\bar b} \left(4 b^2 \
{\bar b}^2-c {\bar c}\right) z {\bar a}
\right.\nonumber\\
&+&\left. 2 a^2 (8 {\bar a} {\bar b}-3 \
{\bar c}) z \left({\bar a}^2+{\bar b}^2 z\right)+z \left(-c z \
{\bar b}^4-6 b^2 {\bar c} {\bar b}^2+c {\bar c}^2\right)
\right.\nonumber\\
&+&\left. 4 a b \left(3 \
{\bar a}^4+10 {\bar b}^2 z {\bar a}^2-4 {\bar b} {\bar c} z {\bar a}+z \
\left(3 z {\bar b}^4+{\bar c}^2\right)\right)\right)
\nonumber\\
{\tilde \delta}_{L,2}^{(
8
)}&=&
\frac{8}{z^2} \left({\bar a}^2+{\bar b}^2 z\right) \left(-\left(4 b^3-2 a^2 z \
b+a c z\right) {\bar a}^3+{\bar b} z \left(6 z a^3-8 b^2 a-b c\right) \
{\bar a}^2
\right.\nonumber\\
&+&\left. z \left(-{\bar c} z a^3+8 b {\bar b}^2 z a^2+3 b^2 {\bar c} \
a+{\bar b}^2 c z a-6 b^3 {\bar b}^2+b c {\bar c}\right) \
{\bar a}
\right.\nonumber\\
&+&\left. {\bar b} z \left({\bar c} b^3-2 a {\bar b}^2 z \
b^2+\left({\bar b}^2 c-3 a^2 {\bar c}\right) z b+a z \left(4 a^2 \
{\bar b}^2 z-c {\bar c}\right)\right)\right)
\nonumber\\
{\tilde \delta}_{L,2}^{(
10
)}&=&
\frac{2}{z^2} \left({\bar a}^2+{\bar b}^2 z\right)^2 \left(({\bar c}-16 \
{\bar a} {\bar b}) z^2 a^4+8 b z \left({\bar a}^2-{\bar b}^2 z\right) \
a^3
\right.\nonumber\\
&+&\left. 2 z \left(3 c {\bar a}^2+16 b^2 {\bar b} {\bar a}+b^2 {\bar c}-3 \
{\bar b}^2 c z\right) a^2+\left(-8 {\bar a}^2 b^3+8 {\bar b}^2 z b^3+4 c \
{\bar c} z b\right) a
\right.\nonumber\\
&-&\left. 16 {\bar a} b^4 {\bar b}-6 {\bar a}^2 b^2 c+b^4 \
{\bar c}+6 b^2 {\bar b}^2 c z+c^2 {\bar c} z\right)
\nonumber\\
{\tilde \delta}_{L,2}^{(
12
)}&=&
-\frac{8 \left(b^2-a^2 z\right) \left({\bar a}^2+{\bar b}^2 z\right)^3 \
\left({\bar b} z \left(z a^3-3 b^2 a-b c\right)+{\bar a} \left(b^3-3 a^2 z \
b-a c z\right)\right)}{z^3}
\nonumber\\
{\tilde \delta}_{L,2}^{(
14
)}&=&
\frac{2 (4 a b+c) \left(b^2-a^2 z\right)^2 \left({\bar a}^2+{\bar b}^2 \
z\right)^4}{z^3}
\nonumber\\
{\tilde \delta}_{L,2}^{(
16
)}&=&
0
\nonumber\\
{\tilde \delta}_{L,3(0)}^{(
1
)}&=&
4 {\bar b}
\nonumber\\
{\tilde \delta}_{L,3(0)}^{(
3
)}&=&
4 \left(a ({\bar c}-4 {\bar a} {\bar b})+b \left(\frac{3 \
{\bar a}^2}{z}+{\bar b}^2\right)\right)
\nonumber\\
{\tilde \delta}_{L,3(0)}^{(
5
)}&=&
\frac{4}{z} \left(c {\bar a}^3-15 b^2 {\bar b} {\bar a}^2+6 b^2 {\bar c} \
{\bar a}+3 {\bar b}^2 c z {\bar a}-8 a b \left({\bar a}^2+{\bar b}^2 \
z\right) {\bar a}-3 b^2 {\bar b}^3 z
\right.\nonumber\\
&-&\left. {\bar b} c {\bar c} z+a^2 z \left(z \
{\bar b}^3+9 {\bar a}^2 {\bar b}-4 {\bar a} {\bar c}\right)\right)
\nonumber\\
{\tilde \delta}_{L,3(0)}^{(
7
)}&=&
-\frac{1}{z^2}\(4 \left(\left(11 b^3-3 a^2 z b+4 a c z\right) {\bar
  a}^4+4 {\bar b} \ 
z \left(4 z a^3-2 b^2 a+b c\right) {\bar a}^3
\right.\right.\nonumber\\
&+&\left.\left. z \left(-6 {\bar c} z a^3+22 \
b {\bar b}^2 z a^2+14 {\bar b}^2 c z a+6 b^3 {\bar b}^2-3 b c \
{\bar c}\right) {\bar a}^2
\right.\right.\nonumber\\
&+&\left.\left. 4 {\bar b} z \left(2 {\bar c} b^3-2 a \
{\bar b}^2 z b^2+\left({\bar b}^2 c-4 a^2 {\bar c}\right) z b+a z \left(4 \
a^2 {\bar b}^2 z-c {\bar c}\right)\right) {\bar a}
\right.\right.\nonumber\\
&+&\left.\left. z \left(\left(3 \
{\bar b}^4 z-2 {\bar c}^2\right) b^3+z \left(9 a^2 z {\bar b}^4-3 c \
{\bar c} {\bar b}^2+4 a^2 {\bar c}^2\right) b
\right.\right.\right. \nonumber\\
&+& \left.\left.\left. a z \left(6 c z \
{\bar b}^4-6 a^2 {\bar c} z {\bar b}^2+c \
{\bar c}^2\right)\right)\right)\)
\nonumber\\
{\tilde \delta}_{L,3(0)}^{(
9
)}&=&
\frac{1}{z^2}\( 4 \left({\bar a}^2+{\bar b}^2 z\right) \left(z^2 \left(11 z \
{\bar b}^3+19 {\bar a}^2 {\bar b}-4 {\bar a} {\bar c}\right) a^4
\right.\right.\nonumber\\
&+& \left.\left. 8 b z \
\left({\bar a}^3+3 {\bar b}^2 z {\bar a}-{\bar b} {\bar c} z\right) \
a^3
\right.\right.\nonumber\\
&+&\left.\left. z \left(6 c {\bar a}^3-34 b^2 {\bar b} {\bar a}^2+2 \left(7 {\bar c} \
b^2+6 {\bar b}^2 c z\right) {\bar a}-3 {\bar b} \left(2 b^2 {\bar b}^2+c \
{\bar c}\right) z\right) a^2
\right.\right.\nonumber\\
&+&\left.\left. 4 b \left(-2 b^2 {\bar a}^3+\left(c {\bar c} \
z-6 b^2 {\bar b}^2 z\right) {\bar a}+2 {\bar b} z \left({\bar c} \
b^2+{\bar b}^2 c z\right)\right) a+15 {\bar a}^2 b^4 {\bar b}-6 \
{\bar a}^3 b^2 c
\right.\right.\nonumber\\
&+&\left.\left. {\bar b} z \left(3 {\bar b}^2 b^4+3 c {\bar c} b^2+2 \
{\bar b}^2 c^2 z\right)+{\bar a} \left(-6 {\bar c} b^4-12 {\bar b}^2 c z \
b^2+c^2 {\bar c} z\right)\right)\)
\nonumber\\
{\tilde \delta}_{L,3(0)}^{(
11
)}&=&
\frac{4}{z^3} \left(b^2-a^2 z\right) \left({\bar a}^2+{\bar b}^2 z\right)^2 \
\left(\left(b^3+7 a^2 z b+4 a c z\right) {\bar a}^2
\right.\nonumber\\
&+& \left. 4 {\bar b} z \left(3 z \
a^3-b^2 a+b c\right) {\bar a}
\right.\nonumber\\
&+&\left. z \left(-{\bar c} z a^3+9 b {\bar b}^2 z \
a^2-b^2 {\bar c} a+6 {\bar b}^2 c z a+3 b^3 {\bar b}^2-b c \
{\bar c}\right)\right)
\nonumber\\
{\tilde \delta}_{L,3(0)}^{(
13
)}&=&
\frac{4 \left(b^2-a^2 z\right)^2 \left(3 {\bar b} z a^2-b^2 \
{\bar b}+{\bar a} c\right) \left({\bar a}^2+{\bar b}^2 z\right)^3}{z^3}
\nonumber\\
{\tilde \delta}_{L,3(0)}^{(
15
)}&=&
-\frac{4 b \left(b^2-a^2 z\right)^3 \left({\bar a}^2+{\bar b}^2 \
z\right)^4}{z^4}
\nonumber\\
{\tilde \delta}_{L,3(1)}^{(
1
)}&=&
2 {\bar b}
\nonumber\\
{\tilde \delta}_{L,3(1)}^{(
3
)}&=&
\frac{6 b {\bar a}^2}{z}+10 b {\bar b}^2-2 a {\bar c}
\nonumber\\
{\tilde \delta}_{L,3(1)}^{(
5
)}&=&
-\frac{2}{z} \left(-c {\bar a}^3-9 b^2 {\bar b} {\bar a}^2-3 {\bar b}^2 c z \
{\bar a}-9 b^2 {\bar b}^3 z+{\bar b} c {\bar c} z+a^2 z \left(3 z \
{\bar b}^3+7 {\bar a}^2 {\bar b}-2 {\bar a} {\bar c}\right)
\right.\nonumber\\
&+&\left. 4 a b \
\left({\bar a}^3-{\bar b}^2 z {\bar a}+{\bar b} {\bar c} \
z\right)\right)
\nonumber\\
{\tilde \delta}_{L,3(1)}^{(
7
)}&=&
\frac{1}{z^2}\( 2 \left(\left(b^3-5 a^2 z b-2 a c z\right) {\bar
  a}^4+8 {\bar b} z \ 
\left(z a^3+b^2 a+b c\right) {\bar a}^3
\right.\right.\nonumber\\
&+&\left.\left. z \left(6 {\bar b}^2 b^3-6 a \
{\bar c} b^2-3 c {\bar c} b-2 a^2 {\bar b}^2 z b+2 a {\bar b}^2 c \
z\right) {\bar a}^2
\right.\right.\nonumber\\
&+&\left.\left. 4 {\bar b} z^2 \left(2 {\bar b}^2 z a^3-2 b {\bar c} \
a^2+2 b^2 {\bar b}^2 a-c {\bar c} a+2 b {\bar b}^2 c\right) {\bar a}
\right.\right.\nonumber\\
&+&\left.\left. z^2 \
\left(5 b^3 {\bar b}^4-6 a b^2 {\bar c} {\bar b}^2+a c {\bar c}^2+b \
\left(-5 a^2 z {\bar b}^4-3 c {\bar c} {\bar b}^2+2 a^2 \
{\bar c}^2\right)\right)\right)\)
\nonumber\\
{\tilde \delta}_{L,3(1)}^{(
9
)}&=&
-\frac{1}{z^2}\( 2 \left({\bar a}^2+{\bar b}^2 z\right) \left(z^2 \left(z \
{\bar b}^3-3 {\bar a}^2 {\bar b}+2 {\bar a} {\bar c}\right) a^4-4 b z \
\left(2 {\bar a}^3+{\bar b} {\bar c} z\right) a^3
\right.\right.\nonumber\\
&+&\left.\left. z \left(-6 b^2 z \
{\bar b}^3+6 {\bar a} c z {\bar b}^2-10 {\bar a}^2 b^2 {\bar b}-3 c \
{\bar c} z {\bar b}+2 {\bar a} b^2 {\bar c}\right) a^2
\right.\right.\nonumber\\
&+&\left.\left. 4 b \left(2 b^2 \
{\bar a}^3-2 {\bar b} c z {\bar a}^2+c {\bar c} z {\bar a}+b^2 \
{\bar b} {\bar c} z\right) a+{\bar a} c \left(c {\bar c}-6 b^2 \
{\bar b}^2\right) z
\right.\right.\nonumber\\
&+&\left.\left. b^2 {\bar b} \left(5 b^2 {\bar b}^2+3 c \
{\bar c}\right) z+{\bar a}^2 {\bar b} \left(5 b^4-2 c^2 \
z\right)\right)\)
\nonumber\\
{\tilde \delta}_{L,3(1)}^{(
11
)}&=&
-\frac{2}{z^3} \left(b^2-a^2 z\right) \left({\bar a}^2+{\bar b}^2 z\right)^2 \
\left(\left(3 b^3+a^2 z b+2 a c z\right) {\bar a}^2+8 a {\bar b} z \
\left(b^2-a^2 z\right) {\bar a}
\right.\nonumber\\
&+&\left. z \left({\bar c} z a^3-9 b {\bar b}^2 z \
a^2+b^2 {\bar c} a+9 b^3 {\bar b}^2+b c {\bar c}\right)\right)
\nonumber\\
{\tilde \delta}_{L,3(1)}^{(
13
)}&=&
-\frac{2 \left(b^2-a^2 z\right)^2 \left(-3 {\bar b} z a^2+4 {\bar a} b a+5 \
b^2 {\bar b}+{\bar a} c\right) \left({\bar a}^2+{\bar b}^2 \
z\right)^3}{z^3}
\nonumber\\
{\tilde \delta}_{L,3(1)}^{(
15
)}&=&
-\frac{2 b \left(b^2-a^2 z\right)^3 \left({\bar a}^2+{\bar b}^2 \
z\right)^4}{z^4}
\nonumber\\
\delta_{R,1}^{(
1
)}&=&
4 {\bar a}
\nonumber\\
\delta_{R,1}^{(
3
)}&=&
-4 \left(5 a {\bar a}^2-b {\bar c}+3 a {\bar b}^2 z\right)
\nonumber\\
\delta_{R,1}^{(
5
)}&=&
\frac{4}{z} \left(-3 \left(b^2-3 a^2 z\right) {\bar a}^3+{\bar b} (4 a b+3 c) \
z {\bar a}^2
\right. \nonumber\\
&-& \left. z \left(7 b^2 {\bar b}^2-9 a^2 z {\bar b}^2+4 a b \
{\bar c}+c {\bar c}\right) {\bar a}
\right.\nonumber\\
&+&\left. {\bar b} z \left(2 {\bar c} b^2-4 a \
{\bar b}^2 z b+{\bar b}^2 c z\right)\right)
\nonumber\\
\delta_{R,1}^{(
7
)}&=&
-\frac{1}{z}\( 4 \left(z \left(5 {\bar a}^4+6 {\bar b}^2 z {\bar
  a}^2+{\bar b}^4 \ 
z^2\right) a^3+2 b (4 {\bar a} {\bar b}-3 {\bar c}) z \
\left({\bar a}^2+{\bar b}^2 z\right) a^2
\right.\right.\nonumber\\
&-&\left.\left.\left(5 b^2 {\bar a}^4-8 \
{\bar b} c z {\bar a}^3+\left(2 b^2 {\bar b}^2+3 c {\bar c}\right) z \
{\bar a}^2+8 {\bar b} z \left(b^2 {\bar c}-{\bar b}^2 c z\right) \
{\bar a}
\right.\right.\right.\nonumber\\
&+&\left.\left.\left. z \left(\left(5 {\bar b}^4 z-2 {\bar c}^2\right) b^2+3 \
{\bar b}^2 c {\bar c} z\right)\right) a
\right.\right.\nonumber\\
&+&\left.\left. b \left(8 b^2 {\bar b} \
{\bar a}^3+2 {\bar b}^2 c z {\bar a}^2+\left(8 b^2 {\bar b}^3 z-4 \
{\bar b} c {\bar c} z\right) {\bar a}+c z \left({\bar c}^2-2 {\bar b}^4 \
z\right)\right)\right)\)
\nonumber\\
\delta_{R,1}^{(
9
)}&=&
-\frac{1}{z^2}\(4 \left({\bar a}^2+{\bar b}^2 z\right) \left(\left(5
z^2 a^4-6 b^2 \ 
z a^2+b^4\right) {\bar a}^3+6 {\bar b} c z \left(b^2-a^2 z\right) \
{\bar a}^2
\right.\right.\nonumber\\
&-&\left.\left. z \left(3 {\bar b}^2 b^4+4 a {\bar c} b^3+\left(10 a^2 z \
{\bar b}^2+3 c {\bar c}\right) b^2+\left(8 a {\bar b}^2 c z-4 a^3 \
{\bar c} z\right) b
\right.\right.\right.\nonumber\\
&+&\left.\left.\left. z \left({\bar b}^2 \left(2 c^2-5 a^4 z\right)-3 a^2 c \
{\bar c}\right)\right) {\bar a}
\right.\right.\nonumber\\
&+&\left.\left. {\bar b} z \left(2 {\bar c} b^4-8 a \
{\bar b}^2 z b^3+2 a^2 {\bar c} z b^2+4 a z \left(2 a^2 z {\bar b}^2+c \
{\bar c}\right) b+c^2 {\bar c} z\right)\right)\)
\nonumber\\
\delta_{R,1}^{(
11
)}&=&
\frac{4}{z^2} \left(a^2 z-b^2\right) \left({\bar a}^2+{\bar b}^2 z\right)^2 \
\left(3 z \left(3 {\bar a}^2+{\bar b}^2 z\right) a^3+b (8 {\bar a} \
{\bar b}+{\bar c}) z a^2
\right.\nonumber\\
&+&\left. \left(-9 {\bar a}^2 b^2+{\bar b}^2 z b^2+c \
{\bar c} z\right) a-8 {\bar a} b^3 {\bar b}+b^3 {\bar c}+2 b {\bar b}^2 \
c z\right)
\nonumber\\
\delta_{R,1}^{(
13
)}&=&
-\frac{4 \left(b^2-a^2 z\right)^2 \left({\bar a}^2+{\bar b}^2 z\right)^3 \
\left({\bar b} (4 a b+c) z+{\bar a} \left(5 a^2 z-3 \
b^2\right)\right)}{z^3}
\nonumber\\
\delta_{R,1}^{(
15
)}&=&
\frac{4 a \left(a^2 z-b^2\right)^3 \left({\bar a}^2+{\bar b}^2 \
z\right)^4}{z^3}
\nonumber\\
\delta_{R,2}^{(0)}
&=&
0
\nonumber\\
\delta_{R,2}^{(
2
)}&=&
2 c
\nonumber\\
\delta_{R,2}^{(
4
)}&=&
\frac{8 \left({\bar b} z \left(z a^3+b^2 a+b c\right)-{\bar a} \
\left(b^3+a^2 z b+a c z\right)\right)}{z}
\nonumber\\
\delta_{R,2}^{(
6
)}&=&
-\frac{2}{z} \left((12 {\bar a} {\bar b}+{\bar c}) z^2 a^4+16 b z \
\left({\bar b}^2 z-{\bar a}^2\right) a^3
\right.\nonumber\\
&+& \left. 2 z \left(-3 c {\bar a}^2-20 b^2 \
{\bar b} {\bar a}+b^2 {\bar c}+3 {\bar b}^2 c z\right) a^2
\right.\nonumber\\
&+&\left. 4 b \left(4 \
{\bar a}^2 b^2-4 {\bar b}^2 z b^2-4 {\bar a} {\bar b} c z+c {\bar c} \
z\right) a+6 {\bar a}^2 b^2 c+b^4 {\bar c}-6 b^2 {\bar b}^2 c z
\right. \nonumber\\
&+& \left. c^2 \
{\bar c} z+4 {\bar a} {\bar b} \left(3 b^4-c^2 z\right)\right)
\nonumber\\
\delta_{R,2}^{(
8
)}&=&
-\frac{8}{z^2} \left(b^2-a^2 z\right) \left(\left(4 b^3-6 a^2 z b-a c
z\right) \ 
{\bar a}^3+{\bar b} z \left(2 z a^3-8 b^2 a-3 b c\right) {\bar a}^2
\right.\nonumber\\
&+&\left. z \
\left({\bar c} z a^3-8 b {\bar b}^2 z a^2+b^2 {\bar c} a-3 {\bar b}^2 c \
z a+2 b^3 {\bar b}^2+b c {\bar c}\right) {\bar a}
\right.\nonumber\\
&+&\left. {\bar b} z \left(4 \
{\bar b}^2 z^2 a^3+b {\bar c} z a^2-6 b^2 {\bar b}^2 z a+c {\bar c} z \
a+b^3 {\bar c}-b {\bar b}^2 c z\right)\right)
\nonumber\\
\delta_{R,2}^{(
10
)}&=&
\frac{2}{z^2} \left(b^2-a^2 z\right)^2 \left(c {\bar a}^4+8 b^2 {\bar b} \
{\bar a}^3-2 \left(3 {\bar c} b^2+{\bar b}^2 c z\right) {\bar a}^2+4 \
{\bar b} \left(2 b^2 {\bar b}^2+c {\bar c}\right) z {\bar a}
\right.\nonumber\\
&+&\left. 16 a b \
\left({\bar a}^2+{\bar b}^2 z\right)^2+2 a^2 (4 {\bar a} {\bar b}-3 \
{\bar c}) z \left({\bar a}^2+{\bar b}^2 z\right)
\right. \nonumber\\
&+& \left. z \left(c z \
{\bar b}^4-6 b^2 {\bar c} {\bar b}^2-c {\bar c}^2\right)\right)
\nonumber\\
\delta_{R,2}^{(
12
)}&=&
\frac{1}{z^3}\(8 \left(b^2-a^2 z\right)^3 \left({\bar a}^2+{\bar b}^2
z\right) \ 
\left(b {\bar a}^3+3 a {\bar b} z {\bar a}^2+\left(3 b {\bar b}^2-a \
{\bar c}\right) z {\bar a}
\right.\right.\nonumber\\
&+& \left.\left. {\bar b} z \left(a {\bar b}^2 z-b \
{\bar c}\right)\right)\)
\nonumber\\
\delta_{R,2}^{(
14
)}&=&
\frac{2 (4 {\bar a} {\bar b}-{\bar c}) \left(b^2-a^2 z\right)^4 \
\left({\bar a}^2+{\bar b}^2 z\right)^2}{z^3}
\nonumber\\
\delta_{R,2}^{(
16
)}&=&
0
\nonumber\\
\delta_{R,3(0)}^{(
1
)}&=&
4 b
\nonumber\\
\delta_{R,3(0)}^{(
3
)}&=&
-4 \left(3 {\bar b} z a^2+4 {\bar a} b a-b^2 {\bar b}+{\bar a} c\right)
\nonumber\\
\delta_{R,3(0)}^{(
5
)}&=&
\frac{4}{z} \left(\left(-b^3+9 a^2 z b+4 a c z\right) {\bar a}^2+8 a {\bar b} \
z \left(a^2 z-b^2\right) {\bar a}
\right.\nonumber\\
&+&\left. z \left({\bar c} z a^3+15 b {\bar b}^2 \
z a^2-3 b^2 {\bar c} a+6 {\bar b}^2 c z a-3 b^3 {\bar b}^2-b c \
{\bar c}\right)\right)
\nonumber\\
\delta_{R,3(0)}^{(
7
)}&=&
-\frac{1}{z}\( 4 \left(z^2 \left(11 z {\bar b}^3+3 {\bar a}^2 {\bar b}+4 \
{\bar a} {\bar c}\right) a^4+4 b z \left(4 {\bar a}^3+2 {\bar b}^2 z \
{\bar a}+{\bar b} {\bar c} z\right) a^3
\right.\right.\nonumber\\
&+&\left.\left. z \left(6 c {\bar a}^3+22 b^2 \
{\bar b} {\bar a}^2-14 b^2 {\bar c} {\bar a}-6 b^2 {\bar b}^3 z+3 \
{\bar b} c {\bar c} z\right) a^2
\right.\right.\nonumber\\
&-&\left.\left. 4 b \left(4 b^2 {\bar a}^3-4 {\bar b} c \
z {\bar a}^2+\left(2 b^2 {\bar b}^2+c {\bar c}\right) z \
{\bar a}+{\bar b} z \left(b^2 {\bar c}-2 {\bar b}^2 c z\right)\right) \
a-6 {\bar a}^3 b^2 c
\right.\right.\nonumber\\
&+&\left.\left. {\bar a} {\bar c} \left(6 b^4-c^2 \
z\right)+{\bar a}^2 \left(4 {\bar b} c^2 z-9 b^4 \
{\bar b}\right)+{\bar b} z \left(3 {\bar b}^2 b^4-3 c {\bar c} b^2+2 \
{\bar b}^2 c^2 z\right)\right)\)
\nonumber\\
\delta_{R,3(0)}^{(
9
)}&=&
\frac{1}{z^2}\( 4 \left(a^2 z-b^2\right) \left(\left(-11 b^3+19 a^2 z
    b+4 a c z\right) \ 
{\bar a}^4
\right.\right.\nonumber\\
&+&\left.\left. 8 {\bar b} z \left(-z a^3+3 b^2 a+b c\right) {\bar a}^3
\right.\right.\nonumber\\
&+&\left.\left. z \
\left(6 {\bar c} z a^3+34 b {\bar b}^2 z a^2-12 b^2 {\bar c} a+14 \
{\bar b}^2 c z a-6 b^3 {\bar b}^2-3 b c {\bar c}\right) {\bar a}^2
\right.\right.\nonumber\\
&-&\left.\left. 4 \
{\bar b} z \left(2 {\bar c} b^3-6 a {\bar b}^2 z b^2-2 {\bar b}^2 c z \
b+a z \left(2 a^2 z {\bar b}^2+c {\bar c}\right)\right) {\bar a}
\right.\right.\nonumber\\
&+&\left.\left. z \
\left(\left(2 {\bar c}^2-3 {\bar b}^4 z\right) b^3-12 a {\bar b}^2 \
{\bar c} z b^2+3 {\bar b}^2 z \left(5 a^2 {\bar b}^2 z-c {\bar c}\right) \
b \right.\right.\right. \nonumber\\
&+& \left.\left.\left. 
a z \left(6 c z {\bar b}^4+6 a^2 {\bar c} z {\bar b}^2+c \
{\bar c}^2\right)\right)\right)\)
\nonumber\\
\delta_{R,3(0)}^{(
11
)}&=&
-\frac{4}{z^2} \left(b^2-a^2 z\right)^2 \left({\bar a}^2+{\bar b}^2 z\right) \
\left(c {\bar a}^3+9 b^2 {\bar b} {\bar a}^2-\left(6 {\bar c} \
b^2+{\bar b}^2 c z\right) {\bar a}
\right.\nonumber\\
&-& \left. 3 b^2 {\bar b}^3 z+{\bar b} c \
{\bar c} z
\right.\nonumber\\
&+&\left. a^2 z \left(z {\bar b}^3-7 {\bar a}^2 {\bar b}+4 {\bar a} \
{\bar c}\right)+4 a b \left(3 {\bar a}^3+{\bar b}^2 z {\bar a}+{\bar b} \
{\bar c} z\right)\right)
\nonumber\\
\delta_{R,3(0)}^{(
13
)}&=&
\frac{4 \left(a^2 z-b^2\right)^3 \left({\bar a}^2+{\bar b}^2 z\right)^2 \
\left(3 b {\bar a}^2+b {\bar b}^2 z+a {\bar c} z\right)}{z^3}
\nonumber\\
\delta_{R,3(0)}^{(
15
)}&=&
-\frac{4 {\bar b} \left(b^2-a^2 z\right)^4 \left({\bar a}^2+{\bar b}^2 \
z\right)^3}{z^3}
\nonumber\\
\delta_{R,3(1)}^{(
1
)}&=&
2 b
\nonumber\\
\delta_{R,3(1)}^{(
3
)}&=&
2 \left(-3 {\bar b} z a^2+5 b^2 {\bar b}+{\bar a} c\right)
\nonumber\\
\delta_{R,3(1)}^{(
5
)}&=&
\frac{2}{z} \left(\left(3 b^3-7 a^2 z b-2 a c z\right) {\bar a}^2+4
  {\bar b} z \ 
\left(z a^3+b^2 a+b c\right) {\bar a}
\right.\nonumber\\
&+&\left. z \left({\bar c} z a^3-3 b^2 \
{\bar c} a+9 b^3 {\bar b}^2-b \left(9 a^2 z {\bar b}^2+c \
{\bar c}\right)\right)\right)
\nonumber\\
\delta_{R,3(1)}^{(
7
)}&=&
\frac{1}{z}\(2 \left(z^2 \left(z {\bar b}^3+5 {\bar a}^2 {\bar b}-2 {\bar a} \
{\bar c}\right) a^4+8 b z \left({\bar a}^3-{\bar b}^2 z \
{\bar a}+{\bar b} {\bar c} z\right) a^3
\right.\right.\nonumber\\
&-&\left.\left. z \left(6 b^2 z {\bar b}^3+6 \
{\bar a} c z {\bar b}^2+2 {\bar a}^2 b^2 {\bar b}-3 c {\bar c} z \
{\bar b}+2 {\bar a} b^2 {\bar c}\right) a^2
\right.\right.\nonumber\\
&-&\left.\left. 4 b \left(2 b^2 {\bar a}^3-2 \
{\bar b} c z {\bar a}^2-2 b^2 {\bar b}^2 z {\bar a}+c {\bar c} z \
{\bar a}+2 b^2 {\bar b} {\bar c} z\right) a+b^2 {\bar b} \left(5 b^2 \
{\bar b}^2-3 c {\bar c}\right) z
\right.\right.\nonumber\\
&+&\left.\left. {\bar a} c \left(6 b^2 {\bar b}^2-c \
{\bar c}\right) z+{\bar a}^2 {\bar b} \left(5 b^4+2 c^2 \
z\right)\right)\)
\nonumber\\
\delta_{R,3(1)}^{(
9
)}&=&
-\frac{1}{z^2}\( 2 \left(b^2-a^2 z\right) \left(\left(b^3+3 a^2 z b+2
a c z\right) \ 
{\bar a}^4-4 {\bar b} z \left(2 z a^3+b c\right) {\bar a}^3
\right.\right.\nonumber\\
&+&\left.\left. z \left(6 \
{\bar b}^2 b^3+6 a {\bar c} b^2+3 c {\bar c} b-10 a^2 {\bar b}^2 z b-2 a \
{\bar b}^2 c z\right) {\bar a}^2
\right.\right.\nonumber\\
&+&\left.\left. 4 {\bar b} z^2 \left(-2 {\bar b}^2 z \
a^3+2 b {\bar c} a^2+c {\bar c} a-b {\bar b}^2 c\right) {\bar a}
\right.\right.\nonumber\\
&+&\left.\left. z^2 \
\left(5 b^3 {\bar b}^4+6 a b^2 {\bar c} {\bar b}^2-a c {\bar c}^2+b \
\left(-5 a^2 z {\bar b}^4+3 c {\bar c} {\bar b}^2-2 a^2 \
{\bar c}^2\right)\right)\right)\)
\nonumber\\
\delta_{R,3(1)}^{(
11
)}&=&
-\frac{2}{z^2} \left(b^2-a^2 z\right)^2 \left({\bar a}^2+{\bar b}^2 z\right) \
\left(c {\bar a}^3+9 b^2 {\bar b} {\bar a}^2-{\bar b}^2 c z {\bar a}+8 \
a b \left({\bar a}^2+{\bar b}^2 z\right) {\bar a}
\right.\nonumber\\
&+&\left. 9 b^2 {\bar b}^3 \
z+{\bar b} c {\bar c} z+a^2 z \left(-3 z {\bar b}^3+{\bar a}^2 \
{\bar b}-2 {\bar a} {\bar c}\right)\right)
\nonumber\\
\delta_{R,3(1)}^{(
13
)}&=&
\frac{2 \left(a^2 z-b^2\right)^3 \left({\bar a}^2+{\bar b}^2 z\right)^2 \
\left(3 b {\bar a}^2+4 a {\bar b} z {\bar a}+5 b {\bar b}^2 z-a \
{\bar c} z\right)}{z^3}
\nonumber\\
\delta_{R,3(1)}^{(
15
)}&=&
-\frac{2 {\bar b} \left(b^2-a^2 z\right)^4 \left({\bar a}^2+{\bar b}^2 \
z\right)^3}{z^3}
\nonumber
\er

\section{Simpler forms of one-soliton of section  \ref{subsec:mass1}}
\label{app:truncation}

Here we present simpler forms of the one-soliton solution discussed in section
\ref{subsec:mass1}, and for which the coefficients of the tau
functions are given in appendix \ref{app:fullsol}. 

The choice of  the parameters $a$, ${\bar a}$, $b$,
${\bar b}$, 
$c$, and ${\bar c}$, appearing the coefficients $\delta^{(l)}$'s  given in
appendix \ref{app:fullsol}, for which the solution simplifies, and the
corresponding non-vanishing tau functions, are given by 
\begin{enumerate}
\item For $b={\bar b}=c={\bar c}=0$, $a=a_1/4$ and ${\bar a}={\bar
    a}_1/4$ we get
\br
\tau_0 &=& \(1-\frac{a_1 \,{\bar a}_1}{16}\; e^{2\;\Gamma_1}\)^8 
\nonu\\
\tau_1 &=& \(1-\frac{a_1 \,{\bar a}_1}{16}\; e^{2\;\Gamma_1}\)^6
\(1+\frac{a_1 \,{\bar a}_1}{16}\; e^{2\;\Gamma_1}\)^2 
\nonu\\
{\tilde \tau}_{L,1} &=& a_1 \; e^{\Gamma_1}\; \(1-\frac{a_1 \,{\bar
    a}_1}{16}\; e^{2\;\Gamma_1}\)^6 
\(1+\frac{a_1 \,{\bar a}_1}{16}\; e^{2\;\Gamma_1}\)
\nonu\\
\tau_{R,1} &=& {\bar a}_1 \; e^{\Gamma_1}\; \(1-\frac{a_1 \,{\bar
    a}_1}{16}\; e^{2\;\Gamma_1}\)^6 
\(1+\frac{a_1 \,{\bar a}_1}{16}\; e^{2\;\Gamma_1}\)
\nonu
\er
\item For $a={\bar a}=c={\bar c}=0$, $b=b_1/2$ and ${\bar b}={\bar
    b}_1/2$ we get
\br
\tau_0 &=& \(1-\frac{b_1 \,{\bar b}_1}{4}\; e^{2\;\Gamma_1}\)^4
\(1+\frac{b_1 \,{\bar b}_1}{4}\; e^{2\;\Gamma_1}\)^4 
\nonu\\
\tau_1 &=& \(1-\frac{b_1 \,{\bar b}_1}{4}\; e^{2\;\Gamma_1}\)^2
\(1+\frac{b_1 \,{\bar b}_1}{4}\; e^{2\;\Gamma_1}\)^6 
\nonu\\
{\tilde \tau}_{L,3(0)} &=& 2 \,{\bar b}_1 \; e^{\Gamma_1}\; 
\(1-\frac{b_1 \,{\bar b}_1}{4}\; e^{2\;\Gamma_1}\)^3 \(1+\frac{b_1
  \,{\bar b}_1}{4}\; e^{2\;\Gamma_1}\)^4 
\nonu\\
{\tilde \tau}_{L,3(1)} &=& {\bar b}_1 \; e^{\Gamma_1}\; 
\(1-\frac{b_1 \,{\bar b}_1}{4}\; e^{2\;\Gamma_1}\) \(1+\frac{b_1
  \,{\bar b}_1}{4}\; e^{2\;\Gamma_1}\)^6 
\nonu\\
\tau_{R,3(0)} &=& 2 \, b_1 \; e^{\Gamma_1}\; 
\(1-\frac{b_1 \,{\bar b}_1}{4}\; e^{2\;\Gamma_1}\)^3 \(1+\frac{b_1
  \,{\bar b}_1}{4}\; e^{2\;\Gamma_1}\)^4 
\nonu\\
\tau_{R,3(1)} &=&  b_1 \;e^{\Gamma_1}\; 
\(1-\frac{b_1 \,{\bar b}_1}{4}\; e^{2\;\Gamma_1}\) \(1+\frac{b_1
  \,{\bar b}_1}{4}\; e^{2\;\Gamma_1}\)^6 
\nonu
\er
\item For $a={\bar a}=b={\bar b}=0$, $c=c_1/2$ and ${\bar c}={\bar
    c}_1/2$ we get
\br
\tau_0 &=& \(1-\frac{c_1 \,{\bar c}_1}{4}\; e^{4\;\Gamma_1}\)
\(1+\frac{c_1 \,{\bar c}_1}{4}\; e^{4\;\Gamma_1}\) 
\nonu\\
\tau_1 &=& \(1-\frac{c_1 \,{\bar c}_1}{4}\; e^{4\;\Gamma_1}\)^2
\nonu\\
{\tilde \tau}_{L,2} &=& {\bar c}_1 \; e^{2\;\Gamma_1} \; \(1-\frac{c_1
  \,{\bar c}_1}{4}\; e^{4\;\Gamma_1}\) 
\nonu\\
\tau_{R,2} &=& c_1 \; e^{2\;\Gamma_1} \; \(1-\frac{c_1 \,{\bar
    c}_1}{4}\; e^{4\;\Gamma_1}\) 
\nonu
\er
\item For $b={\bar b}=0$, $a= a_1/4$, ${\bar a}= {\bar a}_1/4$
  $c=c_1/2$ and ${\bar c}={\bar c}_1/2$ we get
\br
\tau_0 &=&\left[\(1-\frac{a_1 {\bar a}_1}{16} \; e^{2\;\Gamma_1}\)^4 -
  \frac{c_1 {\bar c}_1}{4}\; e^{4\;\Gamma_1}\right] 
\left[\(1-\frac{a_1 {\bar a}_1}{16} \; e^{2\;\Gamma_1}\)^4 + \frac{c_1
    {\bar c}_1}{4}\; e^{4\;\Gamma_1}\right] 
\nonu\\
\tau_1 &=&
\left[\(1-\frac{a_1 {\bar a}_1}{16} \; e^{2\;\Gamma_1}\)^3
  \(1+\frac{a_1 {\bar a}_1}{16} \; e^{2\;\Gamma_1}\)-  
\frac{c_1 {\bar c}_1}{4}\; e^{4\;\Gamma_1}\right]^2
\nonu\\
{\tilde \tau}_{L,1} &=& a_1 \; e^{\Gamma_1}\(1-\frac{a_1 {\bar
    a}_1}{16} \; e^{2\;\Gamma_1}\)^3  \times \nonumber\\
&& \left[\(1-\frac{a_1 {\bar a}_1}{16} \; e^{2\;\Gamma_1}\)^3
  \(1+\frac{a_1 {\bar a}_1}{16} \; e^{2\;\Gamma_1}\)-  
\frac{c_1 {\bar c}_1}{4}\; e^{4\;\Gamma_1}\right]
\nonu\\
{\tilde \tau}_{L,2} &=& e^{2\;\Gamma_1}\( {\bar c}_1 
\left[\(1-\frac{a_1 {\bar a}_1}{16} \; e^{2\;\Gamma_1}\)^4 - \frac{c_1
    {\bar c}_1}{4}\; e^{4\;\Gamma_1}\right]  
\right. \nonu\\
&+& \left.  
\frac{c_1 \, {\bar a}_1^4}{256 \, z} \; e^{4\;\Gamma_1} 
\left[\(1-\frac{a_1 {\bar a}_1}{16} \; e^{2\;\Gamma_1}\)^4 + \frac{c_1
    {\bar c}_1}{4}\; e^{4\;\Gamma_1}\right]\) 
\nonu\\
{\tilde \tau}_{L,3(0)} &=& \frac{1}{2} e^{3\;\Gamma_1}\( a_1 {\bar c}_1 
\left[\(1-\frac{a_1 {\bar a}_1}{16} \; e^{2\;\Gamma_1}\)^4 - \frac{c_1
    {\bar c}_1}{4}\; e^{4\;\Gamma_1}\right] 
\right. \nonu\\
&+& \left. \frac{{\bar a}_1^3 c_1}{16 z} \; e^{2\;\Gamma_1} 
\left[\(1-\frac{a_1 {\bar a}_1}{16} \; e^{2\;\Gamma_1}\)^4 + \frac{c_1
    {\bar c}_1}{4}\; e^{4\;\Gamma_1}\right]\) 
\nonu\\
{\tilde \tau}_{L,3(1)} &=& -\frac{1}{4}\; e^{3\;\Gamma_1}\(a_1 {\bar
  c}_1 - \frac{c_1 {\bar a}_1^3}{16 z} e^{2\;\Gamma_1}\) 
\times \nonumber\\ 
&& \left[\(1-\frac{a_1 {\bar a}_1}{16} \; e^{2\;\Gamma_1}\)^3
  \(1+\frac{a_1 {\bar a}_1}{16} \; e^{2\;\Gamma_1}\)-  
\frac{c_1 {\bar c}_1}{4}\; e^{4\;\Gamma_1}\right]
\nonu\\
\tau_{R,1} &=& {\bar a}_1 e^{\Gamma_1}\(1-\frac{a_1 {\bar a}_1}{16} \;
e^{2\;\Gamma_1}\)^3 \times \nonumber\\
&& \left[\(1-\frac{a_1 {\bar a}_1}{16} \; e^{2\;\Gamma_1}\)^3
  \(1+\frac{a_1 {\bar a}_1}{16} \; e^{2\;\Gamma_1}\)-  
\frac{c_1 {\bar c}_1}{4}\; e^{4\;\Gamma_1}\right]
\nonu\\
\tau_{R,2} &=& e^{2\;\Gamma_1}\( c_1 
\left[\(1-\frac{a_1 {\bar a}_1}{16} \; e^{2\;\Gamma_1}\)^4 - \frac{c_1
    {\bar c}_1}{4}\; e^{4\;\Gamma_1}\right] 
\right. \nonu\\
&-& \left. \frac{{\bar c}_1 a_1^4 z}{256}\; e^{4\;\Gamma_1}
\left[\(1-\frac{a_1 {\bar a}_1}{16} \; e^{2\;\Gamma_1}\)^4 + \frac{c_1
    {\bar c}_1}{4}\; e^{4\;\Gamma_1}\right]\) 
\nonu\\
\tau_{R,3(0)} &=& -\frac{1}{2} \; e^{3\;\Gamma_1}\( {\bar a}_1 c_1 
\left[\(1-\frac{a_1 {\bar a}_1}{16} \; e^{2\;\Gamma_1}\)^4 - \frac{c_1
    {\bar c}_1}{4}\; e^{4\;\Gamma_1}\right] 
\right. \nonu\\
&-& \left. \frac{a_1^3  {\bar c}_1 z}{16}\; e^{2\;\Gamma_1} 
\left[\(1-\frac{a_1 {\bar a}_1}{16} \; e^{2\;\Gamma_1}\)^4 + \frac{c_1
    {\bar c}_1}{4}\; e^{4\;\Gamma_1}\right]\) 
\nonu\\
\tau_{R,3(1)} &=& \frac{1}{4} \; e^{3\;\Gamma_1}\left[ {\bar a}_1 c_1 + 
\frac{a_1^3  {\bar c}_1 z}{16}\; e^{2\;\Gamma_1}\right]
\times \nonumber\\
&& \left[\(1-\frac{a_1 {\bar a}_1}{16} \; e^{2\;\Gamma_1}\)^3
  \(1+\frac{a_1 {\bar a}_1}{16} \; e^{2\;\Gamma_1}\)-  
\frac{c_1 {\bar c}_1}{4}\; e^{4\;\Gamma_1}\right]
\nonu
\er
\end{enumerate}

\vspace{1cm}

\noindent {\bf Acknowledgements:} P.E.G. Assis is supported by a FAPESP 
scholarship, and \\
L. A. Ferreira is partially supported by a CNPq grant.

\end{document}